\documentclass{aa}  

\usepackage{graphicx}
\usepackage{txfonts}
\usepackage{lscape}
\usepackage{longtable}
\begin{document}

   \title{Star formation history of Canis Major OB1}

   \subtitle{II. A bimodal X-ray population revealed by {\it XMM-Newton}}

   \author{T. Santos-Silva\inst{1,2,3}, J. Gregorio-Hetem\inst{2}, T. Montmerle\inst{3}, B. Fernandes\inst{2},\inst{3} \and B. Stelzer\inst{4,5}
          }
          
             \offprints{T. Santos-Silva}

\institute{Universidade Federal do Rio Grande do Sul, Brazil,\\
              \email{thais.santos@ufrgs.br}
         \and
             Universidade de S\~ao Paulo, IAG, Departamento de Astronomia, Brazil,
             \and
             Institut d'Astrophysique de Paris, France
         \and
             Eberhard-Karls Universit\"at, Institut f\"ur Astronomie und Astrophysik, Sand 1, D-72076 T\"ubingen, Germany 
         \and INAF - Osservatorio Astronomico di Palermo, Piazza del Parlamento 1, 90134 Palermo, Italy\\
             }


 
  \abstract
  {} 
  { The Canis Major OB1 Association has an intriguing scenario of star formation, especially in the region called Canis Major R1 (CMa R1) traditionally assigned to a reflection nebula, but in reality an ionized region. This work is focused on the young stellar population associated to CMa R1, for which our previous results from ROSAT, optical and near-infrared data had revealed two stellar groups with different ages, suggesting a possible mixing of populations originated from distinct star-formation episodes.}
  {The X-ray data allow the detected sources to be characterized according to hardness ratios, light curves and spectra. Estimates of mass and age were obtained from the {\it 2MASS} catalogue, and used to define a complete subsample of stellar counterparts, for statistical purposes.}
  {A catalogue of 387 {\it XMM-Newton} sources is provided, 78\% being confirmed as members or probable members of the CMa R1 association.  Flares (or similar events) were observed for 13 sources, and the spectra of 21 bright sources could be fitted by a thermal plasma model. Mean values of fits parameters were used to estimate X-ray luminosities. We found a minimum value of log(L$_{X}$[erg/s]) = 29.43, indicating that our sample of low-mass stars ($M_\star \leq 0.5 M_\odot$), being faint X-ray emitters, is incomplete. Among the  250 objects selected as our complete subsample (defining our  ``best sample''),  171 are found to the East of the cloud, near Z CMa and dense molecular gas, 50\% of them being young (< 5 Myr) and 30\% being older (> 10 Myr). The opposite happens to the West, near GU CMa, in areas lacking molecular gas: among 79 objects, 30\% are young and 50\% are older. These findings confirm that a first episode of distributed star formation occurred in the whole studied region $\sim$10 Myr ago and dispersed the molecular gas, while a second, localized episode ($<$ 5 Myr) took place in the regions where molecular gas is still present.}
  {}
%
  {}


\maketitle
%


\section{Introduction}


The efficiency of using X-ray observations to discover large samples of pre-main sequence 
stars  has been demonstrated 
in  many star-forming regions. For instance, thousands of X-ray sources have been 
identified in Orion Nebula by means of  the
{\it Chandra Orion Ultradeep Project -- COUP} 
\citep{2005ApJS..160..319G,2005ApJS..160..353G} and several hundreds were detected in 
the Taurus Molecular Cloud by means of  the {\it XMM-Newton Extended Survey of Taurus - XEST} \citep{2007A&A...468..353G}.


\begin{figure*}[ht]
\begin{center}
\includegraphics[width=2.0\columnwidth, angle=0]{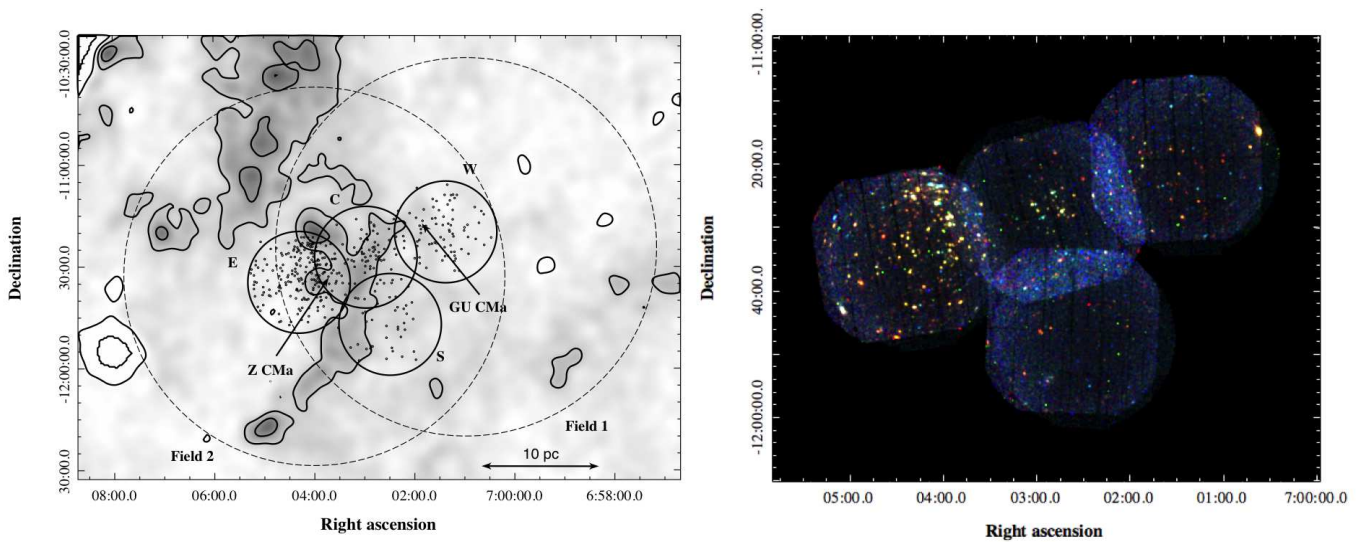}

\end{center}

\caption{ {\it Left}: Spatial distribution of X-ray sources detected by {\it XMM-Newton} (fields E,C,S and W - full lines) on the A$_V$ map (Cambrésy, private communication) with A$_{V}$ = 2.0 and 4.0 mag contours, compared with {\it ROSAT} fields from Gregorio-Hetem et al. (2009) (dashed lines). {\it Right}: Mosaic of images of {\it XMM-Newton} EPIC PN, MOS 1 and MOS 2 for the same CMa fields combined in three different energy bands. The filters red, green and blue present the soft band (0.5-1.0 keV), medium band (1.0-2.0 keV) and hard band (2.0-7.3 keV) respectively.}

\label{fig1}

\end{figure*}


The method is now familiar: the X-rays come from high-mass stars owing to their strong 
wind shocks, and also from the numerous young, low-mass stars ($M_\star < 2 M_\odot$: 
T Tauri stars), owing to their magnetic activity and/or  accretion, 
which may be present up to a relatively old age ($\sim$10 Myr, depending on their mass). Therefore, once 
they have lost their accretion disks (no IR excess, no H$\alpha$ emission: 
 ages $\geq$ 10 Myr), X-rays are the only way to identify the stars as being young. 
 The ages are then derived using colour-magnitude diagrams 
and theoretical Pre-Main Sequence (PMS) evolutionary tracks.

 The discovery of large samples of PMS stars has shown that
many star-forming regions seem to have complex star formation histories, typically involving a sequence of individual  episodes that created different sub-groups or clusters. For an understanding of the star formation process, it is therefore of essential importance to look for young stars not only around the center of a prominent cluster, but also to investigate its environment, in order to detect and identify \textit{distinct stellar populations} that often are found in separate clusters or groups, having different ages.

One prominent example is the Scorpius-Centaurus OB association, where the individual sub-groups have ages between $\sim$5~Myr and $\sim$15~Myr \citep{2008hsf2.book..235P}.
A similar situation, although with smaller age differences of just a few Myr, is seen in the sub-groups of the Ori OB1 association \citep{2008hsf1.book..459B}. The $\sim$4~Myr stars of the Ori OB1c sub-group are seen in projection directly in front of the famous Orion Nebula Cluster and contaminate the cluster.

Another spectacular illustration of the diversity of young populations,  not only in age,  but also in space, is provided by the {\sl Chandra} 22-field mosaic, $\sim$1.4 sq. deg. survey of the Carina nebula \citep{2011ApJS..194....1T,2011ApJS..194....9F}, 
where the $\sim$10,000 X-ray sources detected as young stars clearly belong to two distinct, equally numerous groups: $(i)$ a {\it clustered} population, with centrally concentrated distributions of stars (including cases of ``clusters of clusters''), and a fairly homogeneous  {\it distributed} fainter population. This is consistent with two main modes of star formation: localized cluster formation from dense molecular cores, and more widely spread star formation from smaller condensations (e.g., ``pillars'' eroded by the hydrodynamical feedback from winds and/or supernova explosions from massive stars ($M_\star > 8 M_\odot$), as in the Eagle Nebula \citep{2010A&A...521A..61G,2011A&A...531A..51F}. The various clusters, as well as the distributed young stars  in Carina nebula, also show a large spread in ages ($\sim$1-10 Myr depending on the location within the nebula).

Therefore, it is now clear, in particular from X-ray observations, that a {\it large-scale} picture of young clusters, and not only of a small area around their brightest members, is essential to reliably disentangle age and membership effects -- in particular the  spatial mixing of stars having different formation histories (i.e., originating in distinct clusters or born via various feedback mechanisms at different epochs).

The above discussion raises crucial issues on star formation in clusters: 
 What is the ``real'' membership of a cluster? And, which are the youngest and oldest member stars? 
The answer to the former is needed to build a reliable mass distribution and the 
latter to determine the global age distribution, and hence the duration of star formation 
 in a given region.

These issues are not restricted to massive star forming regions, but may apply to young clusters in general, hence fuel the ongoing debate on the duration of the star formation process \citep[e.g., ][]{2011EAS....51..245P}. Such a statement can be inferred for the CMa R1 star-forming region ($d \sim 1$ kpc) because it is a young complex, with a lot of molecular material still around, and star formation both finished and still ongoing, according to our previous results based on  
the wide field of {\it ROSAT} on the famous arc-shaped ionized nebula Sh2-296 (\citealp{2009A&A...506..711G}, hereafter Paper I ). This nebula, of as yet unclear origin, comprises known young ($\sim$1-5 Myr) clusters, including around the famous binary (FU Ori-type + Herbig) star Z CMa.  Figure \ref{fig1} shows the two fields, covering more than $\sim$2.7 deg$^{2}$, where 98 X-ray sources were detected 
 with {\it ROSAT}. A detection limit of   F$_{X}$ = 5.0 x 10$^{-15}$ erg s$^{-1}$ cm$^{-2}$ (log L$_{X}$ = 29.78) was achieved in Field 1 (exposure time of 20ks),  while observations in Field 2 were less sensitive because of a much shorter exposure (5 ks) (F$_{X}$ = 8.4 x 10$^{-15}$ erg s$^{-1}$ cm$^{-2}$ ; log L$_{X}$ = 30).

 Paper I report the discovery of a previously unknown cluster of  low-mass stars having ages up to $\sim$10 Myr, called the ``GU CMa cluster'', after the name of its brightest member. It is located away from dense molecular material, indicating that star formation has now ceased in its vicinity. 
 In contrast, the existence of several younger clusters on the other side of the CMa R1 molecular cloud, including around Z CMa, has been known for some time as a result of previous surveys at various wavelengths  \citep[see review by][]{2008hsf2.book....1G}. With respect to the molecular cloud, the GU CMa cluster therefore lies opposite to the Z CMa cluster, in a region devoid of molecular gas. The age gradient in each cluster, and the similarity of the fraction of the intermediate-age populations (see Fig. 10 of Paper I) strongly suggests some degree of mixing between the two clusters. The young population in-between Z CMa and GU CMa (the ``inter-cluster'' region) could be really mixed (i.e., coming from two distinct clusters, mixing at their edges), or perhaps be indicative of the existence  of an as yet unnoticed, distributed population on a large scale.

 Results from Paper I motivated our group to propose new X-ray observations of CMa R1, more sensitive than those previously obtained with {\it ROSAT} in order to improve the identification of the entire young stellar population in this region. We have focused on the ``inter-cluster'' region between the GU CMa and Z CMa sub-groups (see {\it XMM-Newton} fields in Fig. \ref{fig1}) aiming to investigate if there is a mixing of their populations.  Besides getting a more complete sample of the CMa R1 young stellar population we  use the properties of X-ray and near-infrared emission of these sources to identify their nature and disentangle the scenario of star formation in this region.

The outline of the paper is as follows. In the next section, we describe the source detection and data reduction of the {\it XMM-Newton} observations, and the X-ray general properties are presented in Sect. 3. Section 4 is dedicated to the infrared analysis based on {\it 2MASS}  and {\it WISE} data. A comparative analysis between X-rays and parameters derived from  the infrared is performed in Sect. 5. Section 6 summarizes the results of this work. The general picture of star formation in CMa R1, extending the results from Paper I, is discussed in Sect. 7 and the main conclusions are presented in Sect. 8. Finally, Appendix A gives details about the analysis of the {\it XMM-Newton} data  and Appendix B gives the complete catalogue of X-ray sources.


\section{Observational data}


For this work, four fields (each about 30 arcmin diameter with some overlap)  were observed with the {\it XMM-Newton} satellite. 
 These fields are located (Fig. \ref{fig1}): inside the arc-shaped ionized nebula, next to Z CMa - {\it Field E} (East); around GU CMa - {\it Field W} (West); and between both - {\it Field C} (Center) and {\it Field S} (South). The central coordinates from each field are given in Table \ref{tab1}. These observations were performed with the EPIC cameras (MOS1, MOS2 and PN) in full frame mode with medium filter. 
The C, W and S fields had  an exposure time without background corrections of about 30\,ks while  field E 
had 40\,ks.

These observations were analyzed with the {\it XMM-Newton Scientific Analysis System} (SAS) version 11.0.0 
software. The calibrated and concatenated events lists were obtained by {\it epproc} and {\it empproc} tasks applied to PN and MOS raw data, respectively. Fields E, W and S were affected by high background activity, 
mainly in observations of the PN camera. In order to maximize the signal-to-noise ratio of weak sources we created lists of 
clean events using the standard procedure for EPIC cameras that  filters the background, removes flares, as well as bad pixels, bad events and reduces  noise\footnote[1]{See the SAS thread at  \it{http://xmm.esac.esa.int/sas/current/documenta-tion/threads/EPIC\_ filterbackground.shtml}.}.
The good time intervals (GTI) of these observations are presented in Table \ref{tab1}.


\begin{table}[h] 
\caption{Observing log for the {\it XMM-Newton} observations of CMa R1.} 
\begin{center}
{\scriptsize
\begin{tabular}{|l|c|c|c|c|}

\hline 
 Field name& ID Obs. & $\alpha$ & $\delta$  &  GTI (ksec) $^{(a)}$  \\ 
 & & (J200) & (J2000) &  PN - MOS  \\ \hline \hline

CMa cluster East & 0654880201 & 07 04 18.3 & -11 27 24.0 & 32 - 35 \\
CMa cluster Center & 0654880101 & 07 02 58.4 & -11 34 44.7 & 30 - 32\\
CMa cluster South & 0654880401 & 07 02 29.5 & -11 47 12.4 &  28 - 32\\
CMa cluster West & 0654880301 & 07 01 23.0 & -11 19 56.6 &  28 - 32 \\\hline

\end{tabular}
}

\label{tab1}

\end{center}
{\scriptsize
(a) Good time interval from {\it XMM-Newton} observations.
}
\end{table}


The source detection was performed in  two steps: individually for each of the three EPIC cameras, and merging them. In both cases, the detection was made in the three energy bands defined by \citet{2011A&A...526A..21B}: soft - S$_{B2011}$ = 0.5 - 1.0 keV, medium - M$_{B2011}$ = 1.0 - 2.0 keV and hard - H$_{B2011}$ = 2.0 - 7.3 keV. In order to distinguish young stars from field objects, different energy bands were tested and these were chosen because they are more efficient  for the detection of (thermal) stellar sources. 
  The images from the three cameras were created in each of these bands and also in  the full energy range (0.5 - 7.3 keV). The right panel of the Fig. \ref{fig1} shows a combined image obtained with EMOSAIC task, illustrating the three energy bands detections in PN+MOS images: S$_{B2011}$ (red), M$_{B2011}$ (green) and H$_{B2011}$ (blue).

As a first step, the data from the three EPIC cameras were analyzed separately, by using the SAS metatask EDETECT\_CHAIN for source detection in each detector. This metatask creates exposure maps used to correct the images for the quantum efficiency, filter transmission and mirror vignetting. It also provides detection maps that are used in order to perform sliding box detection using locally estimated background. These sources are masked and background maps are created. Then, the metatask performs a second detection of sources using the background map and derives the parameters by a maximum likelihood method, for each source.  In order to explore softer energy bands, this procedure was also applied only to the PN data considering the energy bands suggested by  \citet{2001A&A...365L..45H}  as more efficient to detect (non-thermal) extragalactic sources: S$_{H2001}$ = 0.2 - 0.5 keV, M$_{H2001}$ = 0.5 - 2.0 keV and H$_{H2001}$ = 2.0 - 4.5 keV.  However, the tests exploring these softer energy bands  turned out to be less efficient for the detection of (thermal) stellar sources. For this reason we adopted 0.5 keV as lower limit.

As a second step, the images (corrected by exposure map) were divided by the effective area of the respective detector, in order to  take into account the differences in efficiency  of the EPIC-PN and EPIC-MOS detectors. (Then, with the EMOSAIC task, a combined image was created from these images, for each energy band.) Finally, the source detection was performed  from combined image using the EMOSAICPROC metatask, that works similarly to the EDETECT\_CHAIN taking into account the merged data from different observations and, instruments improving the  source statistical significance and enabling detections of weak sources. However, this task considers only sources present in all combined instruments. A double check was applied in the individual images, searching for sources that were missed in the second step due to placement problems, like  falling outside of the field-of-view (FOV),  inside a bad column or near to a gap of one or more of the instruments. In both cases the detection threshold was of maximum likelihood ML $\geq$ 15. Some sources were detected in two fields,  because fields E, W and S are overlapping field C, and in this case we chose the sources with  the highest signal-to-noise ratio.

The detection procedure has provided a catalogue containing 387 sources: 
84 are in Field C, 187 in Field E, 79 in Field W, and 37 in Field S. 
351 of them were detected using the merged images of all three cameras, 
and 36 were detected by only one or two cameras. Table B.1 (Appendix B) lists the X-ray 
detections and parameters for all sources in the four fields studied. These parameters 
are: coordinates (J2000), maximum likelihood (ML) of source detection, count rate (CR) in 
the total energy band (0.5 - 7.3 keV), and hardness ratios 
HR1$_{i}$ $=$ $(M_{i} - S_{i})/(M_{i} + S_{i})$ and 
HR2$_{i}$ = $(H_{i} - M_{i})/(H_{i} + M_{i})$. 
 Here, S$_{i}$, M$_{i}$ and H$_{i}$ are the energy bands 
defined above,  where the index $i$ is related to the energy 
band ranges defined by B2011 and H2001.

 Light curves and spectra were extracted only from EPIC-PN data, using the standard SAS routines\footnote[2]{\it{http://xmm.esac.esa.int/sas/current/documentation/threads/timing. shtml}}$^{,}$\footnote[3]{\it{http://xmm.esac.esa.int/sas/current/documentation/threads/PN\_spe-ctrum\_thread.shtml}}. There are 47 sources in MOS 1/2 FOVs only, but none is bright enough for extraction of spectra and light curves.
 In both procedures the source and background regions were chosen by visual inspection, for each source. For the light curves we adopted the B2011 full energy range  (0.5 -- 7.3 keV) and time bins of 1000 sec. The spectra were obtained in the full standard energy range (0.2 -- 10 keV) suggested by SAS and were analyzed using XSPEC version 12.7.1.


\section{ Results on X-ray properties}


 The X-ray data give us information related to the nature of the source, inferred from 
hardness ratios (HRs) diagrams; parameters derived from spectral analysis, and 
from  the light curve.  Appendix A gives more details on how these parameters were 
obtained  from EPIC/PN data.
 
From HRs diagrams analysis, based on B2011 and H2001 energy bands, we could classify 194 sources by comparing their X-ray emission with two model grids: thermal plasma, APEC (Astrophysical Plasma Emission Code, \citealp{2001ApJ...556L..91S}) and a power-law (PWL)\footnote[4]{https://heasarc.gsfc.nasa.gov/xanadu/xspec/manual/XSmodelPo-werlaw.html} distribution, both multiplied by an absorption photoelectric model (PHABS)\footnote[5]{https://heasarc.gsfc.nasa.gov/xanadu/xspec/manual/XSmodelPha-bs.html}.
The sources that are compatible only with the APEC grid were classified as stellar  objects, only 1/181 (0.5\%) of them do not have IR counterpart. The sources compatible with a power-law may have another origin, probably extragalactic or perhaps compact objects. 
 There are 195 sources that we call “undefined”, 47 of them can be fitted by either model (APEC or PWL), and 148 by neither. According to the results from the infrared ({\it 2MASS+WISE}) data analysis (see Sect. 5), we suggest that 84 sources remain as undefined objects: 23  probably are foreground stars, and 17  have counterparts that are too faint (bad quality data), without confident classification. The other 44 undefined sources (11\% of the {\it XMM-Newton} sample)  do not have {\it 2MASS} data,  probably are background objects, which is in agreement with the expected  10\% of contamination, at this level of sensitivity, mainly due to extragalactic sources \citep{2005ApJS..160..319G}.


\begin{table}[ht] 
\caption{Source classification based on X-ray emission.} 
\begin{center}
{
\begin{tabular}{|l|c|c|c|c|}

\hline

	& 	Stellar$^{a}$		 & 	Other$^{b}$		 & 	\multicolumn{2}{c|}{Undefined}	  	 	\\ \hline 
Field & APEC	 & 	PWL		 & 	(APEC $+$ PWL)$^{c}$ & None$^{d}$\\ \hline \hline
E	 &	102	 &	7	  & 22 &	 56	\\

C	 &	39	 &	-	  & 12 &	 33	\\

S	 &	9	 &	1	  & 6 &	21	\\  

W	 &	31	 &	3	  & 7 &	38	\\ \hline

Total	 &	181	 &	11	 & 47 &	148		\\ \hline

\end{tabular}
}
\label{tab2}
\end{center}
{\scriptsize
(a) Sources fitted by APEC models; 
(b) Sources fitted by  Power-Law models; 
(c) Undefined sources: fitted by both models; 
(d) Undefined sources: fitted by neither model.

}
\end{table}



\subsection{Light Curves and Spectra}


 Magnetically active stars, including T Tauri stars, typically show 
 variable X-ray emission, for example, due to {\it flare-like} events varying from few minutes ($<$ 30 min) to few hours ($>$ 8.5h) and involving a large release of energy (10$^{32}$ to 10$^{35}$ erg/s, see Table \ref{tabA1}) \citep{1999ARA&A..37..363F,2003SSRv..108..577F,2004A&ARv..12...71G,2007A&A...468..463S}. In young stars, particularly T Tauri stars, the light curves of these events appear with different shapes \citep[e.g.,][]{2005ApJS..160..469F,2005ApJS..160..423W,2007A&A...468..485F,2007A&A...468..463S,2010A&A...524A..97L}.

\begin{figure}[t]

\begin{center}
\includegraphics[width=1.0\columnwidth, angle=0]{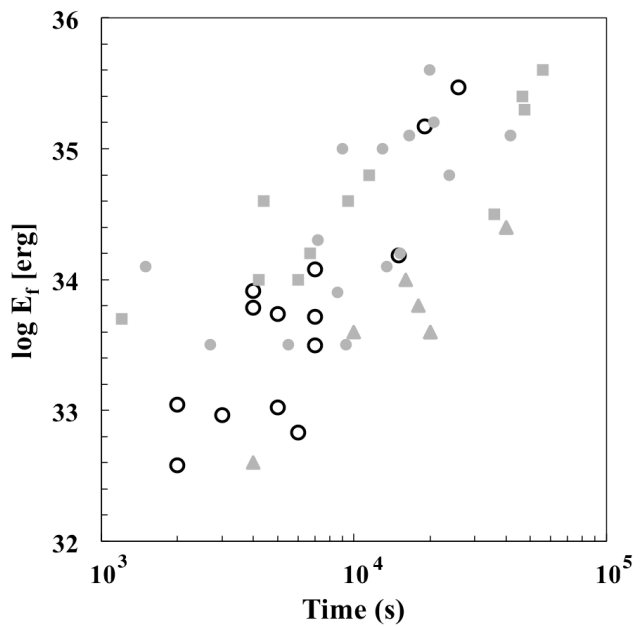}

 \caption{Energy (E$_{f}$) and duration (T$_{f}$) of X-ray flares on 
 CMa R1 sources (open circles),  compared with those observed on 
 members of  {\it Taurus Molecular Cloud} \citep{2007A&A...468..463S}, where 
 Classical-  and Weak- T Tauri are respectively represented by filled circles and 
 squares. Filled triangles show young stellar objects associated with  the 
 ${\eta}$ {\it Chamaleontis cluster} \citep{2010A&A...524A..97L}.}

\label{fig2}
\end{center}
\end{figure}

 Within our four $\sim$ 30 ks exposures, 13 sources presented {\it flare-like} events, 
which we identified using the definition of a flare adopted from \citet{2007A&A...468..485F} and \citet{2005ApJS..160..423W}, as described in Appendix A.2, which also presents all the light curves  and  flare parameters for these sources. The characteristics of {\it flare-like} events in this work are similar to T Tauri stars of the {\it Taurus Molecular Cloud} \citep[e.g.,][]{2007A&A...468..463S}, or young stellar objects from the ${\eta}$ {\it Chamaleontis} cluster \citep{2010A&A...524A..97L}. This similarity is shown in Fig. \ref{fig2} that compares the energy and flare duration for our sample and these  other regions.

In spite of the poor signal-to-noise ratio, we could perform the spectral fits of 
low resolution integrated spectrum for the whole exposure of some sources. 
 The hydrogen column density, plasma temperature and flux for non-flaring 
sources were obtained  by fits of {\sc phabs $\times$ apec} models for 21 sources, adopting 
 metallicity Z = 0.2 Z$_{\odot}$, as appropriate for low-mass stars (see details in Appendix A.1). These spectra and parameters are presented in Appendix A.3. Their average hydrogen column density, N$_{H}$ $=$ 1.8 $\pm$ 1.5 x 10$^{21}$cm$^{-2}$, corresponds to an extinction A$_{V}$ = 0.9 $\pm$ 0.7 mag, by adopting N$_{H}$/A$_{V}$ = 2.1 $\times$ 10$^{21}$ cm$^2$ \citep[e.g.,][]{2003A&A...408..581V}. This value is compatible with  A$_{V}$ = 1.0 mag adopted in Paper I  for CMa R1 and it is inside the range in which most of the sources are compatible with APEC model grids. The coronal temperatures,  varying from 0.5 to 2.1 keV (6.7 <  log T(K) < 7.4), are also compatible with those found in other star formation regions like  $\sigma $ Orionis, $\eta $ Chamaleontis \citep{2008A&A...491..961L,2010A&A...524A..97L} and the {\it Pipe Nebula} \citep{2010ApJ...719..691F}. Note that this means a low-extinction region, if compared with the minimum values for  N$_{H}$ corresponding to A$_{V}$ = 0.4 for the foreground extinction in the direction of CMa R1, computed from the large-scale extinction models by \citet{2005AJ....130..659A}, N$_{H}$ = 5.9 x 10$^{21}$cm$^{-2}$ from LAB Map \citep{2005yCat.8076....0K} and 6.9 x 10$^{21}$cm$^{-2}$ from DL Map \citep{1990ARA&A..28..215D}.

\begin{figure}[t]
\begin{center}
\includegraphics[width=1.0\columnwidth, angle=0]{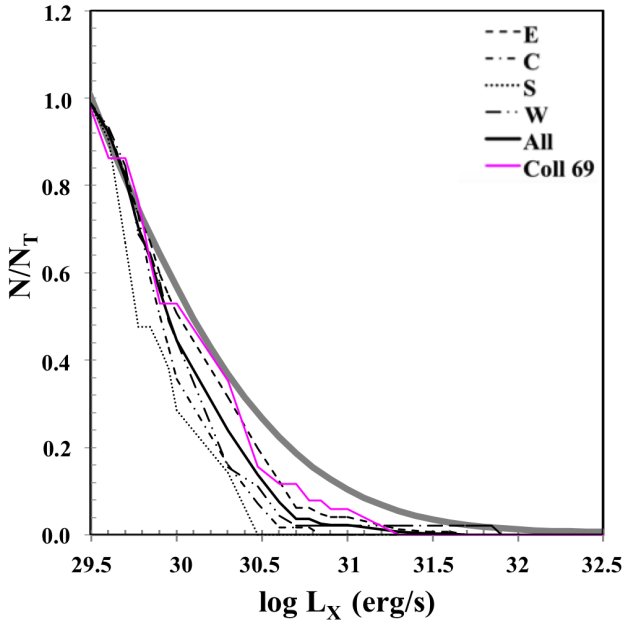}

\caption{Cumulative distribution of X-ray luminosities for sources in CMa R1 (black lines) and  Collinder 69 \citep[][magenta line]{2011A&A...526A..21B}. A thick grey line shows the log-normal distribution with $\mu$= 29.3 and $\sigma$=1 proposed by \citet{2005ApJS..160..379F}.
}

\label{fig3}
\end{center}
\end{figure}


\subsection{X-ray fluxes and luminosities}


Based on the flux derived from the spectral fits of the brightest sources, we obtained the energy conversion factor (ECF) from the correlation between count-rate and flux. The linear fits of these parameters gives a mean value of EFC~=~(1.60~$\pm$~0.04) $\times$ 10$^{-12}$ erg cm$^{-2}$ cts$^{-1}$ compatible with ECF = 1.56 $\times$ 10$^{-12}$ erg cm$^{-2}$ cts$^{-1}$ calculated by 
PIMMS\footnote[6]{http://heasarc.gsfc.nasa.gov/cgi-bin/Tools/w3pimms/w3pimms.pl} for a 1-T thermal model adopting the mean values for plasma temperature and N$_{H}$, derived from the EPIC/PN spectra (see Sect. 3.1).

The X-ray luminosities (L$_{X}$) were derived from the fluxes  for all the 340 sources detected by PN camera, by adopting a distance of $d = 1$ kpc for CMa R1 (\citealp{1999MNRAS.310..210S,2000MNRAS.312..753K},Paper I). Figure \ref{fig3} shows the cumulative $L_{X}$ Function ($L_{X}F$) derived for each field.  As further discussed (Sect.5), the presence of possible non-members contributes with low levels of X-ray emission, not affecting the $L_{X}F$. A thick line shows $L_{X}F$ for all sources, compared to the  5\,Myr-old cluster Collinder 69  \citep{2011A&A...526A..21B} represented by the magenta line in Fig. \ref{fig3}\footnote[7]{ Collinder 69 is a 5 Myr cluster for which count rate and hardness ratios were estimated by using  the same energy range that was adopted by us. This similarity of studying method was the reason for comparing the X-ray properties of our sample with this specific cluster.}.  For both samples, which data have similar sensitivity, the L$_{X}$F cumulative distribution was derived taking into account detected sources with log L$_{X}$(erg/s) $>$ 29.5. As a guidance, we add in Fig. \ref{fig3} the log-normal distribution with $\mu$= 29.3 and $\sigma$=1, the ``Universal X-ray luminosity function'' proposed by
\citet{2005ApJS..160..379F}. The differences on the distribution are found in the high luminosities end, mainly due to the low number of massive stars in our sample. This distribution is similar to the results for other low-mass clusters, as Serpens and NGC 1333, for instance, discussed by \citet{2012AJ....144..101G}.


\section{Analysis of infrared properties}


 The characterization of  the X-ray sources needs to be complemented by observational data obtained in other wavelengths. For instance, \citet{2015MNRAS.448..119F} performed with the {\it Gemini South telescope} a spectroscopic follow-up of optical counterparts  for a partial sample of the {\it XMM-Newton} stellar sources associated with the Sh 2-296 nebula. Among 58 candidates, they found 41 confirmed T Tauri stars and 15 possible PMS stars (including intermediate-mass stars). Almost 50\% of the young stars have less than 1 M$_{\odot}$ and 35\% have masses between 1-2 M$_{\odot}$. While half of  their sample has an age of 1-2 Myr or less, only a small fraction (<10\%) shows evidence of IR excess indicating the presence of circumstellar disks. In comparison with other young star-forming regions \citep[e.g.,][]{2001ApJ...553L.153H,2008ApJ...686.1195H,2010A&A...510A..72F}, this is a very low
fraction of disk-bearing stars.

 In order to expand the search for disk candidates among the X-ray sources associated with CMa R1, we analyse the near-infrared (NIR) counterparts of these sources using available data in the {\it 2MASS} and {\it Wide-field Infrared Survey Explorer} \citep[{\it WISE};][]{2010AJ....140.1868W}  Catalogues. The method used to estimate the infrared properties is described in Sect. 4.1, where we identify the NIR counterparts based on {\it 2MASS} data, and use colour-colour and colour-magnitude diagrams to determine mass and age of the candidates. In the remainder of this section, we describe the IR classification (Sect. 4.2) determined with data from the {\it AllWISE} catalogue; the selection of a ``best sample'', by adopting mass and age criteria (Sect. 4.3); the analysis of mass function (Sect. 4.4) and age distribution (Sect. 4.5).


\subsection{Near-infrared counterparts}

\begin{figure*}[t]
\begin{center}
\includegraphics[width=2.0\columnwidth, angle=0]{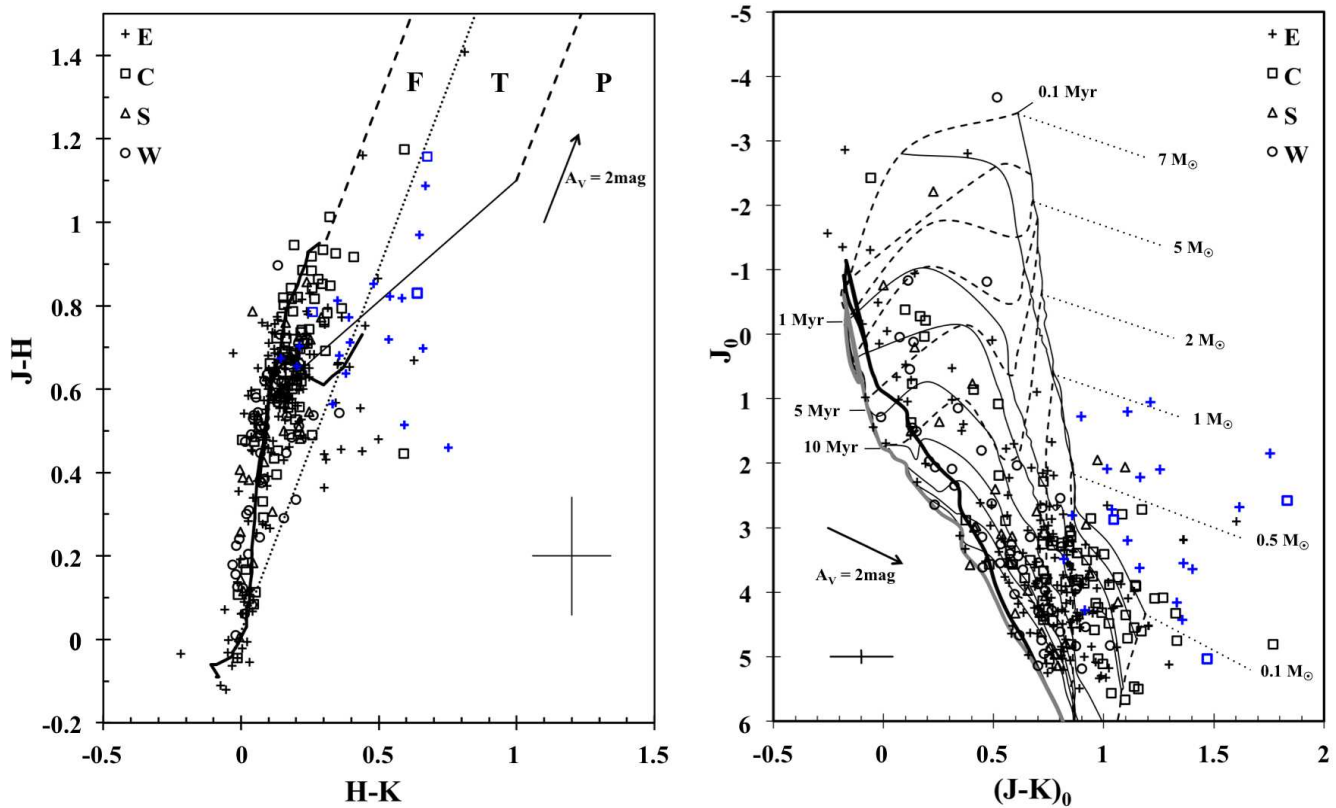}

\caption{{\it Left}: Colour-colour diagram for all 340 {\it 2MASS} counterparts of X-ray sources found in fields E, C, S and W.  Blue symbols represent disk-bearing (Class I and II) stars classified in Sect. 4.2. The ZAMS and the locus of giant stars  are indicated by thick lines, while the locus of T Tauri stars  is represented by a thin line. Dotted and dashed lines show reddening vectors. {\it Right}: Colour-magnitude diagram showing the isochrones 0.2, 1, 5, 10, 15, 20 Myr and ZAMS (full lines), early Main Sequence (grey thick line) and evolutionary pre-MS tracks 0.1, 0.5, 1, 2, 3, 4, 5, 6 and 7 M$_{\odot}$ (dashed line) from  \citet{2000A&A...358..593S}.}

\label{figB1}
\end{center}
\end{figure*}

 We selected NIR counterparts by searching the {\it 2MASS} catalogue \citep{2003yCat.2246....0C} for candidates located less than $10''$ away from the nominal X-ray source positions. No counterpart was found for 45 sources. Candidates for which the distance seems to be incompatible with the cloud were disregarded. In the colour-magnitude diagram (see Fig. \ref{figB1}), these sources appear below the Main Sequence, indicating they probably are field stars.

 The complete list of NIR counterpart candidates is given in Table B.2, but we consider as reliable only those with AAA flags in the {\it 2MASS} catalogue, {\it i. e.,} magnitudes with high signal-noise ratio (S/N $>$ 10), low errors ($<$ 0.1 mag)  and above the completeness limit (J $<$ 15.8, H $<$ 15.1 and K$_{s}<$ 14.7) ensuring good photometric quality \citep{2005ApJ...624..808L}. Table \ref{tab_count} gives the number of X-ray sources for each field and the corresponding number of their  reliable NIR counterparts. Almost all are found less than $5''$ away from the centroid of the X-ray emission. This value has been used typically as a good estimate of the effective radius within $\sim$90\% confidence of the uncorrected positions \citep{2008A&A...491..961L,2003AN....324...89W}.


\begin{table}[h] 
\caption{Number of sources with one (1) or more (2, 3) {\it 2MASS} counterparts.} 
\begin{center}
{

\begin{tabular}{|l|c|c|c|c|c|c|}

\hline 
Field	 & 	1	 & 	2	 & 	3	  & 	Total X$^{a}$	 & 	Total IR$^{b}$	\\ \hline\hline
E	 & 	121	 & 	30	 & 	2	 & 	153	 & 	187	\\
C	 & 	53	 & 	10	 & 	1	& 	64	 & 	76	\\
S	 & 	23	 & 	1	 & 	1	& 	25	 & 	28	\\
W	 & 	47	 & 	1	 & 	-     & 	48	 & 	49	\\ \hline
Total	 & 	244	 & 	42	 & 	4 & 	290	 & 	340	\\\hline

\end{tabular}
}
\label{tab_count}
\end{center}
{\scriptsize
(a) Total number of X-ray sources with {\it 2MASS} counterparts for each observed field; 
(b) The sum of NIR counterparts, by considering single and multiple candidates.}

\end{table}


 In total, we selected 340 reliable NIR counterparts to 290 X-ray sources, among them 46 have multiple counterpart candidates. Following \citet{2012A&A...547A.107S}, we compare their positions in the colour-colour diagram (Fig. \ref{figB1} left) with theoretical curves of the zero age main-sequence (ZAMS) from \citet{2000A&A...358..593S}\footnote[8]{http://www.astro.ulb.ac.be/$\sim$siess/pmwiki/pmwiki.php/WWWTools/ Plots}; giants \citep{1988PASP..100.1134B}. Figure \ref{figB1} (left) also includes an arrow that represents the reddening vector of A$_V$ =2 mag \citep{1985ApJ...288..618R}, the T Tauri stars locus \citep{1997AJ....114..288M}, and the regions defined by \citet{2011MNRAS.411.2530J} according to the NIR excess: stars with accretion disks are expected to be found in region ``T'' (Classical T Tauri stars: CTTS) and Class I protostars appear in region ``P''. Field stars and diskless T Tauri stars (Weak T Tauri stars: WTTS), having little or no excess, are mainly located in region ``F''. It can be noted that most of the counterparts  are found in region F and/or near the ZAMS, indicating a low level of NIR excess for these sources.

 Figure \ref{figB1} (right) shows the colour-magnitude diagram (CMD) with reddening corrections made according to the extinction law from \citet{1989ApJ...345..245C}  for A$_{V}$ = 0.9 mag which is the mean value in the visual extinction map\footnote[9]{L. Cambrésy, private communication; \citep[see][]{2002AJ....123.2559C}.} and corresponds to the average hydrogen column density (N$_H$ = 1.9 $\times$ 10$^{21}$cm$^{-2}$), obtained from the X-ray spectrum fits (see Sects. 3.2 and A.3).The absolute magnitudes were estimated by using the distance modulus m$_{J}$ – M$_{J}$ = 10 mag., according to the cloud distance adopted in Sect. 3.1  (d = 1 kpc). 

  The CMD shows theoretical isochrones for 0.2 to 20 Myr,  the ZAMS, and 0.1 to 7 M$_{\odot}$ evolutionary tracks. We also included the ``Early Main Sequence''  model that  represents the last stage of evolution in the \citet{2000A&A...358..593S} calculations (see Sect. 4.3). Mass and ages were estimated  from comparison with the theoretical curves by interpolating the models, except for candidates appearing  rightwards of the models in the CMD.  According to the {\it WISE} data (analysed in Sect. 4.2), several of these candidates are Class I or Class II objects (see blue symbols in Fig. \ref{figB1}).

\begin{figure}[t]
\begin{center}
\includegraphics[width=0.91\columnwidth, angle=0]{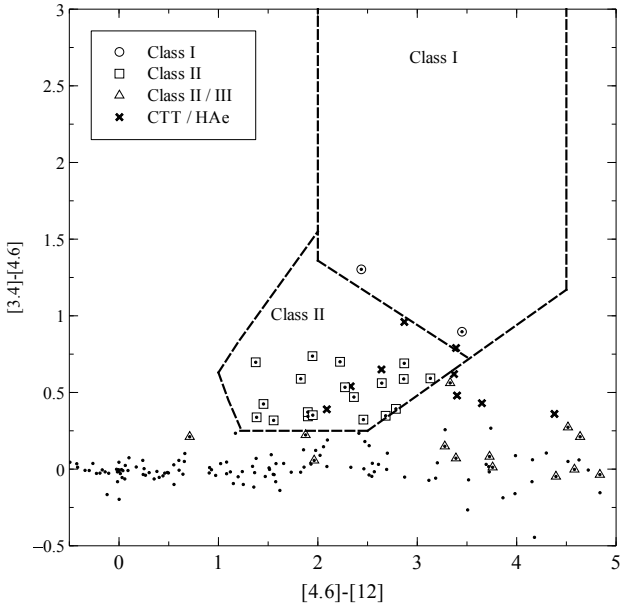}

\caption{{\it WISE} colours of counterparts of X-ray sources (dots), compared with the locus of Class I and Class II proposed by \citet{2014ApJ...791..131K}. Open symbols show objects classified according to their infrared excess, while crosses indicate known disk-bearing stars \citep{2015MNRAS.448..119F}.} 

\label{figwise}
\end{center}
\end{figure}



\subsection{Classification based on {\it WISE} data}


 Among our list of 340 NIR counterparts (Table \ref{tab_count}), only 272 are also listed in the {\it AllWISE} data release \citep{2013yCat.2328....0C}. However, we focused the infrared classification on 157 objects with reliable {\it WISE} photometry, i.e., 115 sources with errors greater than 0.2 mag in Bands 1 (3.4 $\mu$m) and 2 (4.6 $\mu$m) and upper limits in Band 3 (12 $\mu$m) were not considered. In a first analysis, we looked for IR excess by combining the {\it 2MASS} and {\it WISE} data in the K-[4.6] vs.  H-K diagram.

The distribution of the 157 selected sources in this diagram revealed 34 objects  with K-[4.6] > 0.7mag, which is an indication of IR excess \citep{2011MNRAS.410..227C,2015MNRAS.448..119F} that could be due to the presence of a disk. This excess is in agreement with the colours of T Tauri and Herbig Ae/Be stars associated with Sh2-296 \citep{2015MNRAS.448..119F},which were included in this analysis (indicated by crosses in Fig. \ref{figwise}) as representative of known disk-bearing stars.

A more conclusive infrared classification was obtained by using the criteria presented by \citet{2014ApJ...791..131K} to classify YSOs based on {\it WISE} colours.  Figure \ref{figwise} shows the  [3.4]-[4.6] {\it vs.} [4.6]-[12] diagram and the Class I, II and III regions (dashed lines) defined according to the distribution of objects associated to Taurus Molecular cloud \citep{2010ApJS..186..259R}. This analysis provided the separation of the 34 candidates, which show IR excess (K-[4.6] $>$ 0.7mag),  in different types: 2 Class I; 19 Class II; and 13 Class II/ III. The other 123 NIR counterparts having good quality of data, but not showing IR excess, are considered Class III. Finally, the remaining 115 NIR counterparts with bad quality of {\it WISE} data, to which we could not assign an infrared classification, are marked with ``??'' in the last column of  Table B.2.

As we discuss further in Sect. 5.3,  in spite of the youth of the sample associated with CMa R1, a low fraction (21/157 < 14\%) of disk-bearing stars (Class I and Class II) is found, reinforcing the previous partial results of \citet{2015MNRAS.448..119F}.

\begin{figure}[t]
\begin{center}
\includegraphics[width=1.0\columnwidth, angle=0]{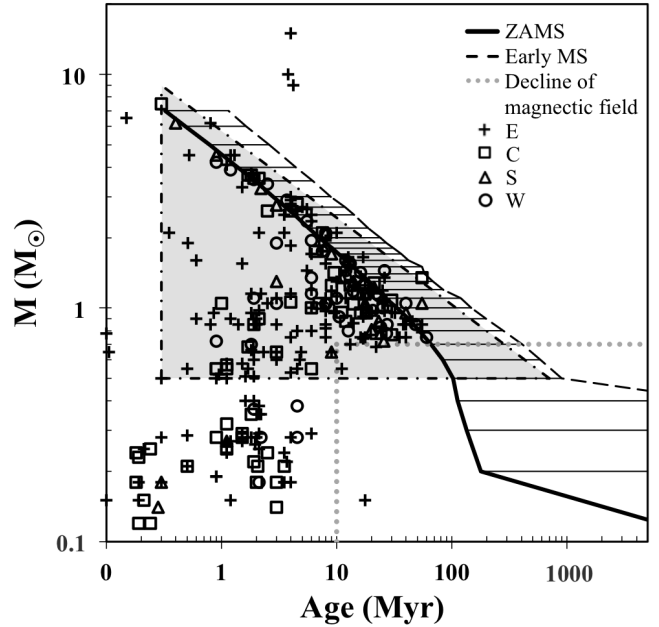}

\caption{``Best sample'' of X-ray sources based on estimates of masses and ages  for 340 {\it 2MASS} counterparts (see Table B.2), which takes into account the incompleteness of our sample below $\sim$ 0.5 M$_{\odot}$, are present in the grey area.  The empty area between dotted lines corresponding to the mass range 0.5 - 0.7 M$_{\odot}$ and age $>$ 10 Myr is interpreted as the decline of magnetic activity of low-mass young stars with age (see Sect. 7.2).} 

\label{fig4}
\end{center}
\end{figure}



\subsection{Defining a ``best sample'' of {\it XMM-Newton} sources}


As a consequence of the dependence of X-ray luminosity on stellar mass, the selection of our sample naturally imposes  a  mass detection threshold. More precisely, due to our {\it XMM-Newton} detection limit (see below, Sect. 5.2), and  in comparison with other X-ray observations of star-forming regions, our list of X-ray sources is incomplete for low-mass stars. In order to statistically improve the analysis of  the 2MASS data, we have adopted conservative criteria searching for NIR sources, which restrict the counterparts selection to the reliable candidates where the source sample is complete, i.e., our so-called ``best sample''.

The selection criteria are based on  Fig. \ref{fig4} that compares  masses and ages of our candidates with theoretical values interpolated from the zero age main-sequence (ZAMS)  and Early Main Sequence evolutionary tracks from \citet{2000A&A...358..593S}.  According to these authors, this last track is more representative of stars with mass > 1.2 M$_{\odot}$ having reached equilibrium after the CNO cycle. This is the last stage of evolution in the Siess calculations, occurring after the ZAMS, and corresponds to the end of deuterium burning, when the nuclear energy production switches to hydrogen burning and starts to provide all the stellar luminosity. 
Considering that our sample is complete only  in the ranges of 0.5 $<$ mass (M$_{\odot}$) $<$ 9 and 0.3 $<$ age (Myr) $<$ ZAMS, when taking into account the {\it XMM-Newton} detection limit for CMa R1 (see Sect. 5.1 and Fig. \ref{fig8}), 
our final  ``best sample'' is highlighted by the hatched region in Fig. \ref{fig4}. By adopting this criterion, the NIR analysis is restricted to 225 counterparts (comprising multiples): 122 of them belong to field E, 37 to field C, 23 to field S, and 43 to field W.

It is important to stress that the other sources (appearing  outside the hatched region of Fig. \ref{fig4}) remain as possible counterparts of the X-ray sources. They were removed of the present analysis only to obtain more conclusive results based on a limited, but complete sample (in a given range of mass and age), rather than on a larger, but incomplete, sample, as explained above.


\begin{figure}[t]

\begin{center}
\includegraphics[width=0.95\columnwidth, angle=0]{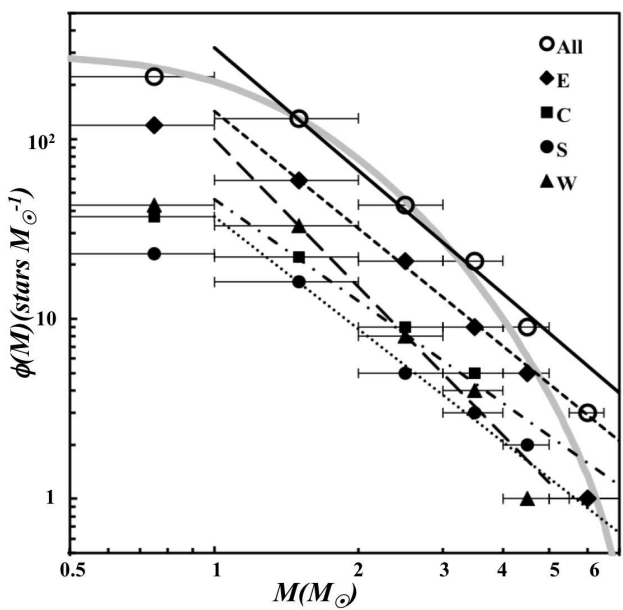}

\caption{Observed mass distribution indicated by different symbols with error bars.  The lines represent the power-law fitting of mass function $\phi$(M) for M $>$ 1 M$_{\odot}$ of individual {\it XMM-Newton} fields: E (dashed), C (dotdashed), S (long dashed) and W (dotted), and considering the entire ``best sample'' (black). The grey line shows the theoretical log-normal curve proposed by \citet{2005ASSL..327...41C} with $\mu$ $=$ 0.3 and $\sigma$ $=$ 0.6.}

\label{fig5}
\end{center}
\end{figure}


\subsection{Mass Function}


Considering the four {\it XMM-Newton} fields together, we find that over 75\% of stars have masses 0.5~$<$~M(M$_{\odot}$)~$<$~2,   together with a few massive stars.
In order to examine differences and similarities in the mass distribution of each field we calculate their cumulative mass function,  which we write in the form $\phi(M) \propto M^{-(1+\chi)}$. According to \citet{2014PhR...539...49K}, for young stellar clusters the mass function is essentially identical to the Initial Mass Function (IMF), so we can directly compare our mass functions with theoretical models of IMF, such as \citet{1955ApJ...121..161S}, \citet{2001MNRAS.322..231K} and \citet{2005ASSL..327...41C}. According to these authors, the IMF for stars with masses larger than 1 M$_{\odot}$  can be represented by a power-law function with a slope $\chi$ $\sim$1.35, while for low-mass stars \citet{2001MNRAS.322..231K} suggests  broken power-law functions with $\chi$ = 1.3 
 (0.5 - 1M$_{\odot}$) and  $\chi$ = 0.3 (0.08 -- 0.5 M$_{\odot}$).  \citet{2005ASSL..327...41C} suggests a log-normal distribution with peak at $\sim$0.2 -- 0.3M$_{\odot}$ and a dispersion of $\sim$0.5 -- 0.6M$_{\odot}$. A good discussion about the differences among these models is presented by \citet{2010ARA&A..48..339B}.

Figure \ref{fig5} shows the observed $\phi(M)$ and the slopes obtained for each {\it XMM-Newton} field.  Aiming to compare our results with the theoretical power-law function (valid for M $>$ 1 M$_{\odot}$), the fit to the slope of $\phi(M)$ does not include stars with 0.5 $<$ M(M$_{\odot}$) $<$ 1.
The mean value of $\chi$ = 1.27 $\pm$ 0.09 obtained for the entire sample is consistent with  Salpeter's IMF and agrees with the  models of \citet{2001MNRAS.322..231K} and \citet{2005ASSL..327...41C}, although it differs somewhat from the values estimated for each field: 1.21 $\pm$ 0.11 (Field E),  0.97 $\pm$ 0.14 (Field C),  1.08 $\pm$ 0.09 (Field S) and 1.73 $\pm$ 0.14 (Field W). Since our ``best sample'' is complete for M $>$ 0.5 M$_{\odot}$, the turnover below  1 M$_{\odot}$ is real and consistent with the log-normal distribution proposed by \citet{2005ASSL..327...41C} as shown in Fig. \ref{fig5}. In this case, the theoretical curve, a log-normal function with $\mu$ $=$ 0.3 and $\sigma$ $=$ 0.6, is compared with the distribution of all sources of the ``best sample''.

Altogether, except for Field W which has a definitely steeper slope, all fields have  mass function slopes comparable to those of other young stellar clusters studied by \citet{2012A&A...547A.107S}, for instance: Collinder 205, Lyng\aa~14, NGC 2362, NGC 2367, NGC 2645, NGC 3572, NGC 3590, NGC 6178, Stock 13 and Stock 16.


\subsection{Age distribution}

\begin{figure}[t]

\begin{center}
\includegraphics[width=1\columnwidth, angle=0]{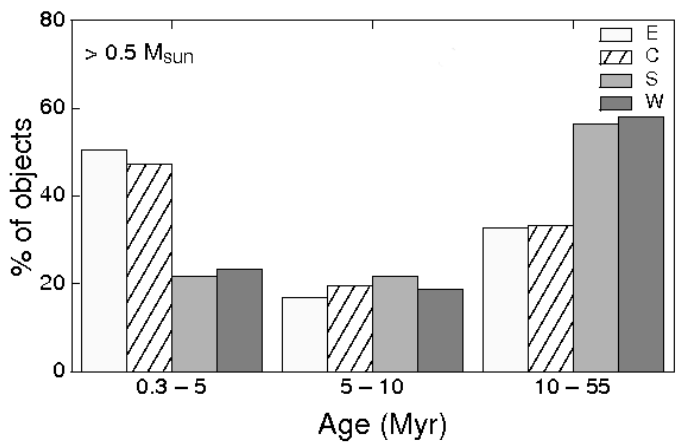}

\caption{ Comparison of age distributions for each field, by considering three ranges of age. }

\label{histo_age}
\end{center}
\end{figure}

 
The histograms in Fig. \ref{histo_age} show, for each field, the distribution of objects 
 in three age ranges. 
The main differences are in Fields E and C  where about 50\% of the sources have ages between 0.3 and 5 Myr, while in Fields S and W more than 55\% of the objects are between $10$ Myr-old and the ZAMS. Even considering the 43 NIR counterparts that were discarded  due to lack of age information (see Sect. 4.1), the distribution follows the same trend, since their NIR-excess suggests they probably are $<$ 0.3 Myr. In this case, the fraction of sources with ages $<$ 5 Myr in Fields E and C would be more than 60\%, 
 not affecting the distribution  of Fields S and W. 

 Note that, although we discuss each field separately, it is important to keep in mind that this division is purely observational, being based on the selection of the {\it XMM-Newton} pointing directions. However, they do reflect to some extent different stellar populations, since the fields were selected on the basis of our previous {\it ROSAT} observations (Paper I). So the large differences apparent in the histogram of Fig. \ref{histo_age} are statistically significant and indicative of real differences in the stellar populations  within the CMa R1 region. 
 In particular, these results fully confirm and extend the age distribution obtained in Paper I, in which, in particular, the group next to CMa GU (age $>$ 10 Myr) appears significantly more evolved than the group near Z~CMa (age $<$ 5 Myr), with a mixture of ages in between.


\section{Comparison of X-ray and NIR properties}


As described in the previous sections, X-ray and NIR data have revealed that most  (79\%) of the {\it XMM-Newton} sources are probable members of CMa R1. The combination of the results from both analyses can confirm their young nature.  On the other hand, 21\% of the {\it XMM-Newton} sample probably are field objects. Among them, ~6\% (23/387) have infrared counterparts that probably are foreground stars, and ~4\% (17/387) have counterparts that are too faint (bad quality data), without reliable classification. The other 11\% of undefined sources (44/387)  do not have {\it 2MASS} data, being classified as possible background objects. We have seen that the {\it XMM-Newton} error boxes may include multiple NIR counterparts, so we restrict the comparative analysis described in this Section to the 158 X-ray sources of our ``best sample'' that are associated with a single NIR counterpart.

\begin{figure}[t]

\begin{center}
\includegraphics[width=0.98\columnwidth, angle=0]{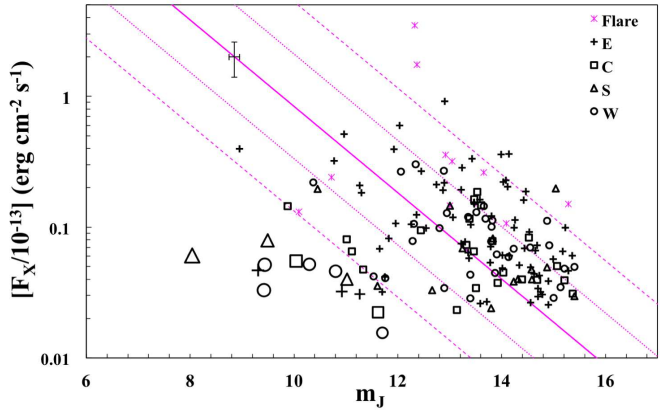}

\caption{ X-ray fluxes (F$_{X}$) {\it vs}  magnitude in {\it 2MASS} J band (m$_{J}$): full line presents the relationship $log (F_{X}) = -9.8 (\pm 0.4) - 0,33m_{J}$ similar to that obtained using {\it ROSAT} data in Paper I.  Larges symbols represent Herbig Ae/Be stars (M > 2 M$_{\odot}$). Dotted and dashed lines show 1$\sigma$ and 2$\sigma$ offset, respectively. }

\label{FXMJ}
\end{center}
\end{figure}


\subsection{X-ray flux vs. $J$-band magnitudes}


Based on {\it ROSAT} sources detected in CMaR1, 
 in Paper I a correlation between X-ray luminosities 
and absolute magnitude in $J$-band (= 1.24$\mu$m) was presented: 
$ log \ L_{X} \ = \ 31 (\pm 0.4) \ - \ 0.33 \ M_{J}$, that is similar to the results for T Tauri stars  associated with nearby clouds like Chamaeleon \citep{1993ApJ...416..623F} and $\rho$ Ophiuchi \citep{1995ApJ...439..752C}. Compact stars and extragalactic objects are expected to appear above this correlation, since their X-ray emission is comparatively much more intense relative to their NIR luminosity. Herbig Ae/Be stars, which have lower levels of X-ray flux, when compared to their NIR emission, show the opposite, appearing below the T Tauri correlation, in spite of the age similarities for these two types of pre-main sequence stars.

Considering that 1 keV (the typical energy peak of our sources) and  the $J$ band coincidently have almost exactly the same extinction cross-section \citep{1996Ap&SS.236..285R}, 
we opted  for a direct comparison between X-ray flux and apparent $J$ magnitude, avoiding inaccuracies due to errors on  the extinction estimation that affects luminosity and absolute magnitude estimation.

In this case, the correlation given in Paper I can be expressed by $log \ F_{X} \ = \ -9.8 (\pm 0.4) \ - \ 0.33 \ m_{J}$,  where $ F_{X}$ is given in erg s$^{-1}$ cm$^{-2}$. A comparison of our sample with this correlation is presented in Figure \ref{FXMJ}, where the sources showing flares are highlighted.  The error-bars shown in the top-left of Fig. \ref{FXMJ} correspond to less than 30\% of the flux and 0.1 in magnitude, which are representative of our ``best sample''.

Most sources of our sample roughly follow the empirical correlation within 2 $\sigma$ deviation. Only seven sources have a larger deviation. but they are not real outliers.  Three of them are associated with flare events, which is not unexpected since flares release a large amount of energy, resulting in a higher X-ray flux than that emitted by the source in quiescence, but are too hot to affect the NIR emission.  The other four objects (three in Field E and one in S) found above the correlation may be unresolved multiple systems with faint companions. In fact, one of them has been identified by  \citet{2015MNRAS.448..119F} as a binary system, in which both companions were classified as WTTS. The fainter star in the binary was not included in our sample of NIR counterparts due to the low  quality of NIR data. Based on the number of X-ray sources with two or more NIR counterpart candidates, we estimate less than 8\% (17/206) of possible (not confirmed) binaries in our ``best sample''. 

On the other hand, 13 sources lie below a 2$\sigma$ deviation from the correlation because they
have low $F_X$ but high values of m$_{J}$. 
Large symbols are used to represent these sources that are more likely coinciding with the  expected region for Herbig Ae/Be stars in Fig. \ref{FXMJ}. This hypothesis is reinforced by the high mass of these objects, that varies from 2.1 to 6.2 M$_{\odot}$.

We conclude that, based on the correlation for T Tauri stars displayed in Fig. \ref{FXMJ}, all the sources in this sample can convincingly be considered as young stellar objects.


\subsection{Masses and X-ray luminosities }


The distribution of $L_{X}$ as a function of mass was compared to that of other young clusters and star-forming regions such as the Orion Nebula Cluster (ONC; \citealp{2005ApJS..160..401P}), the 
Taurus Molecular Clouds(TMC; \citealp{2007A&A...468..353G}) and Collinder 69 \citep{2011A&A...526A..21B}. A summary of lower limits of X-ray luminosities and fluxes, as well as target distances and exposure times of the X-ray observations used in this comparison are presented in Table \ref{tabela4}. 


\begin{table}[ht]
\caption{Limits in X-ray observations}
\begin{center}
{\scriptsize
\begin{tabular}{|l|c|c|c|c|c|}

\hline
    &    Flux $^{(a)}$    &    log (L$_{X}$)$^{(b)}$    &    d $^{(c)}$    &    Exp. Time $^{(d)}$    &    Inst.$^{(e)}$    \\
    &    $ 10^{-15}$erg s$^{1}$cm$^{-2}$    &        &    pc    &    ks    &    \\ \hline \hline
                                   
Field E$_{XMM}$        &    2.5        &    29.48    &    1000        &    32        &    XMM    \\
Field C$_{XMM}$        &    2.2        &    29.43    &    1000        &    30        &    XMM    \\
Field S$_{XMM}$        &    2.4        &    29.46    &    1000        &    28        &    XMM    \\
Field W$_{XMM}$        &    1.6        &    29.27    &    1000        &    28        &    XMM    \\\hline
Field 1$_{ROSAT}$    &    3.6        &    29.78    &    1000        &    20        &    ROSAT    \\
Field 2$_{ROSAT}$    &    8.36        &    30        &    1000        &    5        &    ROSAT    \\\hline
ONC                    &    0.042    &    27.01    &    450        &    838        &    Chandra    \\
TMC                &    1.70        &    27.60    &    140        &    31 - 131    &    XMM    \\
Collinder 69            &    5.22        &    29        &    400        &    28 - 37    &    XMM    \\\hline

\end{tabular}
}
\label{tabela4}
\end{center}
{\scriptsize
(a) Flux limit of X-rays observations;
(b) Luminosity limit;
(c) Distance;
(d) Exposure time;
(e) Instrument used for observations.

}
\end{table}

\begin{figure}[t]

\begin{center}
\includegraphics[width=0.9\columnwidth, angle=0]{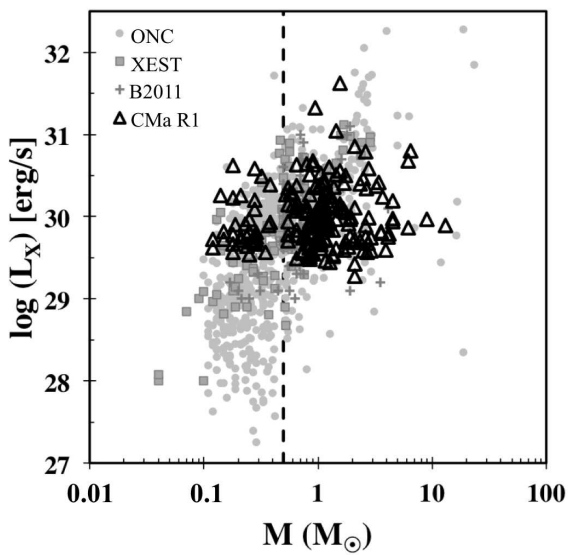}

\caption{ X-ray luminosity vs. stellar mass for the sources in CMa R1 (black triangles) compared with T Tauri  stars in 
other star forming regions: Orion Nebula Cluster (ONC), Taurus Molecular Cloud (TMC)  and Collinder 69 (Coll69).  The dashed line indicates the limit of mass adopted to define our ``best sample''. The gap at 0.4 - 0.5 M$_{\odot}$ is not real, but due to difficulties in interpolating models on this mass range.}

\label{fig8}
\end{center}
\end{figure}

The ONC and TMC are closer than CMa R1, and  some of their X-ray observations were performed with  longer exposure times. Therefore the capability to detect fainter sources results in a more complete sampling of 
the lower mass end of their population (see Table \ref{tabela4}). For comparison, the detection limit of our sample is illustrated in Fig. \ref{fig8}.

The minimum flux detected in the other regions is 10 to 100 times lower than 
 that of CMa R1, clearly suggesting the absence of faint X-ray emitters among our sample, which implies that a considerable number of low-mass stars ($< 0.5~M_{\odot}$) are below our detection limit, especially if we compare the masses and luminosities of sources in our sample with TMC objects. 
 
 It is interesting to note the differences and similarities on the $L_{X}F$ derived in Sect. 3.2 for each field compared to  the Collinder 69 cluster  considering the same mass range (see Fig. \ref{fig3}). 
 The $L_{X}F$ for all the CMa R1 sources has slightly lower X-ray levels than 
 Collinder 69, which has most of  the members with 5 Myr and  
 masses $<$ 2 M$_{\odot}$. In this case, the differences  in 
 X-ray properties depend not only on the larger range of ages of our sample, 
 but also on the masses. As indicated in Fig. \ref{FXMJ},  intermediate-mass young stars tend to 
 have lower X-ray emission when compared with the F$_{X}$ vs. m$_{J}$ correlation found 
 for low-mass stars. This can also be seen by the relatively higher proportion of 
 sources found in Field C ($\sim$19\%), appearing below the 1$\sigma$ correlation 
 in Fig. \ref{FXMJ}, when compared to Field E fraction of sources ($<$ 6\%), 
 which also reflects to the fainter $L_{X}F$ presented by field C. Since the fraction of young stars ($<$ 5 Myr) is similar for both fields, it is not an age effect but differences on mass distribution. As shown in Fig. \ref{fig5}, there is a lack of stars more massive than 4M$_{\odot}$ in field C, giving a less steep slope if compared with the mass distribution in Field E.

\begin{figure*}[t]

\begin{center}
\includegraphics[width=1.70\columnwidth, angle=0]{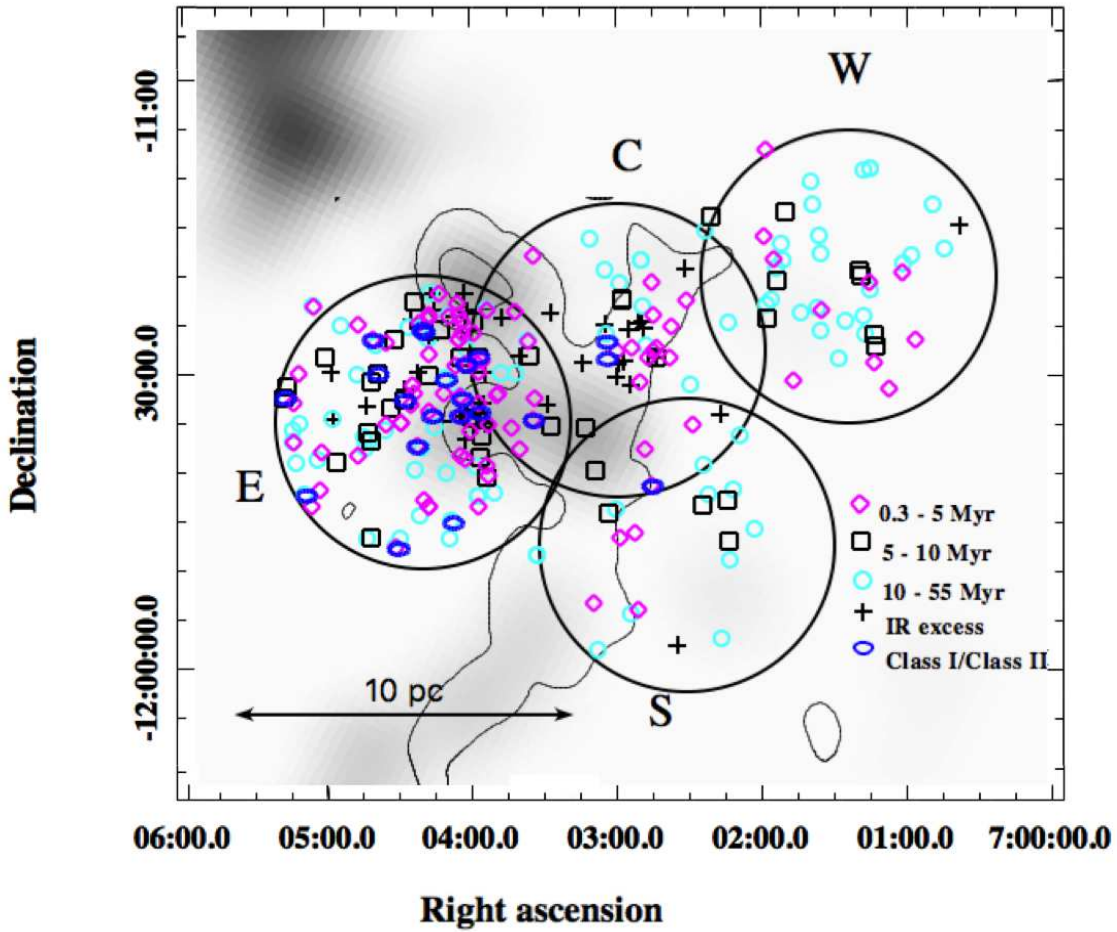}

\caption{ Spatial distribution of  NIR counterpart of X-ray sources  as  function of the age compared  to a  $^{13}$CO map$^{10}$ shown the grey image superimposed on A$_{V}$ > 2 and 4 mag contours (the same as Fig. \ref{fig1}). The diamonds, squares and circles represent objects with less than 5 Myr, between 5 and 10 Myr, and more than 10 Myr respectively. The counterparts found outside of  \citet{2000A&A...358..593S} PMS isochrones are represented by crosses and disk-bearing candidates (see Sect. 4.1) are shown by blue ellipses. Black circles delimit the fields E, C, S and W.}

\label{fig9}
\end{center}
\end{figure*}


\subsection{Comparison with the dust and dense gas distributions}


 As discussed in Sect. 4 there is a low fraction of stars with IR excess, which are those appearing to the right of the vector indicated by the dotted line in the {\it 2MASS} colour-colour diagram (see Fig. \ref{figB1}). When considering the whole sample of {\it 2MASS} counterparts only 8.5\% (29/340) have a K-band excess. The analysis based on {\it WISE} data shows 22\% (34/157)  of the sample with K-[4.6] > 0.7mag that we considered disk-bearing candidates, however only 2 of them were confirmed as Class I objects and 19 are Class II. This small number of Class I protostars and Class II T Tauri stars candidates in the {\it XMM-Newton} fields is also confirmed by  a census of young stellar objects covering the whole CMa OB1 star forming region (10$^{o}$ $\times$ 10$^{o}$) performed by \citet{2016ApJ...827...96F}. These authors follow a slightly modified version of the criteria suggested by \citet{2014ApJ...791..131K}. Despite they use {\it WISE} colours similar to those  adopted by us (Sect. 4.2),  in the list of \citet{2016ApJ...827...96F} a smaller number of disk-bearing stars, 1 Class I and 11 Class II,  is found coinciding with the X-ray sources. Their census is not complete for the {\it XMM-Newton} surveyed area, probably due to their more restrictive criteria in excluding {\it WISE} sources that show any contamination flag.

The spatial distribution of the NIR counterparts is analysed in Fig. \ref{fig9} to look for 
evidence of clustering.  
In spite of the fact that the fields observed with {\it XMM-Newton} were not meant to correspond 
exactly to physically separated clusters or stellar groups, their location turns out to reflect fairly accurately the spatial distribution of the sources as a function of position (in projection). Roughly, Field E contains sources located  on the inner side of the cloud, Fields C and S contain sources distributed 
along the border of the visible nebula, while most of the Field W sources are outside the molecular cloud.  
  
 The position of the sources in Fig. \ref{fig9} is  also compared to the cloud gas distribution, revealed by  $^{13}$CO map\footnote[10]{T. Onishi, private communication; \citep[see][]{2013PASJ...65...78O}.}, where dense areas coincide with the dust distribution, (as seen in Fig. \ref{fig1}) indicated by the A$_{V}$ map$^{9}$. It can be noted that Field W  is dominated by older objects ($>$ 10 Myr) (circles), without preferential distribution. On the other hand, Fields E and C  show a mixing of all  age ranges as well as  objects without well defined age (represented by crosses). 
In Field C, most of the younger objects ($<$ 5 Myr) are within the area with the highest concentration of CO, while in field S younger objects are on the edge or outside the CO contours. In both cases we can see a segregation of ages, in which most of the older sources are to the right side, towards GU CMa, and younger ones are to the left side, near the edge of the nebula. However, young objects are also found in the ``empty'' area around GU CMa, while older ones are also found in the dense area around Z CMa, pointing to an apparent paradox in the history of star formation in the CMa R1 region.

It is interesting to note that about 70\% of NIR counterparts with K-band excess, as well as all disk-bearing candidates  found in \citet{2016ApJ...827...96F}, are distributed in regions with high $^{13}$CO flux (> 20 Jy), which is also the location of dust concentration responsible for the extinction. Since a uniform value A$_{V}$ = 0.9 mag has been adopted in the present work, it is possible that the reddening correction was too small for these sources.
 Several of them are located in the area of the BRC 27 and VDB RN92 clusters that have A$_{V}$ $=$ 6.5 and 4.4 mag, respectively \citep{2002A&A...388..172S,2003A&A...404..217S}. A more detailed estimate of individual values of A$_{V}$ could help us better determine the ages and masses for such embedded objects, and probably increase the sample of {\it bona fide} CMa R1 members. We defer such a study to a future paper.


\section{Summary of the results}


 Our observations performed  with {\it XMM-Newton} resulted in a sample of 387  X-ray
sources (187, 84, 37 and 79 in Fields E, C, S and W respectively), 340 of them having one or more NIR ({\it 2MASS}) counterparts.

In order to characterize the CMa R1 members, we  made complementary use of the X-ray and NIR data:  
 We compared the X-ray hardness ratios to a model grid for a hot thermal plasma and for power-law spectra that we simulated with XSPEC and we compared the NIR photometry to the isochrones from PMS evolutionary models.

Based on results from the X-ray analysis, summarized in Table \ref{tab2}, the sources 
were separated according their HRs: 47\% are well reproduced by  a 
thermal plasma (APEC model), as expected for stars, and 3\% probably are 
 extragalactic sources (power-law model).  The other half of the sample could not be identified 
in one of these categories, because  12\% are reproduced by both models and  38\% are 
 outside the grids.  As described below, the numbers of probable members or field objects were refined in agreement with the analysis of NIR counterparts. Moreover we could obtain more X-ray properties of several 
sources  through their light curves and spectra. About 13 sources show {\it flare-like} 
events and the 21 brightest sources had their  plasma parameters determined by fits of the low resolution  EPIC/PN spectrum with an APEC model. The results from  the spectra of these bright sources were used to determine the energy factor conversion and consequently the X-ray luminosities.

 Among the NIR counterparts, 225 were selected to define our so-called ``best sample''. The fits of the mass distribution $\phi(M)$, assumed to be a power-law of the form $\phi(M) \propto M^{-(1+\chi)}$ gives a slope $\chi$ $\sim$1.3 $\pm$ 0.1,  for stars more massive than 1 M$_{\odot}$, which is consistent with the Salpeter IMF and in agreement with theoretical and observational results from the literature. Almost no difference is found among the observed fields when comparing their individual mass function, except of Field W, which has a lower fraction of stars more massive than 4 M$_{\odot}$, with respect to its fraction of low-mass stars, increasing its slope. Most of the younger sources (< 5 Myr) are present in Fields E and C, while about 60\% of older sources (> 10Myr) are in Fields W and S. All the fields have almost the same proportion ($\sim$20\%) of objects with intermediated ages (5 -- 10 Myr).

 The comparison with the X-ray flux and $J$ magnitude correlation, found for T Tauri 
 stars or Herbig Ae/Be stars, was restricted  to the 158 sources of our ``best sample'' 
 having a single NIR counterpart. Most of them follow the F$_{X}$ $vs.$ m$_{J}$ 
 correlation,  except for 13 below it, probably Herbig Ae/Be stars,  as confirmed by their mass 2 <  M(M$_{\odot}$) < 8,  and for 7 sources that are above this correlation: three are in a flaring state and four have anomalous light curves, meaning that two or more peaks of X-ray flux,  followed by a decline, were detected when compared to the quiescent state.

  Compared with other young stellar clusters, our sample shows a typical 
distribution of X-ray  luminosities as a function of mass. However, the sample is incomplete for low-mass stars, mainly those that are faint X-ray emitters. 
 Our survey could detect only objects having log(L$_{X}$ [erg/s]) $>$ 29.5,  which results in an apparent deficiency of young stars with masses < 0.5 M$_{\odot}$.
Another possible cause for incompleteness is related to the fact that about 10\% of the {X-ray sources may be low-mass stars affected by high levels of visual extinction: most of them are located in the direction of dense regions of the $^{13}$CO map$^{10}$, which prevented estimates of mass and age in this case.


\begin{table}[h] 
\caption{Number of  CMa R1  candidate members.} 
\begin{center}

{
\begin{tabular}{|l|c|c|c|c|}


\hline 
Field	&	M$^{a}$	&	P$^{b}$	&	U$^{c}$	&	Total 	\\ \hline \hline
E	&	70	&	89	&	28	&	187	\\  
C	&	20	&	45	&	19	&	84 \\  
S	&	8	&	17	&	12	&	37	\\  
W	&	20	&	33	&	26	&	79	\\\hline				
Total	&	118	&	185	&	84	&	387	\\\hline

\end{tabular}
}

\label{tab4.1}
\end{center}
{\scriptsize
(a) CMa R1 member;
(b) Possible CMa R1 member;
(c) Undefined sources.

}
\end{table}


By combining the two methods, X-rays and NIR, we propose a classification of the {\it XMM-Newton} sources into 3 groups: (i) CMa R1 members -- these sources are very likely associated to the region because they have X-ray HRs compatible with a thermal plasma model and belong to our ``best sample'' of NIR counterparts; (ii) Possible CMa R1 members - these sources were considered young according to one of the methods only; (iii) undefined sources - objects for which we could not determine their origin by any of the methods, or sources rejected by at least one method. The number of classified sources in each group is shown in Table \ref{tab4.1}.  Among the undefined sources there are 44 (11\%  of the total {\it XMM-Newton} sample) that  we estimate to be extragalactic (background), due to the lack of NIR counterpart; 23 (6\%) objects with NIR colours of field stars (foreground),  while 17 (4\%) with counterparts that have bad quality {\it 2MASS} data, giving them an inconclusive classification.

The optical spectroscopy performed  with {\it Gemini South} by \citet{2015MNRAS.448..119F} covered 40 of our sources in Field E.  Among the CMa R1 members (M) and possible members (P) classified by us, 22 and 18, respectively, were confirmed as PMS objects. Moreover, all Class I, II and III objects,  based on {\it WISE} data, are also found among M and P sources of Table \ref{tab4.1}.

 This agreement also proves the efficiency of the methods adopted in this work to identify the members of CMa R1.


\section{Discussion}


In order to obtain a comparative and wider view of the  young stellar population in CMa\,R1, we include in the present discussion some additional objects selected from Paper I on the basis of their {\it ROSAT} observations.


\begin{table*}[t] 
\caption{Number of X-ray sources as a function of age and mass ranges and their spatial distribution.} 
\begin{center}
{

\begin{tabular}{|l|c|c|c|c|c||c|c|}

\hline 

 &  {\it XMM} $^{a}$  &  {\it ROSAT/XMM} $^{b}$    & {\it ROSAT}\_add $^{c}$   & {\it ROSAT}\_Total$^{d}$ & {\it XMM+ROSAT}\_add$^{e}$  & \multicolumn{2}{c|}{Side of Cloud}      \\ \hline         
Age &     &     &    &      &        &  East$^{f}$   & West$^{g}$\\ \hline \hline      
0.3-5 Myr  &  93  (41\%)  &  20  (53\%)  &  12 (48\%)  & 32  (51\%)  &  105   &  81  (47\%)  &  24  (30\%) \\  
5-10 Myr  &  42  (19\%)  &  5  (13\%)  &  5 (20\%)  & 10  (16\%)  &  47   &  32  (19\%)  &  15  (19\%) \\  
 10-55 Myr  &  90  (40\%)  &  13  (34\%)  &  8 (32\%)  & 21  (33\%)  &  98   &  58  (34\%)  &  40  (51\%) \\\hline   
Mass  &     &     &    &     &     &      &    \\ \hline 
0.5-1 M$_{\odot}$  &  93 (41\%)  &  14 (37\%)  &  1 (4\%)  & 15 (24\%)  &  94   &  79 (46\%)  &  15 (19\%) \\  
1-2 M$_{\odot}$ &  89 (40\%)  &  17 (45\%)  &  14 (56\%)  & 31 (49\%)  &  103   &  58 (34\%)  &  45 (57\%) \\  
2-9 M$_{\odot}$ &  43 (19\%)  &  7 (18\%)  &  10 (40\%)  & 17 (27\%)  &  53   &  34 (20\%)  &  19 (24\%) \\ \hline 
\end{tabular}

}

\label{tab5}
\end{center}
{\scriptsize
Percents computed as a function of total number of sources: 
(a) 225 ({\it XMM-Newton}); 
(b) 38 {\it XMM-Newton} best sample detected by {\it ROSAT}; 
(c) Additional 25 {\it ROSAT} sources; 
(d) 63 (all {\it ROSAT}); 
(e) 250 ({\it XMM+ROSAT}); 
(f) 171 (on dense regions: A$_{V}$ > 2); 
and (g) 79 (outside the cloud: A$_{V}$ < 2).

}

\end{table*}


A simple comparison of the spatial density of detected sources in both surveys gives an 
immediate appreciation on how much the sensitivity improved with 
the present {\it XMM-Newton} observations (yielding $\sim$550 sources deg$^{-2}$) 
over the previous {\it ROSAT}  results ($\sim$36 sources deg$^{-2}$). 
However, the {\it ROSAT} sources were detected in a larger  field-of-view ($\sim$2.7 deg$^2$) while the total area covered by {\it XMM-Newton} for the present work is 5 times smaller ($\sim$ 0.7 deg$^2$).

An overview of the spatial distribution of the total number of 250 young stars selected as our enlarged ``best sample'' (225 {\it XMM-Newton} $+$ 25 {\it ROSAT})  indicates that 171 (68\%) are seen close to Z CMa, superimposed in the East side the of A$_{V}$ > 2 mag contour shown in Fig. \ref{figdisagemass}, while 79 are found in the West side of A$_{V}$ contour, around  GU CMa, in areas devoid of molecular gas (A$_{V}$ $\sim$ 0.5 mag).  In the next sections, the location of the sources  is discussed relative to their ages and masses as obtained in  Sect. 4.1 for our sample  and in Paper I for {\it ROSAT} sources, and to the distribution of dense gas, over a wider area than that covered by {\it XMM-Newton} alone.


\subsection{Spatial distribution vs. stellar ages}


With the purpose of investigating the spatial distribution of our sample of young stars relative to their age, Table \ref{tab5} gives  the number of {\it XMM-Newton} sources supplemented by {\it ROSAT} sources, separated in three age ranges, the same ones as in Fig \ref{histo_age}: 0.3 - 5 Myr (``younger stars''), 5 - 10 Myr  (``intermediate-age stars''), and  10 - 55  Myr (``older stars''). Based on these definitions, the fraction of $\sim$ 40\% of the younger stars observed by {\it XMM-Newton}, 
which is similar as that of the older stars, is thus in agreement with Paper I (their Fig. 10, top left panel), based on {\it ROSAT} observations of a  much smaller source sample (they give 45\%  and 35\% respectively) 
in spite of  the differences in sensitivity and field of view. This means that this age distribution is very robust and gives a good characterization of the X-ray detected young stars as a whole, over a large area.

The spatial distribution of objects was also examined relative to their position with respect to the gas, based on the comparison with the density contours of the CO map shown in Fig. \ref{figdisagemass}. The last two columns of Table \ref{tab5} give the number of objects seen in front of dense parts of the cloud (``East'' side, around Z CMa), or of regions devoid of gas (``West'' side, around GU CMa).

This study yields what seem to be paradoxical results. $(i)$ On the one hand, we can clearly see an expected correlation between the ages of stars and their relation with star formation sites:  the youngest stars are spatially correlated with dense gaseous regions of the cloud, as it is observed in, e.g., ONC and OMC2/3 regions \citep{2012AJ....144..192M,2011ApJ...739...84G}, while older stars are spread where the gas is absent. $(ii)$ But on the other hand, we also find young stars in the empty regions (30\%), and older stars in the dense regions (34\%), while intermediate-age stars (19\%) are found everywhere. To help understand this paradox, let us now turn to the distribution of stellar masses.


\subsection{Spatial distribution vs. stellar masses}


 As in the previous Section, we separate the sources in three ranges of mass as shown in panels {\it d, e} and {\it f} of Fig. \ref{figdisagemass}. The number of objects in each range is given in Table \ref{fig5}. When comparing the overall mass distribution of {\it XMM-Newton} sources with that  obtained from {\it ROSAT} (Paper I), we find comparable fractions of stars in the ranges 1 - 2 M$_{\odot}$ and  $>$2 M$_{\odot}$ , respectively $\sim$55\%  and $\sim$ 30\% of the {\it ROSAT} sources (see Paper I, Fig.10, top right panel). As expected because of their different sensitivities, the main difference is found for low-mass stars (0.5 to 1 M$_{\odot}$), for which the fraction in the {\it ROSAT} list is three times lower (15\%) than in the present work (41\%).

The spatial distribution of the stars found on both sides of the cloud (see the last two columns of Table \ref{tab5}) shows a trend (34/53) for the higher stellar masses to be more concentrated in the dense, East side, in the vicinity of Z CMa. More precisely, Fig. \ref{figdisagemass}$d$ shows the stellar mass interval 2 - 9 M$_{\odot}$ broken down in three mass ranges (2 - 3 M$_{\odot}$, 3 - 5 M$_{\odot}$, and 5 - 9 M$_{\odot}$ for {\it XMM-Newton} sources: for all these ranges, the {\it XMM-Newton} field E, overlapping Z CMa, is more populated than the other fields. On the other hand, the {\it ROSAT} data (for which the masses cannot be determined as accurately as from {\it XMM-Newton} data) show that high-mass stars (2 - 9 M$_{\odot}$), which normally are closer to dense matter since they are younger, do exist also far from the main cloud. Intermediate-mass stars (Fig. \ref{figdisagemass}$e$: 1 - 2M$_{\odot}$) are more evenly distributed on both sides of the cloud, and are quite numerous also in the empty regions of ROSAT field, far from the cloud.  In contrast, Fig. \ref{figdisagemass}$f$ shows a decline of low-mass stars towards the empty, West side of the cloud as detected by {\it XMM-Newton}, and none in the {\it ROSAT} field further out.

This strong effect cannot be entirely due to the comparatively low sensitivity of the {\it ROSAT} observations. It is true that, as mentioned in Table \ref{tabela4}, the {\it ROSAT} observations in this area are 0.3 dex less sensitive than the East field {\it XMM-Newton} observation (a factor of 2 in luminosity), so that the {\it ROSAT} census of low-mass stars (M$_{\star}$ < 1 M$_{\odot}$) cannot be complete. Nevertheless, as shown by  Table \ref{tab5}, where we have indicated the sources detected both by {\it ROSAT} and {\it XMM-Newton}, we see that in total 15 low-mass stars have been successfully detected by {\it ROSAT}, but 4 only ($\sim$ 25\%) have been detected in the area defined by the {\it XMM-Newton} Field W, and none beyond. Since we see in all the other panels of Fig. \ref{figdisagemass} that there are many {\it ROSAT} detections outside of the {\it XMM-Newton} fields, this effect must be real.

 To look in greater detail at this apparently different behaviour in our X-ray detection of low-mass stars compared to higher-mass stars, we have created a set of three figures (Fig. \ref{fig12}$a$, \ref{fig12}$b$, and \ref{fig12}$c$), corresponding to the same late age range as Fig. \ref{figdisagemass}$c$ (10-55 Myr), but broken down in three low-mass ranges (respectively 0.5-1 $M_{\odot}$, 0.5-0.8 $M_{\odot}$, and 0.5-0.7 $M_{\odot}$). We also include, for each panel, the {\it ROSAT} detections, even if they correspond also to {\it XMM-Newton} detections. First, we see a very clear trend for the number of sources to diminish towards lower masses: there are only two {\it XMM-Newton} detections left (and no {\it ROSAT} below 0.7 $M_\odot$, compared with 39 {\it XMM-Newton} and 4 {\it ROSAT} detections below 1 $M_{\odot}$). Second, even though we know the {\it ROSAT} sample is incomplete below 1 $M_{\odot}$, we still have 3 {\it ROSAT} detections (and 14 {\it XMM-Newton}) between 0.5 and 0.8 $M_\odot$. However, irrespective of the (low-mass) range, there is no detection in the {\it ROSAT} field outside of the {\it XMM-Newton} fields. Third, this effect corresponds to the empty area visible in the mass-age scatter plot for our ``best sample'' of {\it XMM-Newton} sources (Fig. \ref{fig4}), for M$_{\star}$ < 0.7 M${_\odot}$ and age $>$ 10 Myr. Such an effect is not expected in the {\it ROSAT} data because of its incompleteness in this mass range.

 In fact, there is a physical reason for this behaviour of low-mass stars. The study by \citet{2005ApJS..160..390P} on the evolution of X-ray emission in young stars spanning the 1 - 10 Myr age range, which covers very young clusters in Orion \citep[COUP:][]{2005ApJS..160..353G}, NGC2264 and Chamaeleon, supplemented by older, open clusters (Pleiades, Hyades) extending the ages to 650 Myr, has shown two regimes. During the PMS phase (up to $\sim$ 10 Myr), the X-ray luminosity is roughly constant for all masses, which corresponds to the ``saturated phase'' in which the young stars are fully convective. Then a radiative core develops, and beyond 10 Myr the X-ray luminosity decreases abruptly (see their Fig. 4) in relation with the stellar spin-down \citep[link with the magnetic field generation by the $\alpha\Omega$ dynamo; see also][]{2014MNRAS.441.2361V}. \citet{2005ApJS..160..390P} show that, for masses M$_{\star}$ = 0.5-0.9 M$_{\odot}$,  the median X-ray luminosity $<$L$_{X}>$ (in erg~s$^{-1}$) declines from $<logL_{X}>$ $\sim$ 30 between 1 and $\sim$ 10 Myr (ONC, NGC 2264 and Chamaeleon), down to $<logL_X>$ $\sim$ 29 at $\sim$ 100 Myr (Pleiades and Hyades). The slope of the decline is steeper for higher masses\footnote[11]{More recent studies indicate a decline at later ages ($> 100$ Myr), but they scale L$_{X}$ with other stellar parameters, e.g., $L_{X}/L_{bol}$ \citep{2012MNRAS.422.2024J}, $L_{X}$ normalized by stellar~surface \citep{2017MNRAS.471.1012B}.}. So in our case, given the sensitivity of our X-ray observations  (log L$_{X}$ $\sim$ 29.3 - 29.5 for {\it XMM}, see Table \ref{tabela4}), we interpret the empty area of Fig. \ref{fig4} as evidence for this decline in magnetic activity at $>$ 10 Myr for M$_{\star}$ $<$ 0.7 M$_{\odot}$.

Consequently, the West side of the cloud very likely contains low-mass stars, but these were formed over 10 Myr ago, in an earlier episode of star formation, and are now too faint to be visible in our observations. In fact, several objects detected by {\it WISE} appear in this side of the cloud,  without IR excess, so may be Class III or field stars.

More precisely, since the {\it WISE} detections imply the existence of faint, evolved circumstellar disks (debris, and/or planet-forming), the corresponding stars are classified Class II/III, and it is worth considering whether their presence affects their detectability in X-rays (i.e., their magnetic activity), in addition to, or instead of, the change in internal structure at ages $\sim 10$ Myr. First, with reference to Fig. \ref{fig12}a (M$_{\star}$ = 0.5-1.0 M$_{\odot}$), we find 14 {\it XMM} sources to the West of the cloud complex ($A_V < 2$): 3 have ages $< 10$ Myr, and 11 have ages $> 10$ Myr. However, of these only 5 have a {\it WISE} classification: below 10 Myr, 2 sources are Class III (no disk), above 10 Myr 1 is Class II/III, and 2 are Class III. Within this small sample, the difference is not significant. To increase the statistics, we can enlarge the sample in mass: for the range $M_\star = 0.5-2.0 M_\odot$, we find 19 sources having a {\it WISE} classification, respectively 1 Class II/III vs. 7 Class III for ages $<10$ Myr, and 1 Class II/III vs. 10 Class III for ages $>10$ Myr, so the difference in diskless stars (7 vs. 10) is again not significant. In other words, the decline in magnetic activity after $\sim$ 10 Myr equally affects the X-ray detectability of low-mass stars, whether or not they are surrounded by (evolved) disks.


\section{Conclusions}


We focus our main conclusions on two issues: (i) a major observational contribution, which has increased the known census of the stellar population associated with CMa R1 by a factor 15, and (ii) the implication of these results to unveil its complex star formation history. By discussing the interplay between star formation and molecular clouds that probably dictated the scenario of this history, we aim to shed some light on the role of stellar feedback on molecular clouds, an open issue on the general context of  star formation in the Galaxy.


\begin{landscape}

\begin{figure}[t]
\begin{center}

\includegraphics[width=0.33\columnwidth, angle=0]{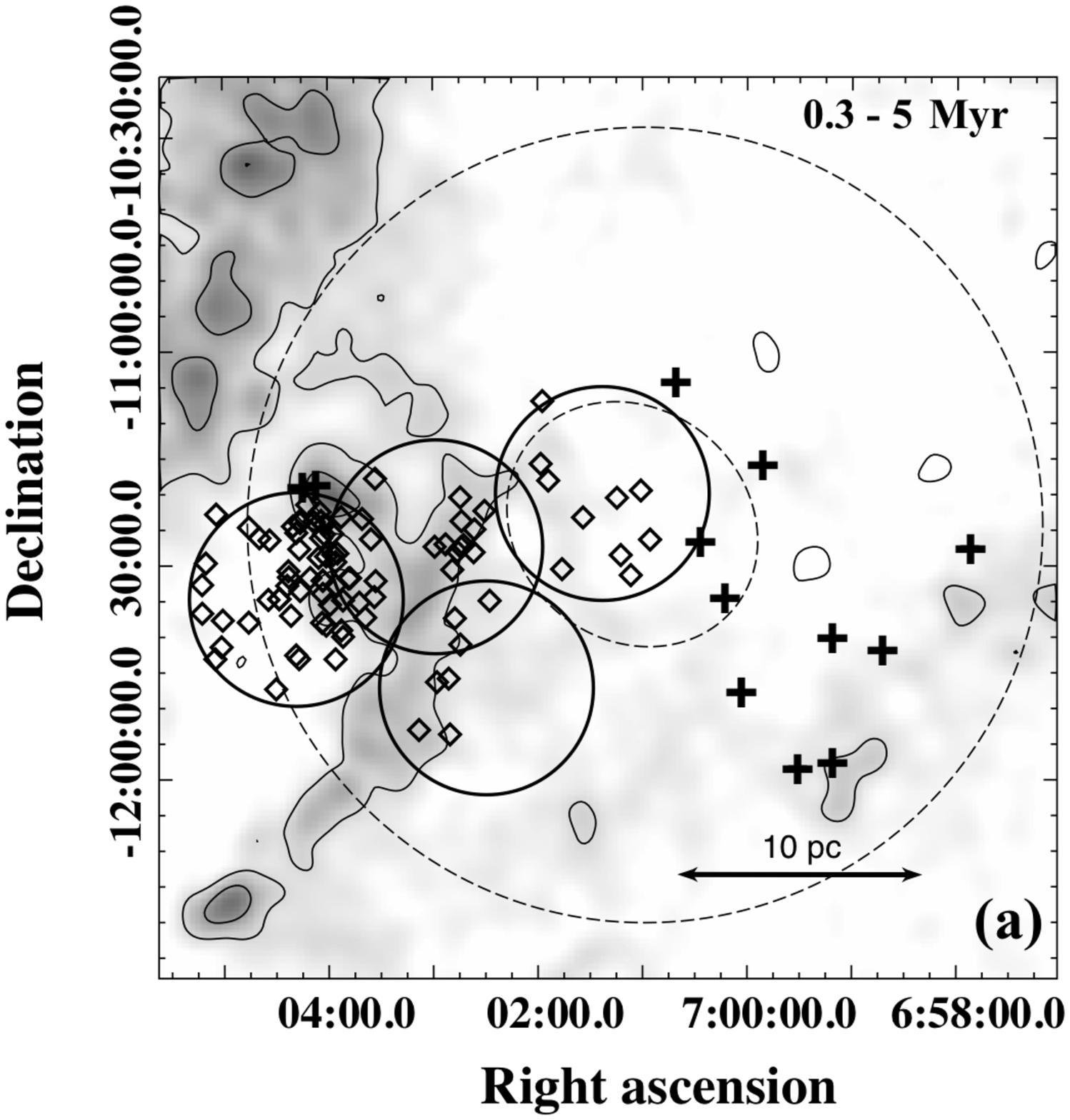}
\includegraphics[width=0.33\columnwidth, angle=0]{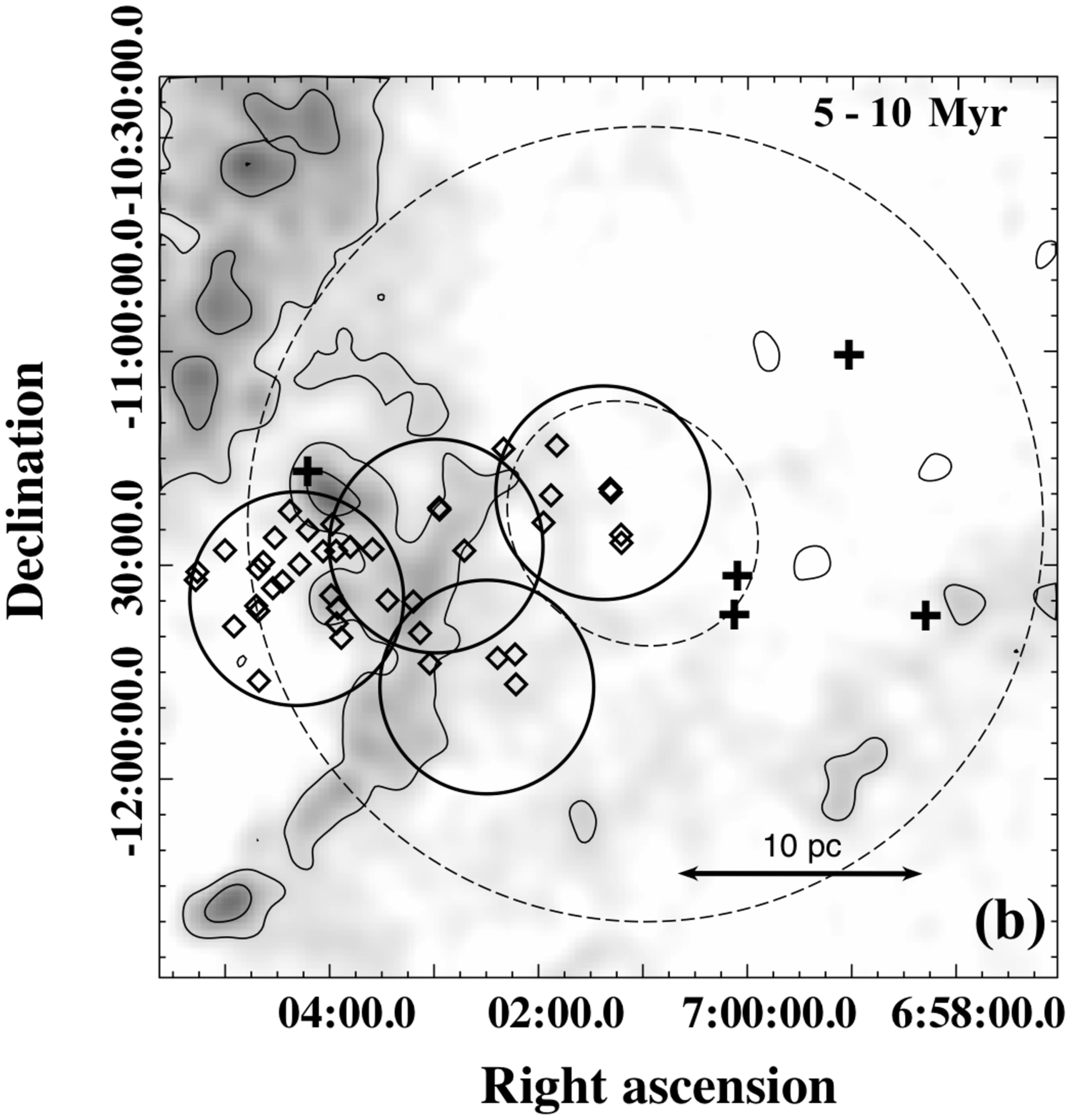}
\includegraphics[width=0.33\columnwidth, angle=0]{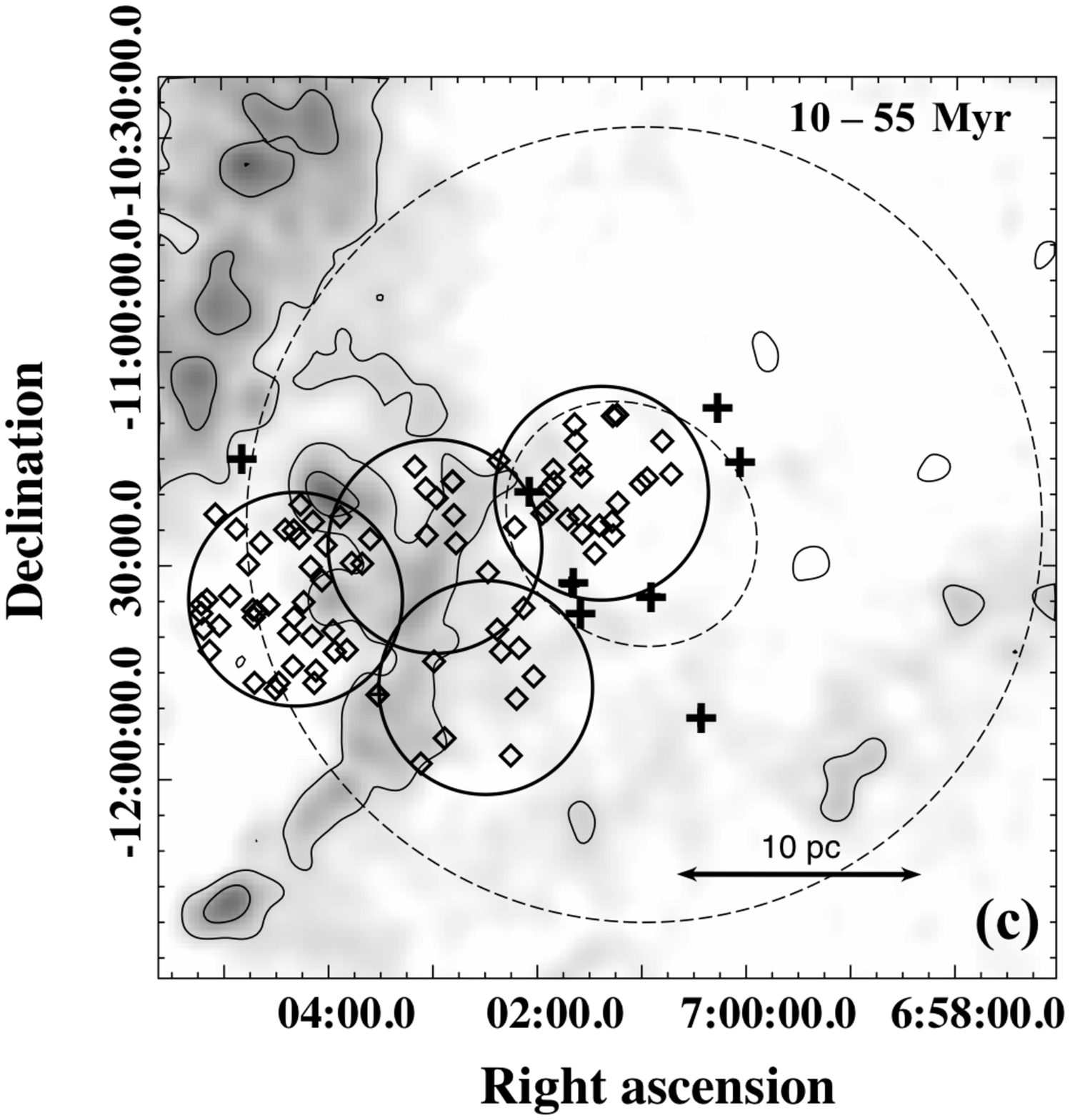}
\includegraphics[width=0.33\columnwidth, angle=0]{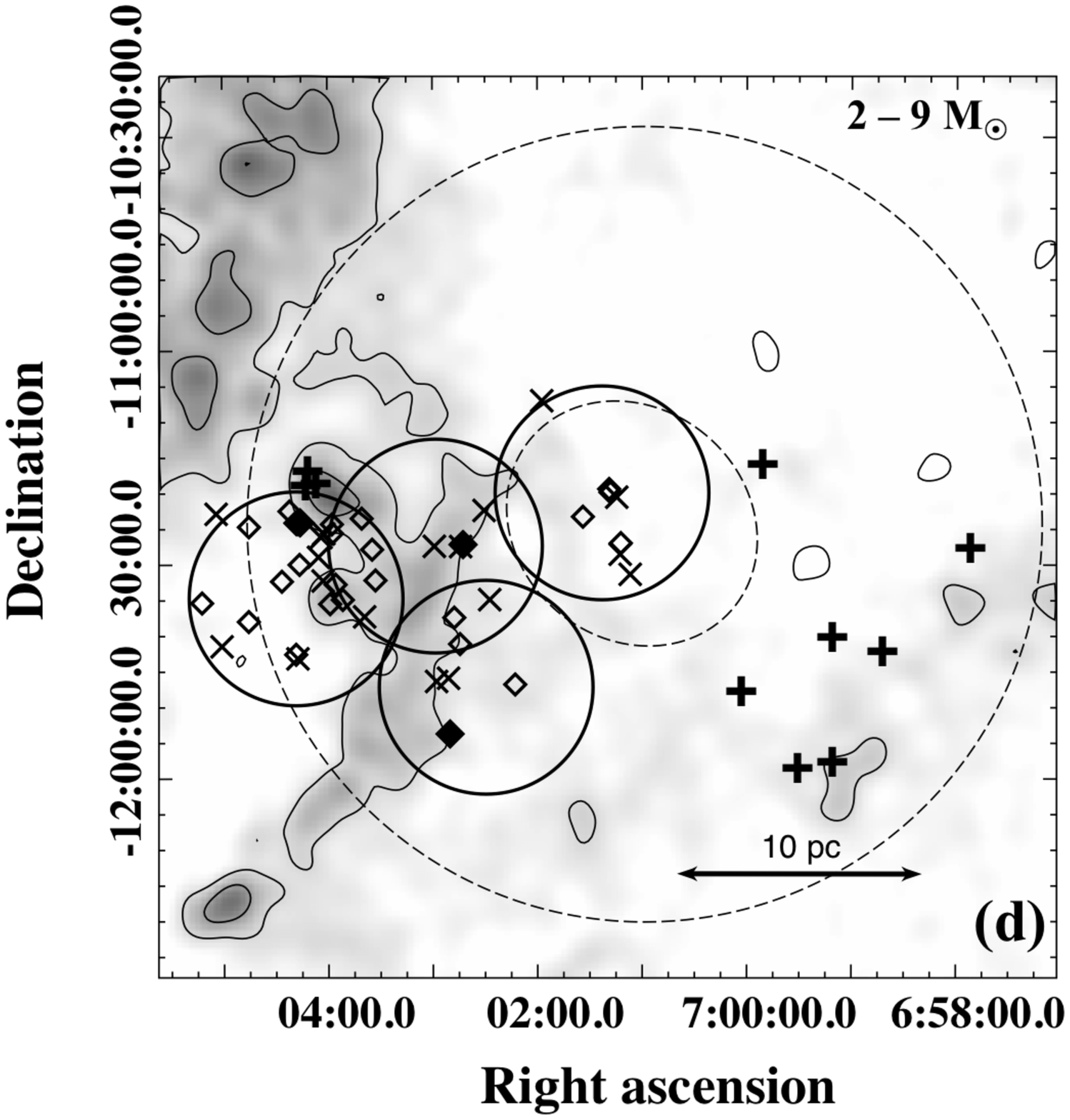}
\includegraphics[width=0.33\columnwidth, angle=0]{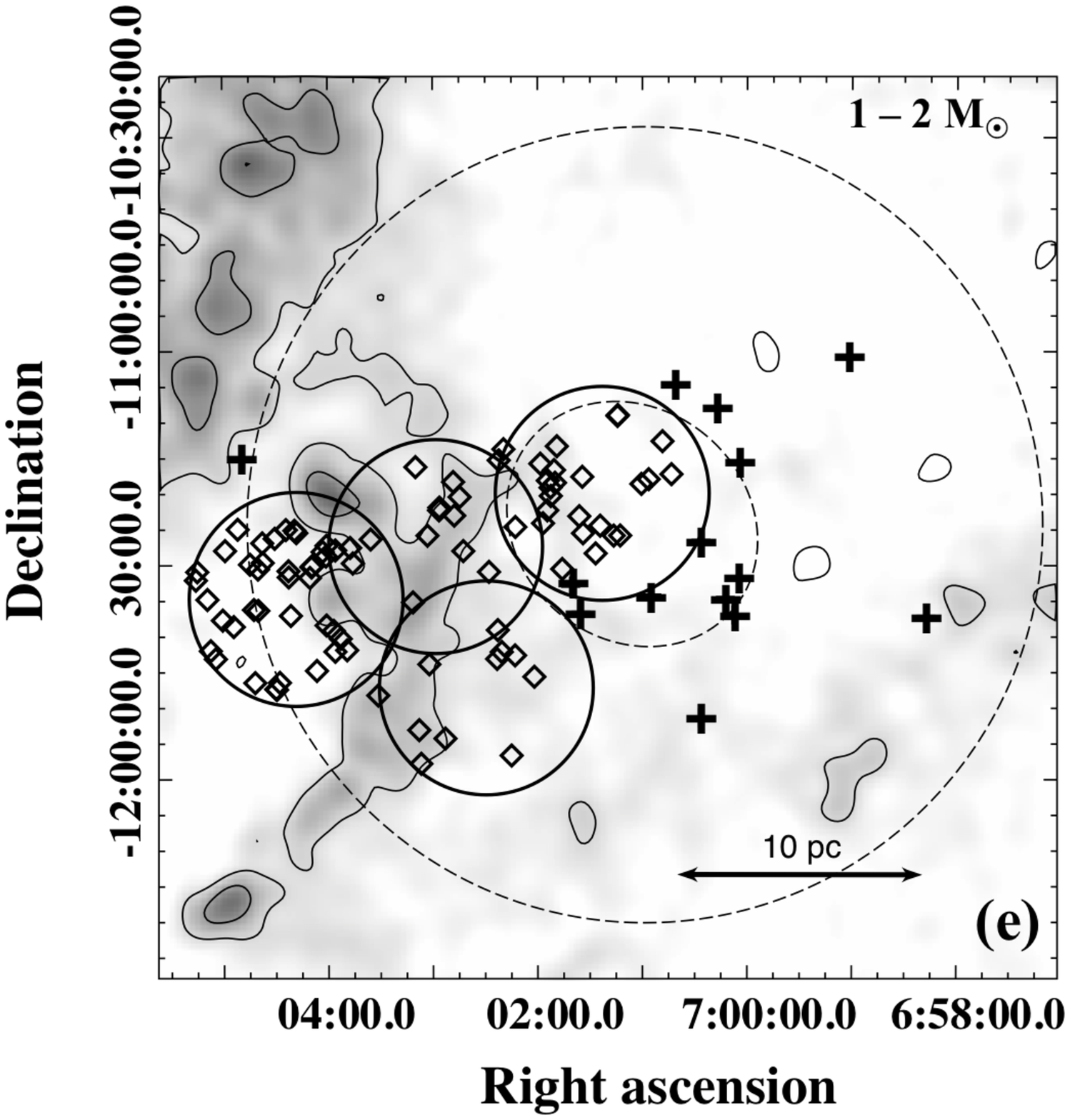}
\includegraphics[width=0.33\columnwidth, angle=0]{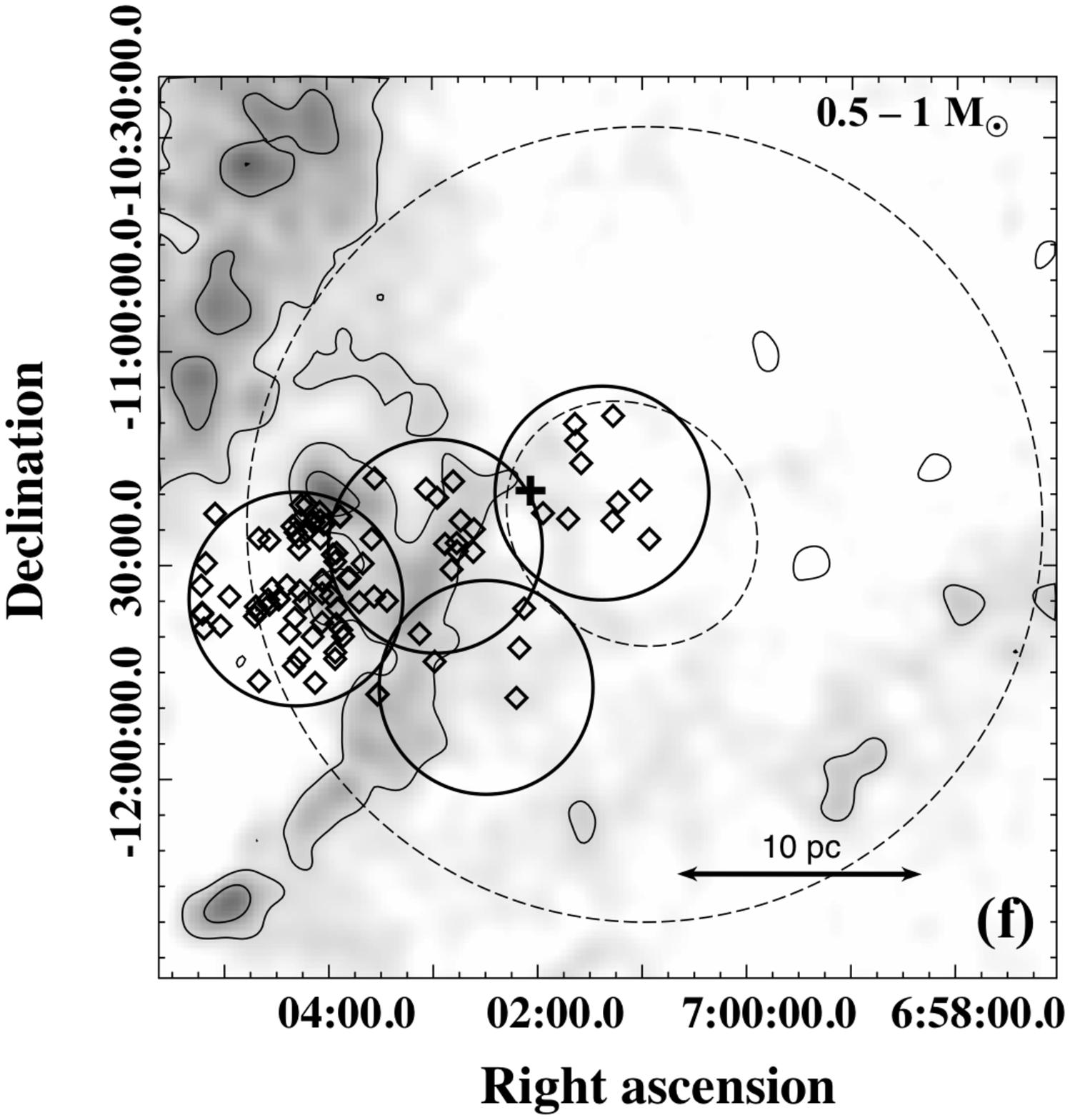}

\caption{ Spatial distribution of NIR counterparts compared with the visual extinction map$^{9}$. The lowest contours indicate a separation of dense regions (A$_{V}$ > 2). The diamonds and crosses represent {\it XMM-Newton} and {\it ROSAT} sources respectively. Top panels a, b and c show objects with less than 5 Myr, 5 to 10 Myr and more than 10 Myr respectively. Bottom panels d, e and f are objects with masses larger than 2 M$_{\odot}$, 1 to 2 M$_{\odot}$ and between 0.5 to 1 M$_{\odot}$, in this panel are highlighted {\it XMM-Newton} objects with masses between 2 and 3 M$_{\odot}$ (diamonds),  3 and 5 M$_{\odot}$ (X) and more than 5 M$_{\odot}$ (full diamonds). Black full lines delimit the fields E, C, S and W and {\it ROSAT} field are by dashed line.}
\label{figdisagemass}
\end{center}
\end{figure}
\end{landscape}

 
 In a study on the reliability of age measurements for YSOs, \citet{1674-4527-12-1-001} points out the debate in the literature concerning the time scale of star formation. Some authors argue that star formation should be a ``fast'', dynamic process \citep[e.g.,][]{2001ApJ...562..852H, 2007ApJ...668.1064E,2010ApJ...723..425D}, while an opposite view of a ``slow'' quasi-static equilibrium process has also been discussed \citep[e.g.,][]{2000ApJ...540..255P,2006ApJ...641L.121T,2007ApJ...666..281H,2011EAS....51..245P}. Testing these theories depends on the estimate of star-forming duration and a good determination of  the age spread of the YSO population.

  The presence of a relatively old population in the whole area (i.e., both the dense gaseous and empty sides of the cloud) suggests that stars were slowly formed in a first episode throughout
the region, > 10 Myr ago. As argued above, the older low-mass stars are not seen in the West side because their X-ray emission has fallen below our detection limit. Massive stars (M$_{\star}$ < 9 M$_{\odot}$) are however present in this area, but are younger than 10 Myr and are absent at older ages. Since these stars have a lifetime $\sim$ 40 Myr, more massive ones having a lifetime $<$ 40-10 $=$ 30 Myr, i.e., M$_{\star}$ > 10 M$_{\odot}$, consistent with the current IMF (Fig. \ref{fig5}), may have existed then exploded, dispersing the molecular clouds and possibly preventing the formation of new stars. Of course, after such a long time (30 Myr ago), one can hardly expect to see any trace of the explosion(s), precisely because the masses and spatial distribution of the molecular clouds must have been very different from what they are now.

We avoid  being affected by these difficulties in 
the age determinations by being able to break down the ages into less precise, but broader ranges. Indeed, we have found a true bimodal distribution of ages, with two clearly distinct groups: one is younger (< 5 Myr) and the other is older (> 10 Myr), with less of 20\% of the objects in the intermediate age range (5 - 10 Myr). Therefore, our results show that at least two star formation episodes took place in the same region, separated by  at least $\sim 5$ Myr.


The older low-mass stars, as well as their putative associated high-mass O stars, are however not seen in the West side: the former because their X-ray emission has fallen below our detection limit, and the latter because they have exploded after having undergone an intense mass loss (Wolf-Rayet) phase, dispersing almost all the CO-emitting material in this area, thus preventing the birth of new stars. 

On the other hand, the presence of a large number of objects < 5 Myr old and some disk-bearing T Tauri stars, as well as Herbig stars, located in the dense part (East side) of the cloud suggests a later episode of concentrated star formation that may have been caused by compression of the gas, perhaps triggered from the West side by the now defunct massive stars of the previous generation.

We can compare this scenario with the conclusions of \citet{2002ApJ...581.1194P}, who developed a picture of star formation history in the Taurus-Auriga cloud complex, based on a comparison between stellar ages and the spatial distribution of the gas. At least 10 Myr ago, a low level of dispersed star-formation occurred over a broad and diffuse gaseous area; due to the quasi-static contraction of the clouds, the material was concentrated in filaments under the gravity action combined with shock dissipation. Recent results based on millimetric observations from {\it Herschel}, for instance, have confirmed the important role of the interstellar filamentary structure on the  formation process of low-mass prestellar cores \citep[e.g.,][]{2010A&A...518L.102A,2014prpl.conf...27A}.
 These filaments have presumably acquired the minimum density required to accelerate the star-formation rate, and a new group of stars was generated in the last few Myr. In our case, however, we attribute the dissipation of gas to the action of massive stars that have now disappeared - such massive stars are not invoked in the Taurus-Auriga picture. \citet{2002ApJ...581.1194P} suggest in this case that  dissipation occurs through the action of low-mass stellar outflows.

\begin{figure}[t]

\begin{center}

\includegraphics[width=0.81\columnwidth, angle=0]{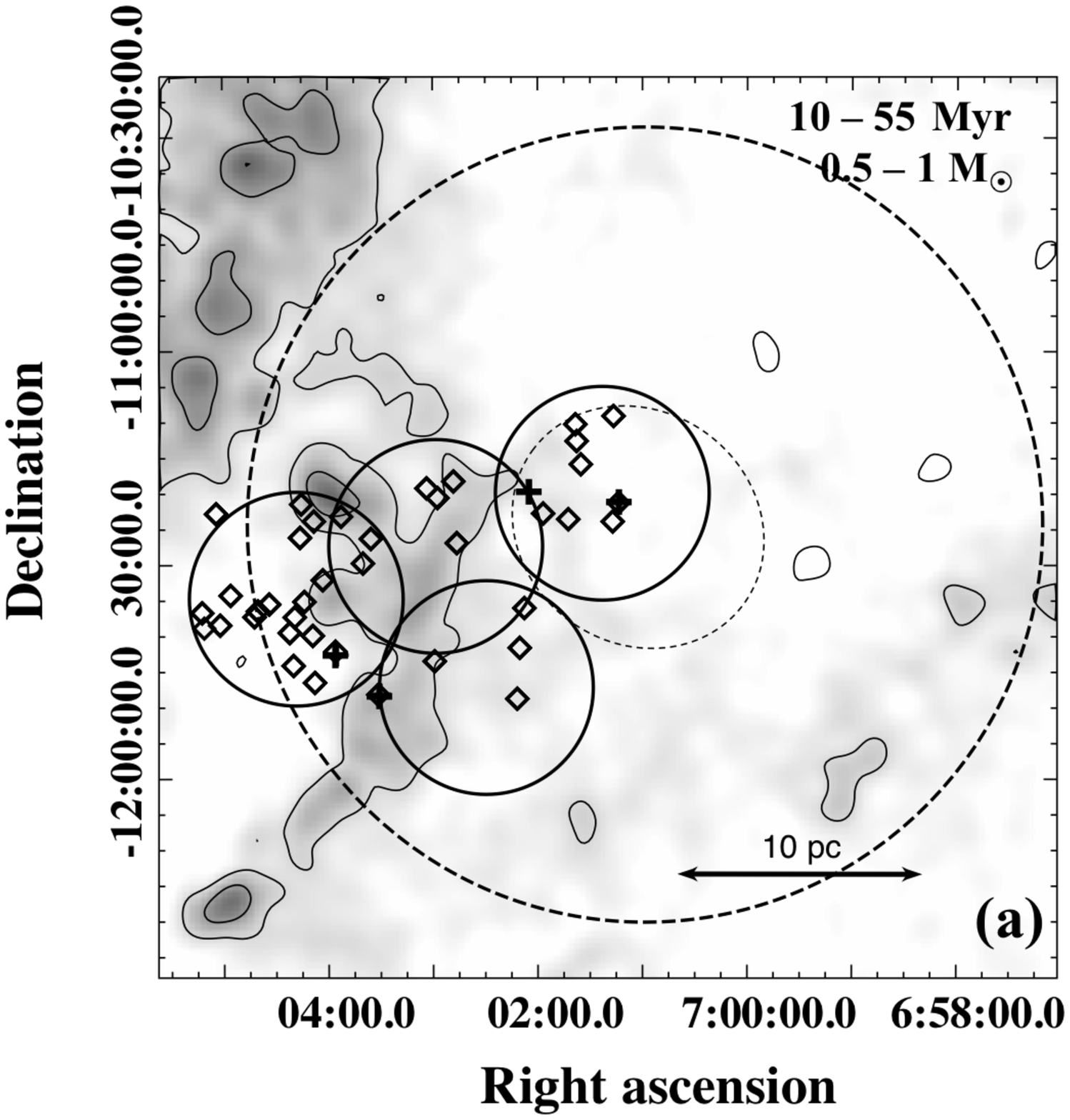}
\includegraphics[width=0.81\columnwidth, angle=0]{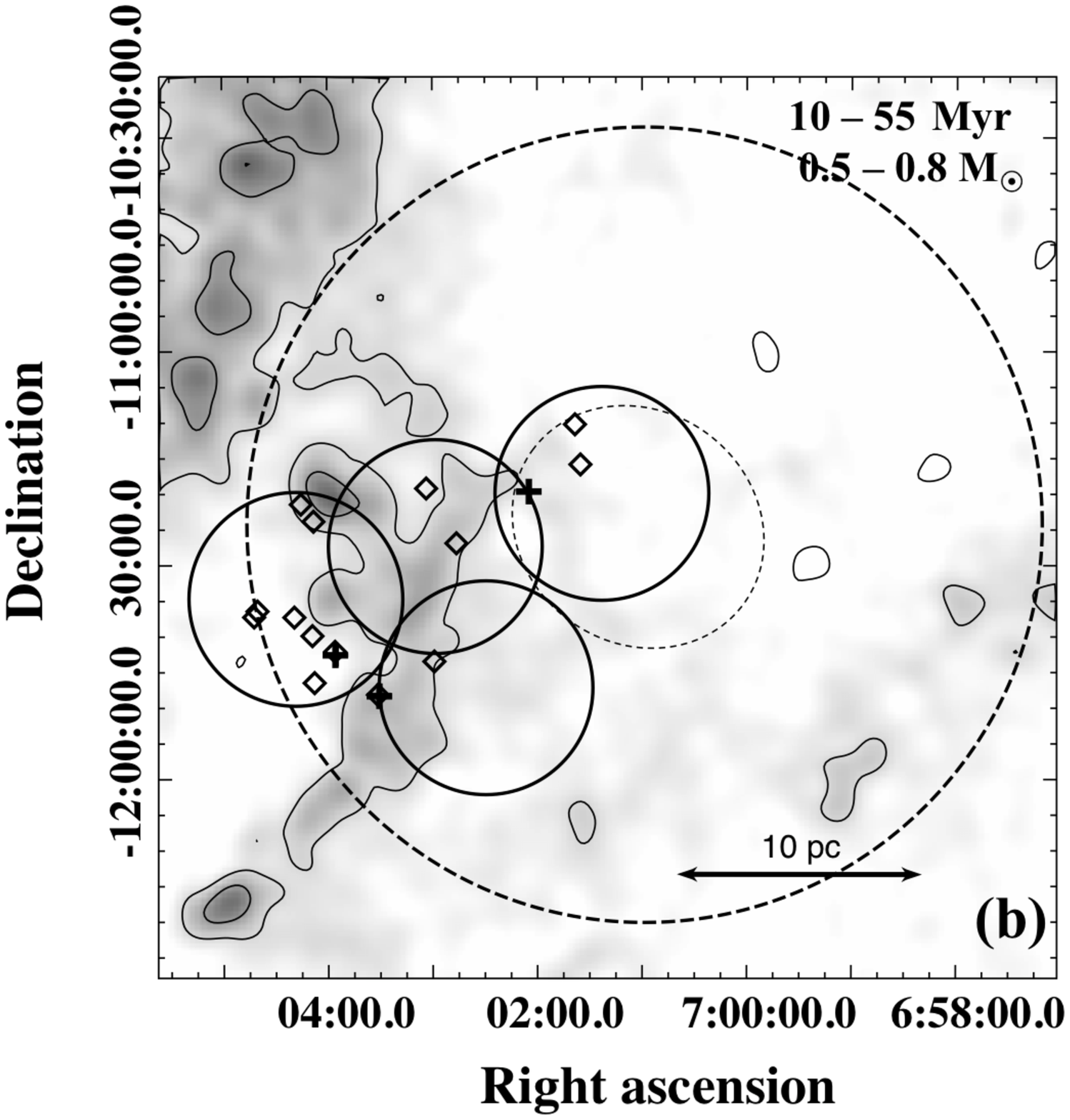}
\includegraphics[width=0.81\columnwidth, angle=0]{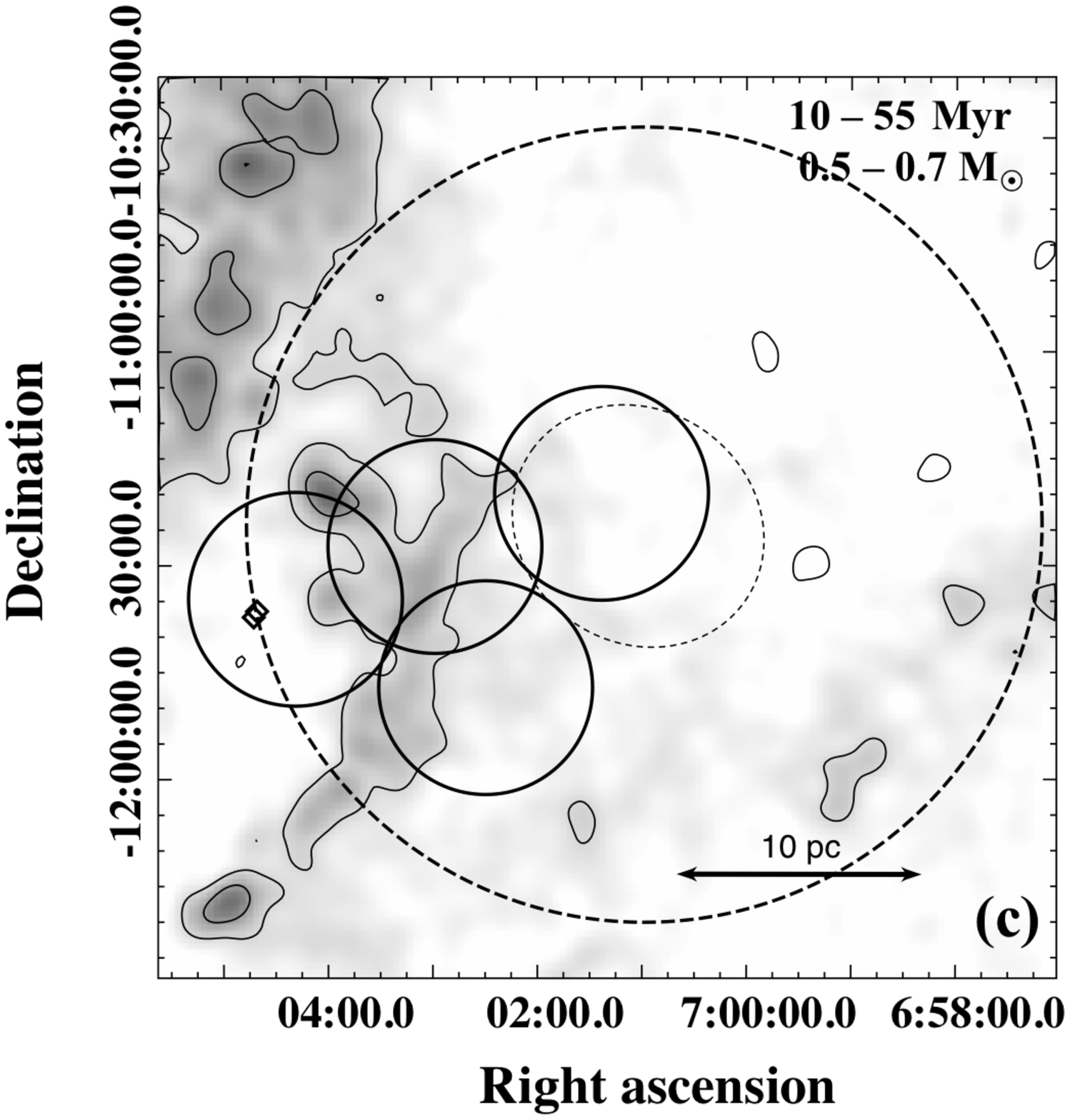}
\caption{Same as Fig. \ref{figdisagemass}c (spatial distribution in the age range 10-55 Myr), but for low-mass stars only.  In the three panels (a, b, c), the stellar masses are broken down in three mass ranges (0.5-1 M$_{\odot}$, 0.5-0.8 M$_{\odot}$, 0.5-0.7 M$_{\odot}$, respectively), diamonds represents {\it XMM-Newton} sources and crosses are {\it ROSAT} sources. As discussed in the text (Sect. 7.2) and as is visible in Fig. \ref{fig4}, the nearly complete absence of stars with M $<$ 0.7 M$_{\odot}$ is due to the decline of their magnetic activity after $\sim$ 10 Myr.
 }

\label{fig12}
\end{center}
\end{figure}

Considering the ages of the CMa R1 members, their spatial distribution, and the masses of the molecular cloud complex, we find indications that this association is going through the final stages of  the star formation process. On a large scale, the cloud material appears dispersed, probably due the evaporation caused by a previous generation of massive stars as argued above.
The list of $^{13}$CO clouds surveyed by \citet{2004PASJ...56..313K} contains only three small clouds (< 10$^{3}$ M$_{\odot}$) near the region covered by our {\it XMM-Newton} and {\it ROSAT} observations.
Projected against our {\it XMM-Newton} fields, the only available matter is provided by the cloud 224.7-02.5 (890 M$_{\odot}$, 21 M$_{\odot}$/pc$^{2}$), which coincides with the area around Z CMa (East side) where the stars are still being formed. Considering the presence of a few stars with $M_\star$ > 8 M$_\odot$, within less than $\sim 10$ Myr from now new supernova explosions will disperse the remaining material, which shall mark the very end of the CMa R1 molecular complex as a star-forming region. 

\begin{acknowledgements}

Part of this work was supported by CAPES/Cofecub Project 712/2011.
TSS acknowledges  financial support from CNPq (Proc. No. 142851/2010-8 and 207433/2014-3) and CAPES (Proj: PNPD20132533).
JGH thanks FAPESP (Proc. No. 2010/50930-6 and 2014/18100-4).
BF thanks CNPq project 150281/2017-0
We thank T. Onishi (Osaka University) for having provided us with his 13CO data in advance of publication.
This work has made use of the VizieR, and Aladin databases operated at CDS, Strasbourg, France.
This publication makes use of data products from the Two Micron All Sky Survey, which is a joint project of the University of Massachusetts and the Infrared Processing and Analysis Center/California Institute of Technology, funded by the National Aeronautics and Space Administration and the National Science Foundation.

\end{acknowledgements}






\begin{appendix} 

\section{X-ray properties}

This appendix describes in detail how the X-ray properties, such as hardness 
ratios (HRs),  variability,  spectra and light curves,  were obtained from the {\it XMM-Newton} PN data  (See Sect.2).

\begin{figure*}[h]
\begin{center}
\includegraphics[width=1\columnwidth, angle=0]{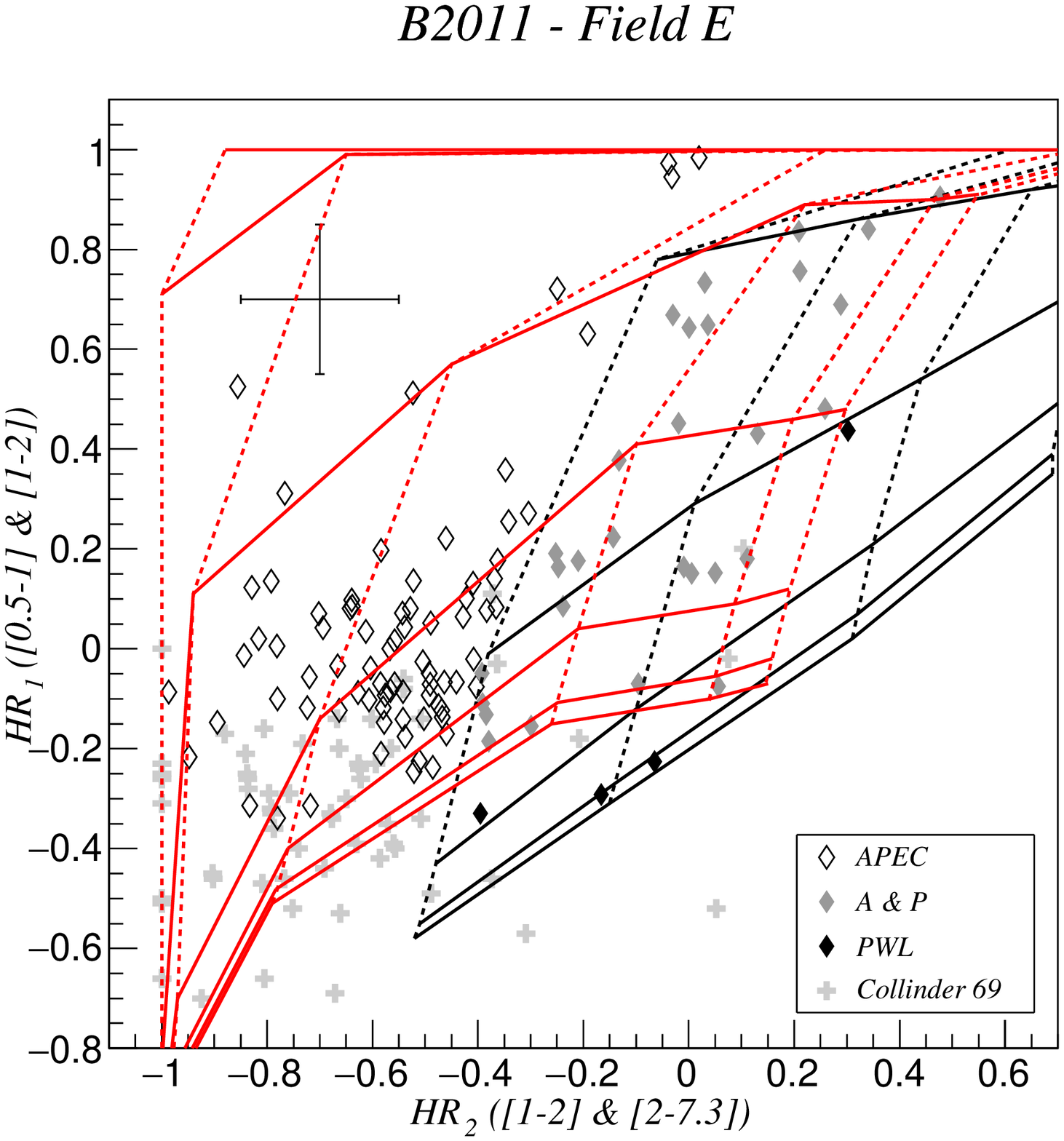}
\includegraphics[width=1\columnwidth, angle=0]{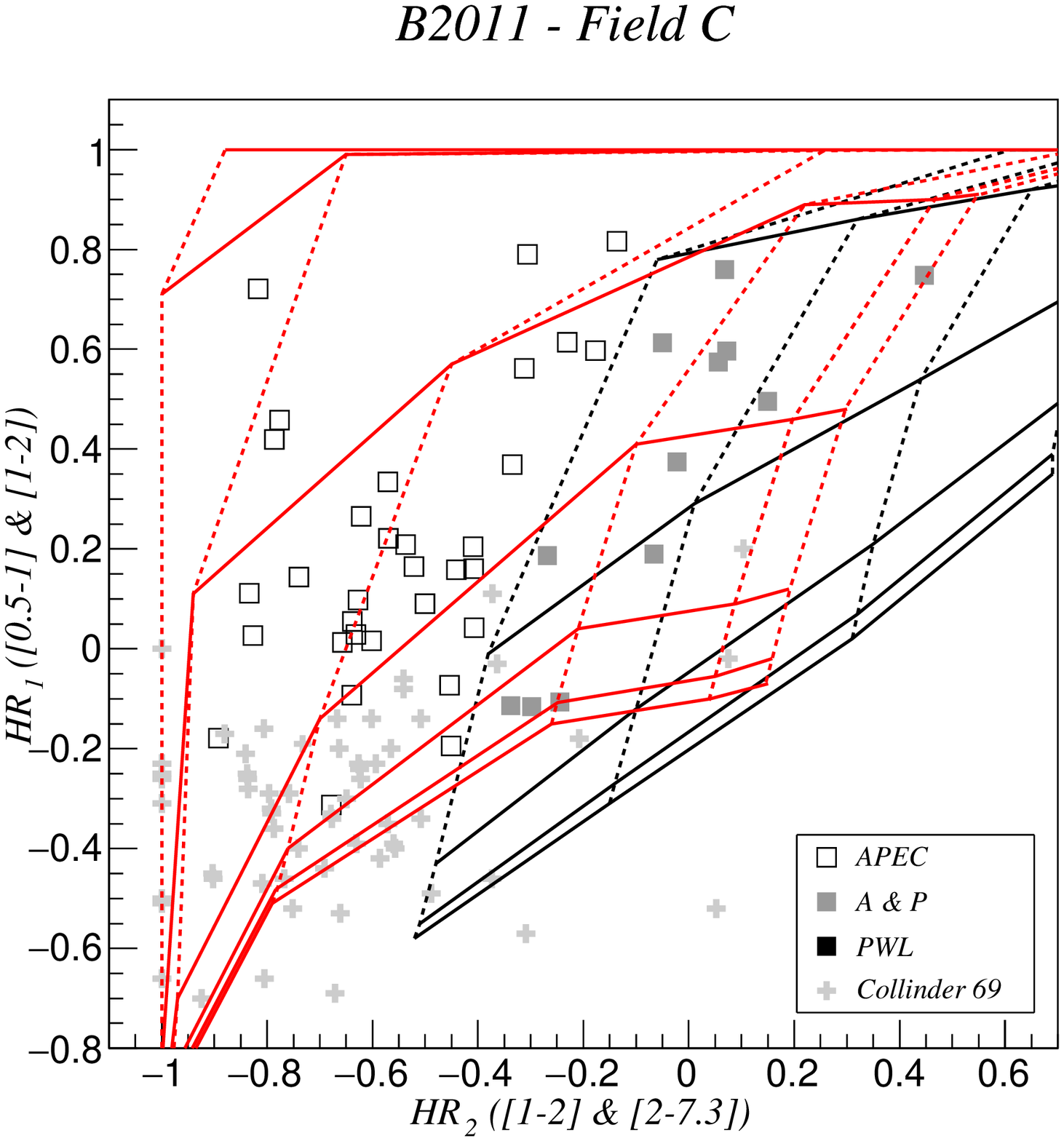}
\includegraphics[width=1\columnwidth, angle=0]{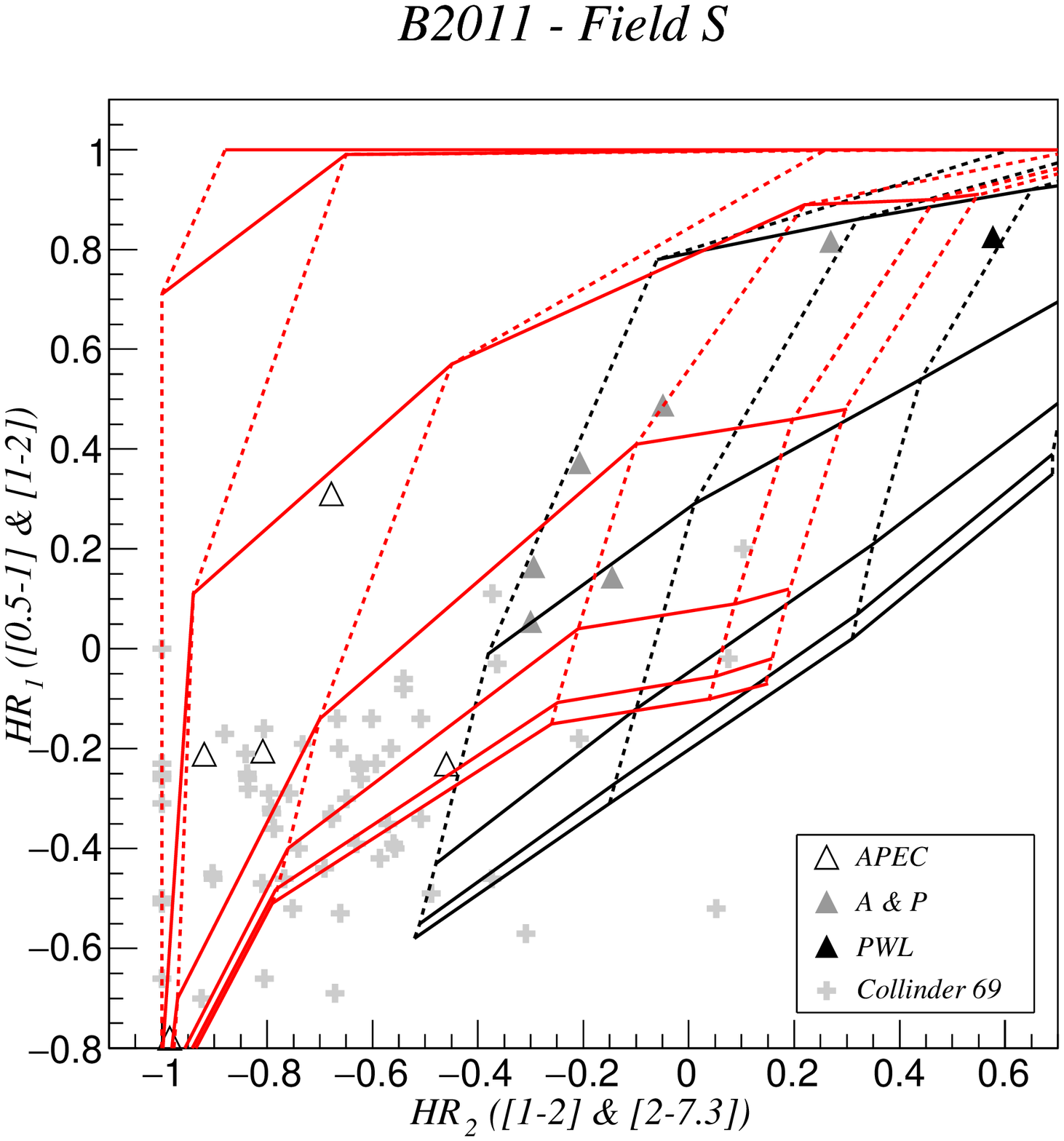}
\includegraphics[width=1\columnwidth, angle=0]{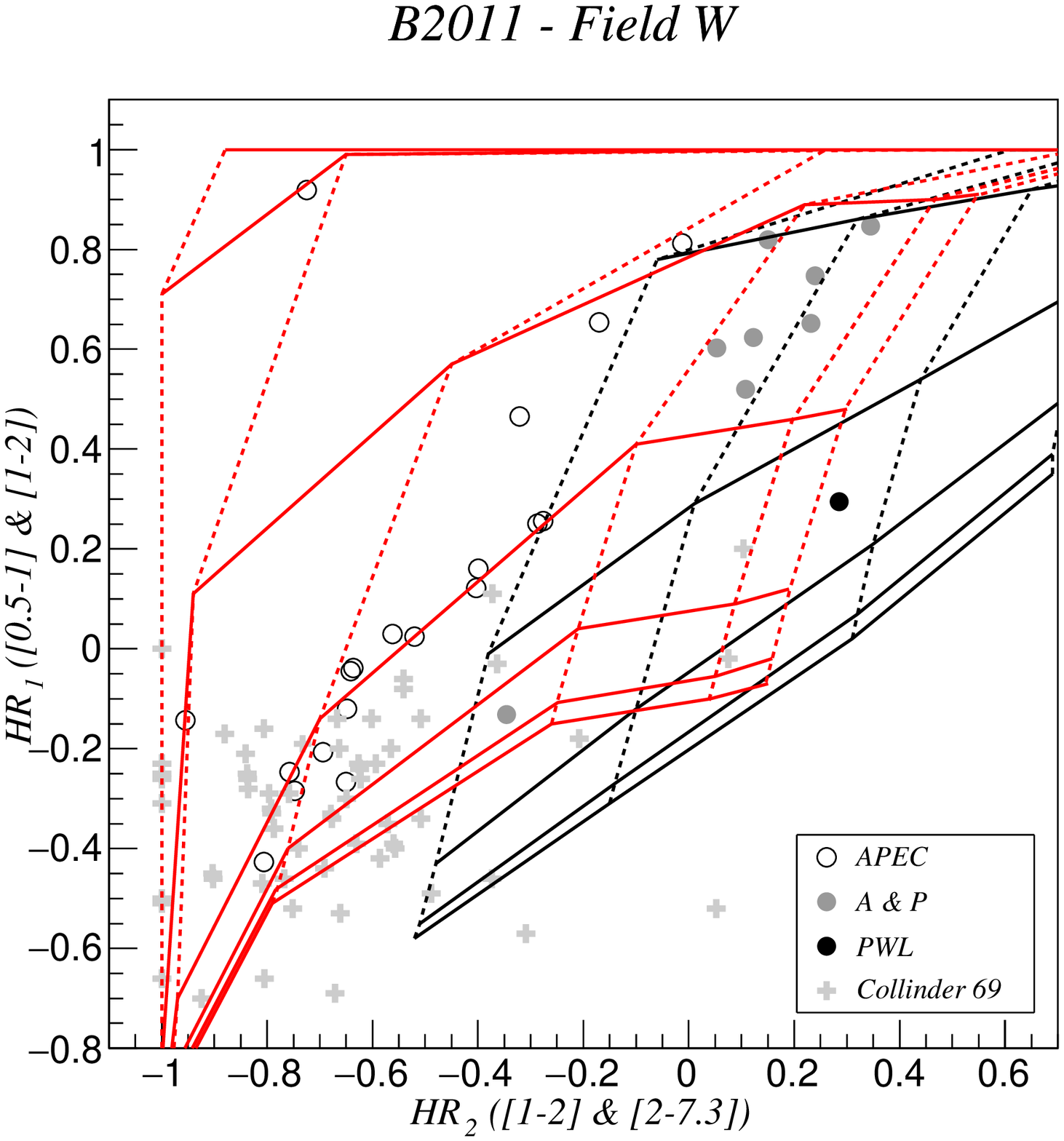}

\caption{Hardness ratio diagrams in 0.5 - 7.3 keV energy band, defined by \citet{2011A&A...526A..21B}, 
comparing our sample with  the stars of the young Collinder 69 cluster (grey crosses). For illustration, error-bars corresponding to 0.15, which is the mean error found in the HR estimates of our sample, are plotted in the first panel. Top panels present the sources from fields E (diamonds) and C (squares), while S (triangles) and W (circles) are shown in the bottom. Grids from the APEC model are displayed in red, by using dotted lines for 6.0 $<$ log T $<$ 8.5 (increasing from left to right, in steps of 0.5).  Power-law grid with $\gamma$ $=$ 0, 1, 2 is shown by black dotted lines (increasing from left to right). Both grids use full lines to represent 20 $<$ log N$_{H}$ $<$ 23, increasing in steps of 0.5, from bottom to top.}

\label{figA.2}
\end{center}
\end{figure*}

\begin{figure*}[ht]
\begin{center}
\includegraphics[width=1\columnwidth, angle=0]{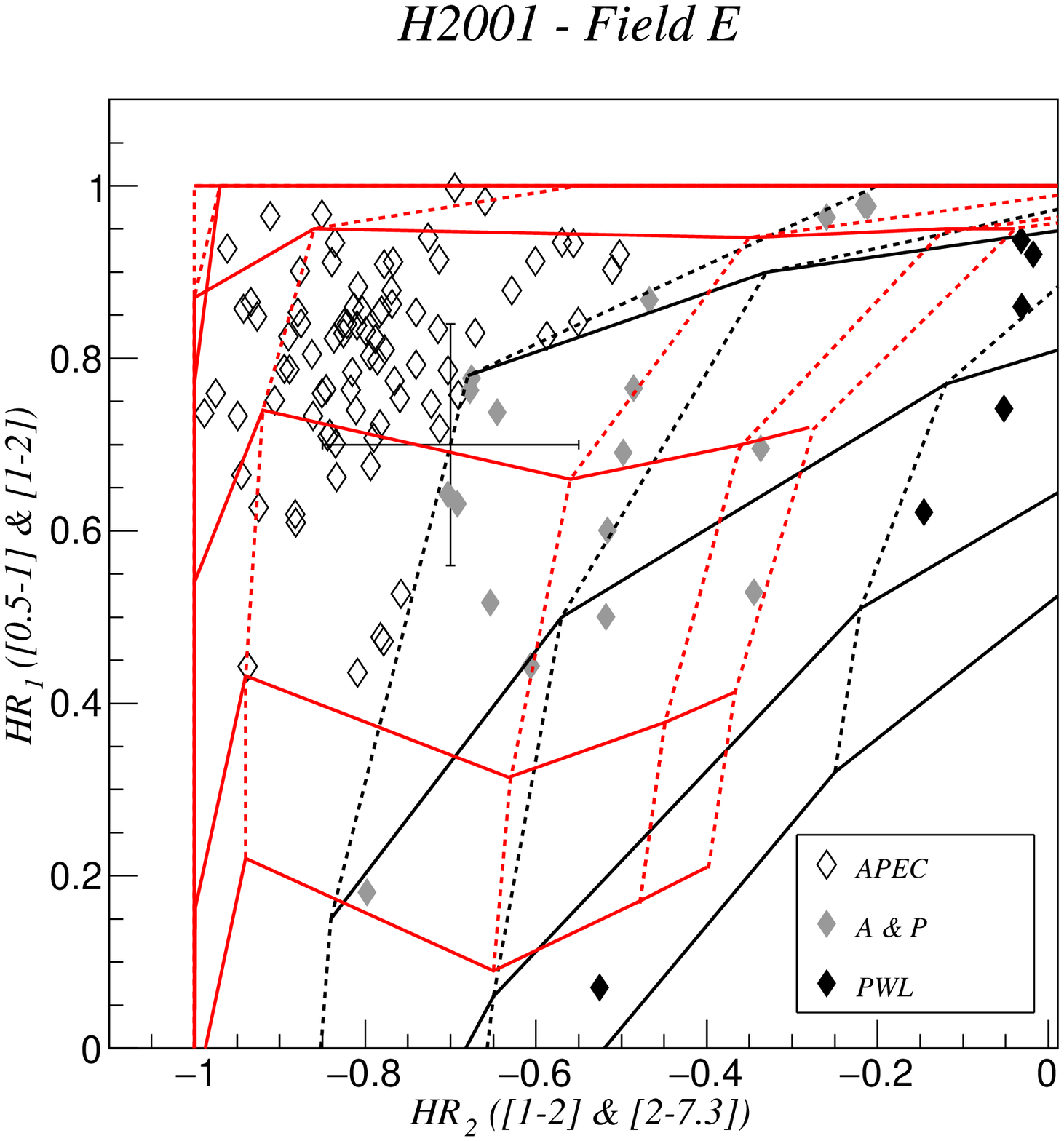}
\includegraphics[width=1\columnwidth, angle=0]{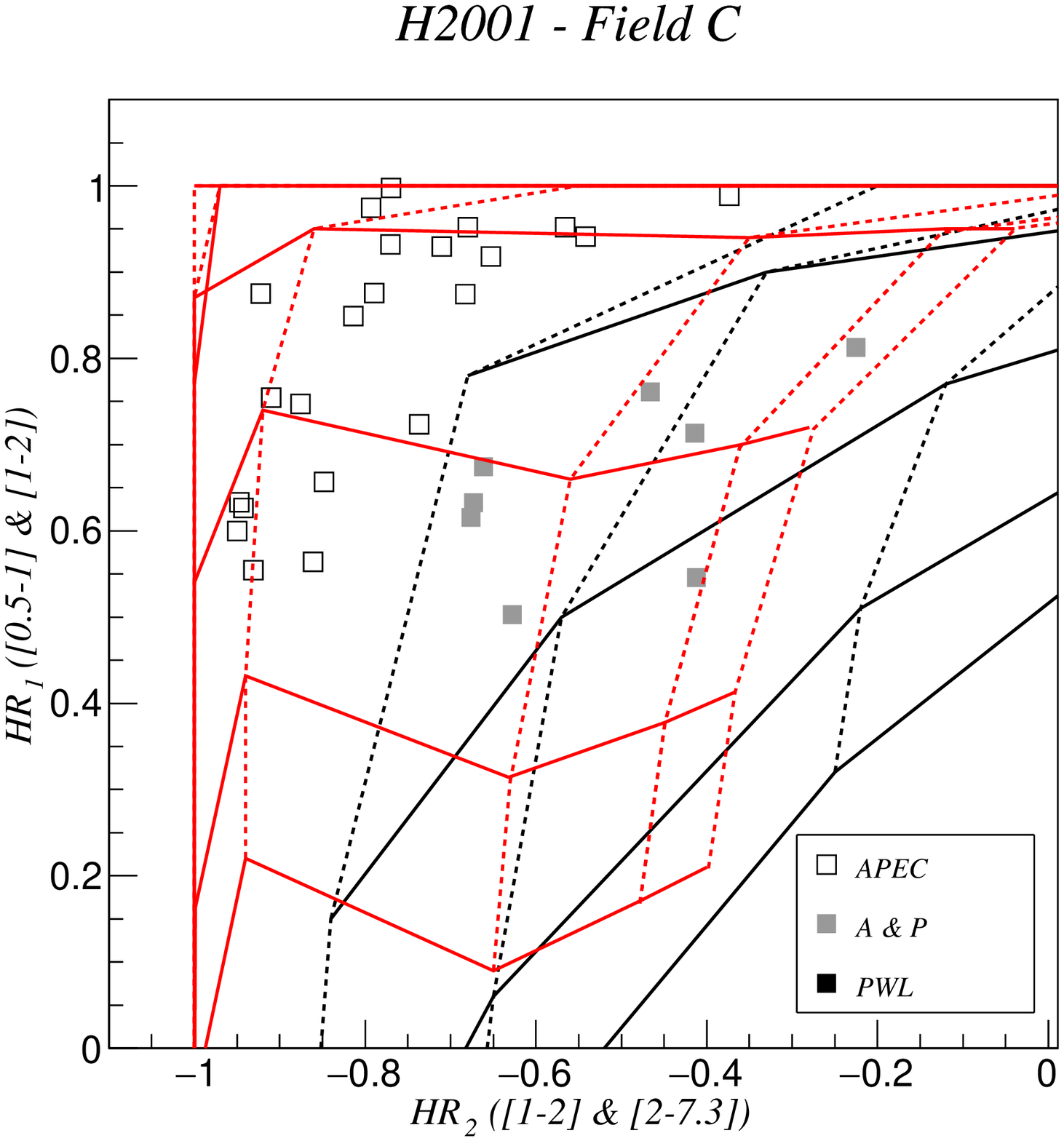}
\includegraphics[width=1\columnwidth, angle=0]{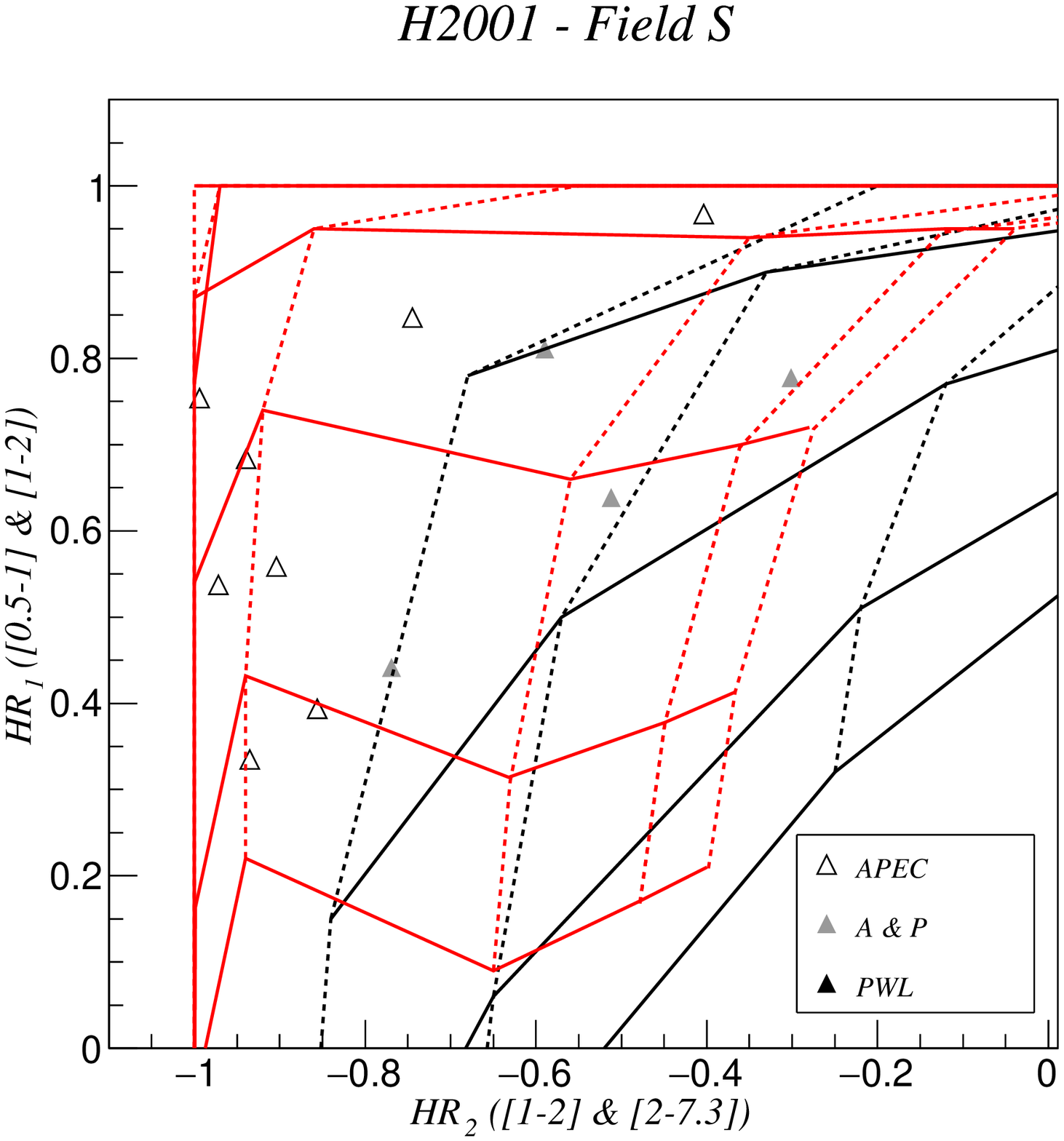}
\includegraphics[width=1\columnwidth, angle=0]{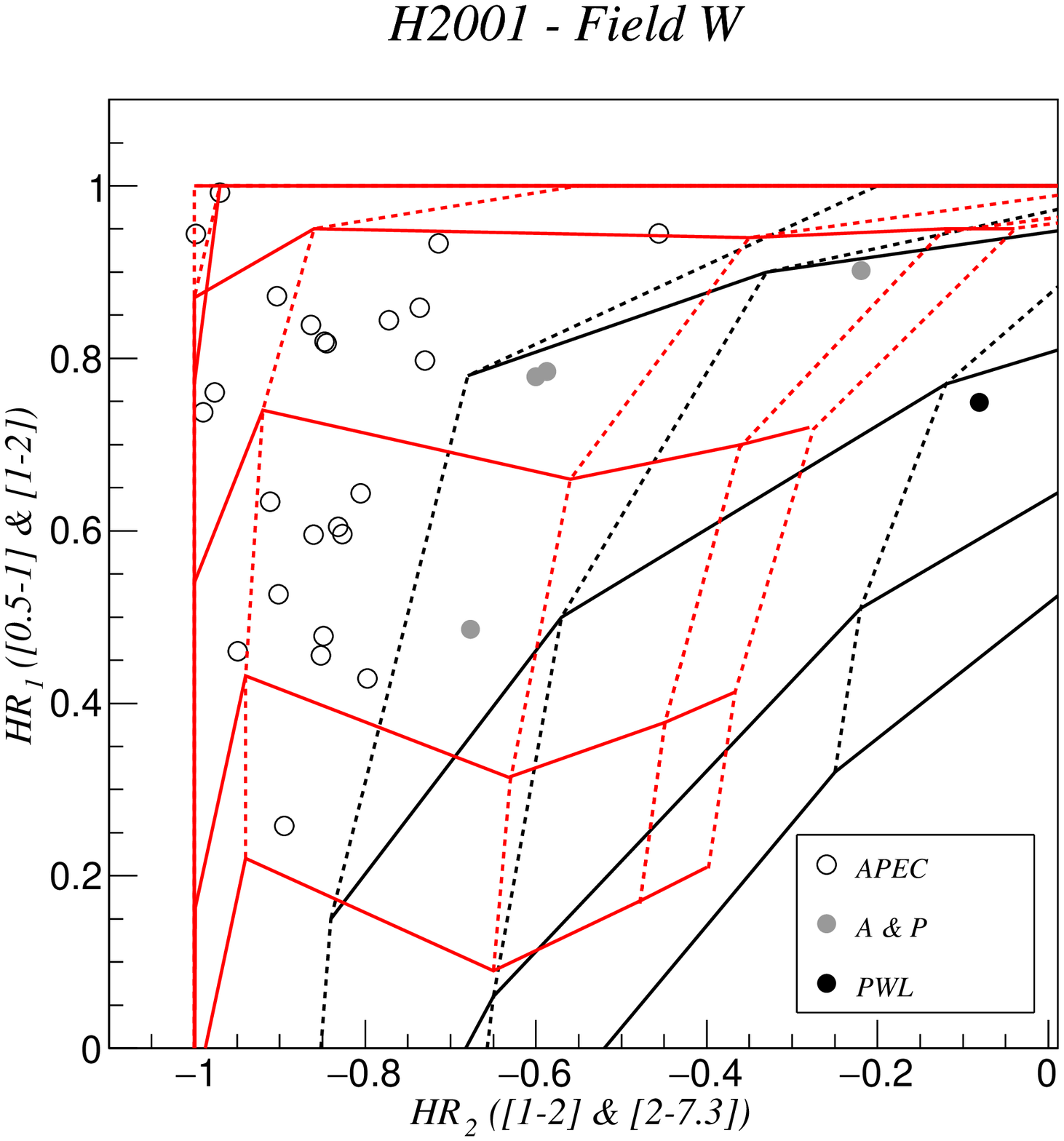}

\caption{The same as Fig. \ref{figA.2} for hardness ratio diagrams based on 0.2 - 4.5 keV energy band suggested by \citet{2001A&A...365L..45H}.}

\label{figA.3}
\end{center}
\end{figure*}


\subsection{Hardness ratios diagrams }


We have adopted the energy bands defined by \citet{2011A&A...526A..21B}  from 0.5 to 7.3 keV and \citet{2001A&A...365L..45H}  from 0.2 to 4.5 keV (see Sect. 2) aiming to distinguish  sources with predominant soft, thermal emission (up to 2 keV),  expected in young stars \citep{1999ARA&A..37..363F}, from sources with hard, non-thermal emission produced by  AGNs\footnote[12]{Typical pulsars, for instance, are very luminous in X-rays, so at the detection level we have here they would have to be very distant, so very improbable.}.

In order to compare the observed HRs with the expected values from a stellar (thermal coronal) emission or extragalactic emission, two models were simulated for the respective energy bands using XSPEC version 12.7.1: a thermal (APEC) and a power-law (PWL) distribution. A multiplicative absorption photoelectric model (PHABS) with log N$_{H}$ between 20 and 23, in steps of 0.5, was applied  to both models. The APEC model was used to simulate the thermal emission for log (T) varying  from 6.0 to 8.5 in steps of 0.5 and  metallicity Z relative to the Sun of Z $=$ 0.2 Z$_{\sun}$. The choice of this metallicity is based on tests that we performed, which provided good fittings with Z varying from 0.04 to 0.34 Z$_{\odot}$. The assumed intermediate value Z $=$ 0.2 Z$_{\odot}$ resulted in differences of less than 10\%  on estimates of  T and N$_{H}$. In the power-law simulations we use the index $\gamma$ = 0, 1, 2 and 3, for F$_{X}$(E) $\propto$ E$^{-\gamma}$. Fig. \ref{figA.2} and \ref{figA.3} show diagrams for Hardness Ratios (HRs) obtained respectively in the bands defined by \citet[][B2011]{2011A&A...526A..21B} and by \citet[][2011]{2001A&A...365L..45H}, described in Sect. 3. The thermal grid is presented by red lines and the power-law grid by black lines. Different colours are used in Fig. \ref{figA.2} and \ref{figA.3} to indicate how the sources were classified as compatible with each grid model.

Our analysis was restricted to the sources with estimates in the range -1 $<$ HR $<$ 1.  These limits avoid sources having soft emission at the same level found in the background (HR $>$ 1), and sources  lacking of emission in the hard bands (HR $<$ 1). Following  \citet{2013A&A...553A..12N}, our sub-sample was also constrained to the sources having HR error-bars smaller than 0.3. Among 340 sources with PN data, only 196 sources with HRs in the B2011 bands (Fig. \ref{figA.2}) fulfil these restriction criteria and could be compared with the grids. Among them, 133 are only compatible with  the APEC grid while 9 with  the PWL grid only. The other 54 coincide with both models. For H2001 HRs diagrams (Fig. \ref{figA.3}) the number of analysed sources is little lower than the subsample studied in the B2011 energy ranges, probably due to  intrinsically low emission of stars in the soft band or high absorption of our observations in the soft band (0.2 to 0.4keV), which gives HR1$_{H2001}$ = -1 for several of the sources. In total, Fig. \ref{figA.3} contains 178 sources, 137 of them are compatible with the APEC grids, 8 for PWL and 33 for both.

Aiming to enlarge the number of classified sources, and to solve inconclusive characterization of sources that coincide with the two model grids, we combined the results from both definitions for energy bands. For each source we adopted the more reliable classification that could be found by comparing its position in both Figs. \ref{figA.2} and \ref{figA.3}. If a given source falls in the models overlapping region of the B2011 HR diagram, for instance, and coincides with the H2001 APEC grid, it is classified as compatible with the APEC model.

We could not discuss the origin of 193 sources because 149 of them have no data available in any HRs diagram and 44 sources remained consistent with both models (APEC and PWL). 
 In total, 194 sources were classified according to their X-ray emission: 
about 6\%  (11/194) of them are compatible only with  the PWL model, so they probably are extragalactic or evolved objects, while 94\% (183/194) may have stellar origin.

The HRs of  $\sim$60\% of the sources are compatible with the APEC  model grid. The majority of them are found in the ranges  21 < log N$_{H}$ < 22 and 6.5 < log T(K) < 7.5, which  corresponds to the plasma temperature expected for low-mass young stars \citep{1999ARA&A..37..363F}. In spite of their larger scattering, in Fig. \ref{figA.2} these objects have HRs comparable with the young stars associated with the  low extinction stellar cluster Collinder 69, studied by \citet{2011A&A...526A..21B}, which are more concentrated in between log N$_{H}$ = 20 - 21.5. The results are discussed in Sect. 3.

\begin{figure}[h]

\begin{center}

\includegraphics[width=0.95\columnwidth, angle=0]{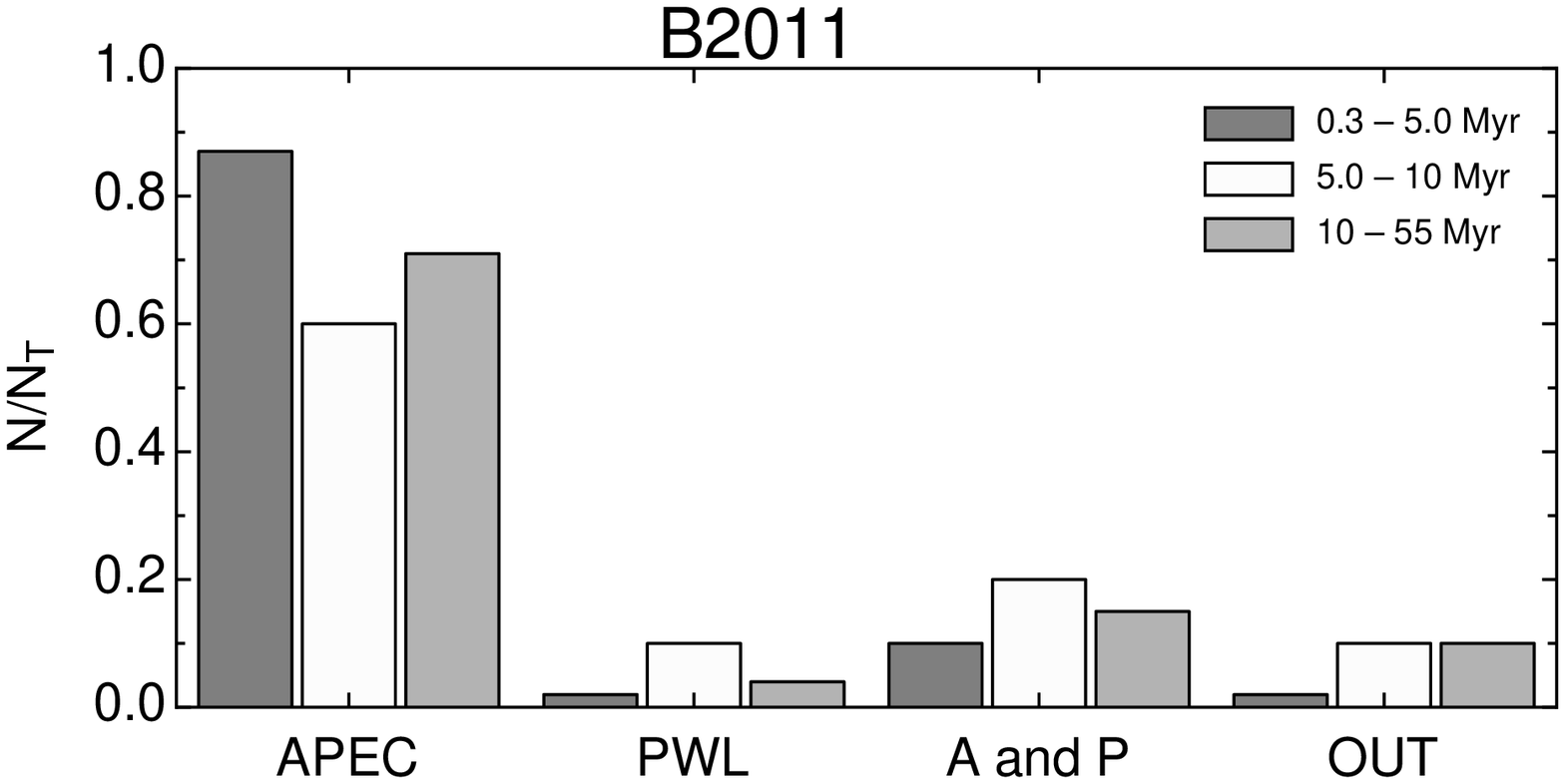}
\includegraphics[width=0.95\columnwidth, angle=0]{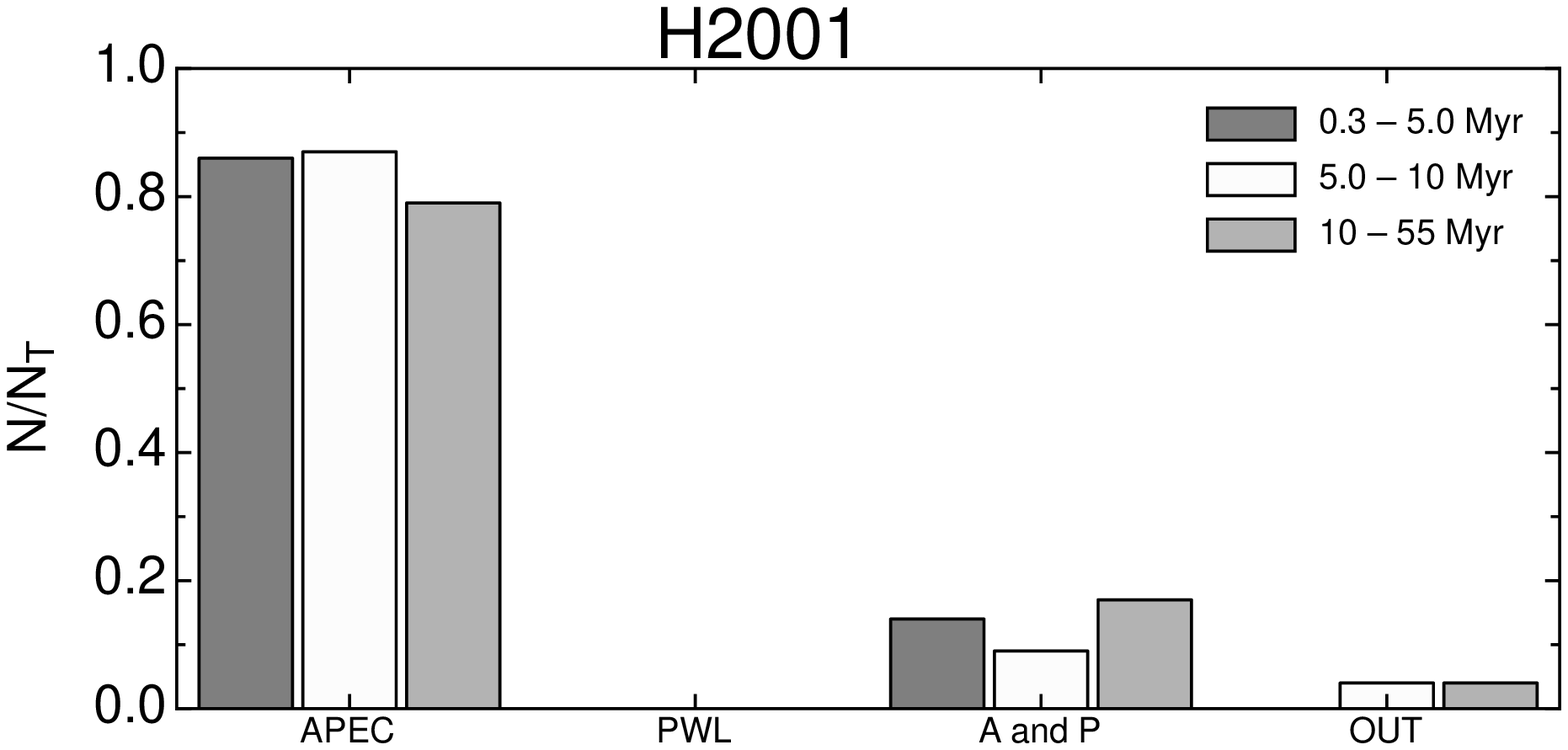}

\caption{ Distribution of sources coinciding with thermal plasma (APEC) or power law (PWL) models, and those appearing in the intersection of both grids (A and P) or 
 outside both grids (OUT).}

\label{figA4}
\end{center}
\end{figure}


\subsubsection{Ages and X-ray emission}


Aiming to discuss the efficiency of adopting the HRs diagrams to select young star candidates, we compare here the results from Sect. 3.1 with  the ages that were determined in Sect. 4.1 and analyzed in Sect. 4.3. In this comparison, we consider only X-ray sources with HRs that are inside the diagrams edges and have NIR counterparts with ages well determined (our ``best sample''). So that 130 and 134 objects were taken into account in B2011 and H2001 diagrams, respectively.

The histograms  in Fig. \ref{figA4} show the distribution of number of sources as a function of age, compared with their position in the HRs diagrams. As expected, most of the objects, in all the PMS ranges of age, correspond to the APEC (stellar emission) model for both HRs diagrams. Only a few sources ($\sim$10\%), mainly for B2011 and those with >5 Myr, coincide with PWL model or appear out of both grids. Finally, Fig. \ref{figA4} also shows 10\% to 20\% of objects with < 10Myr coinciding with both models (APEC and PWL), which is confirmed by both HRs diagrams (B2011 and H2001). This is an indication that these sources must be kept as possible young stars.

 If we consider the sources having infrared excess not included in the ``best sample'', more than 75\% of theses objects are compatible with APEC model in both HRs diagrams, i. e. they may be young stars that are still embedded or have circumstellar disk. Only a few (4\%) coincide with PWL model and the remaining ones (20\%) are compatible with both grids.
 
 These results show that our method, based only on X-ray emission,  is efficient in identifying young star candidates. In fact, the main difficulty of this method is to obtain the HRs in all soft bands.

\begin{figure*}[ht]

\begin{center}
\includegraphics[width=0.78\columnwidth, angle=0]{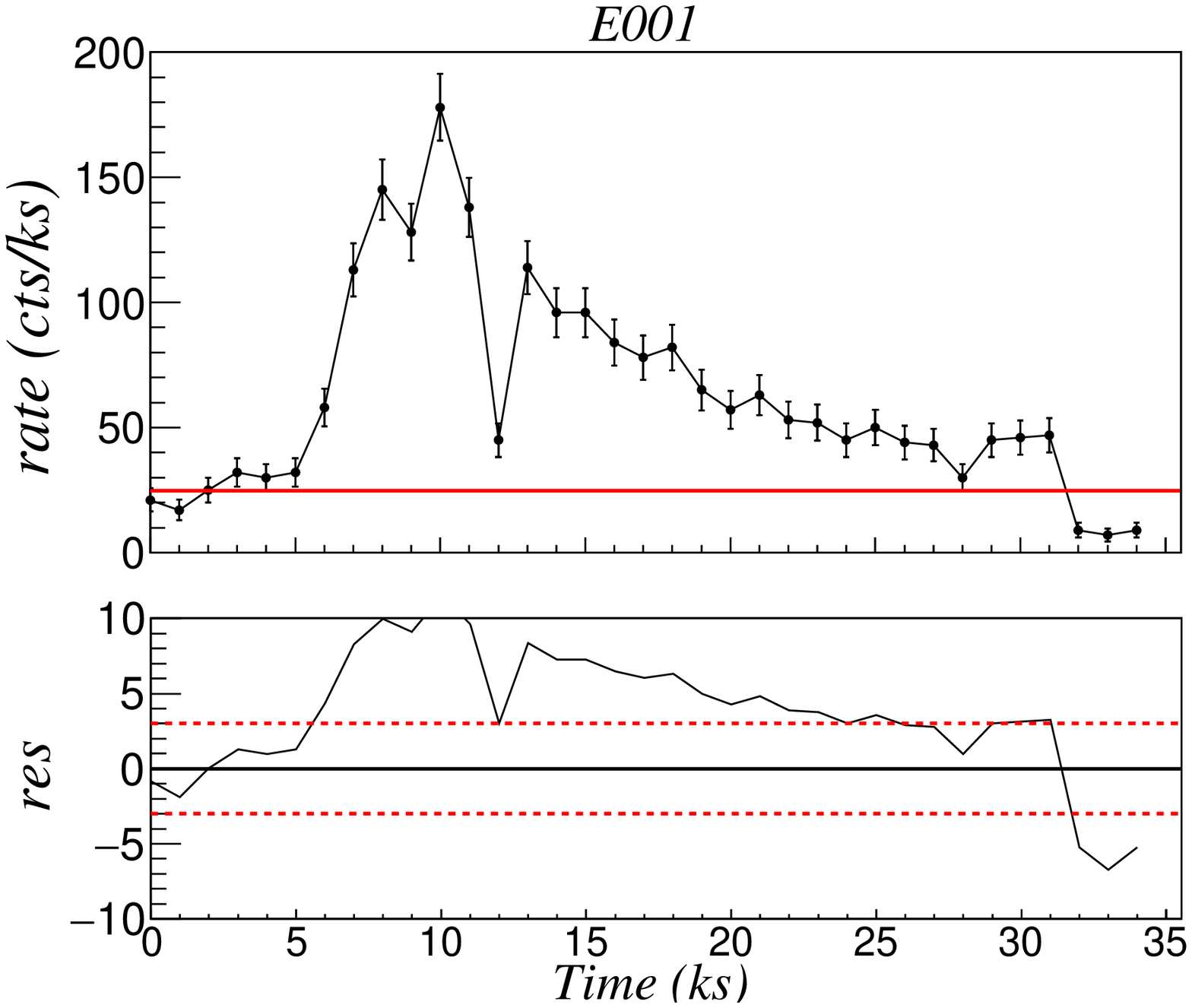}
\includegraphics[width=0.78\columnwidth, angle=0]{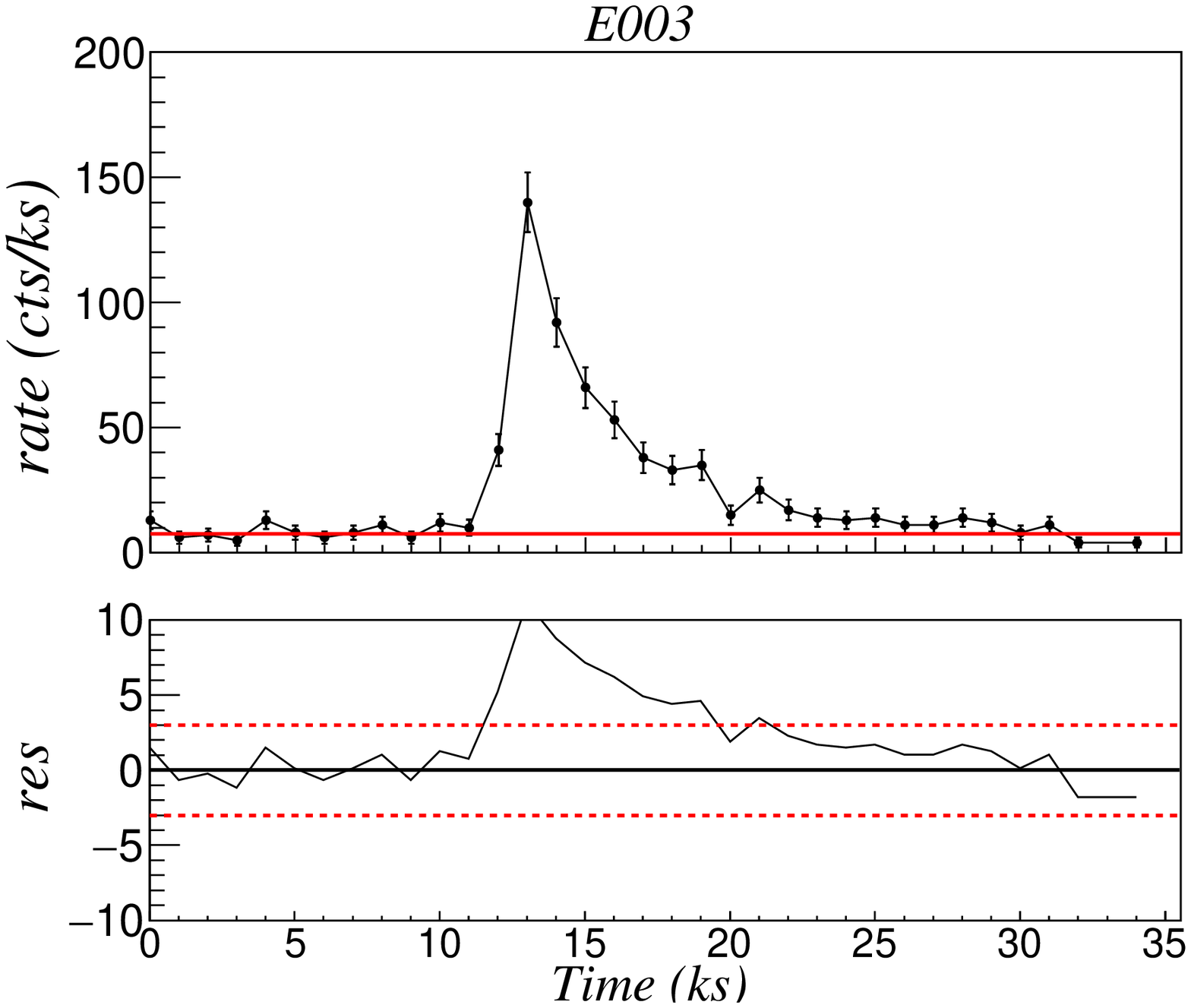}
\includegraphics[width=0.78\columnwidth, angle=0]{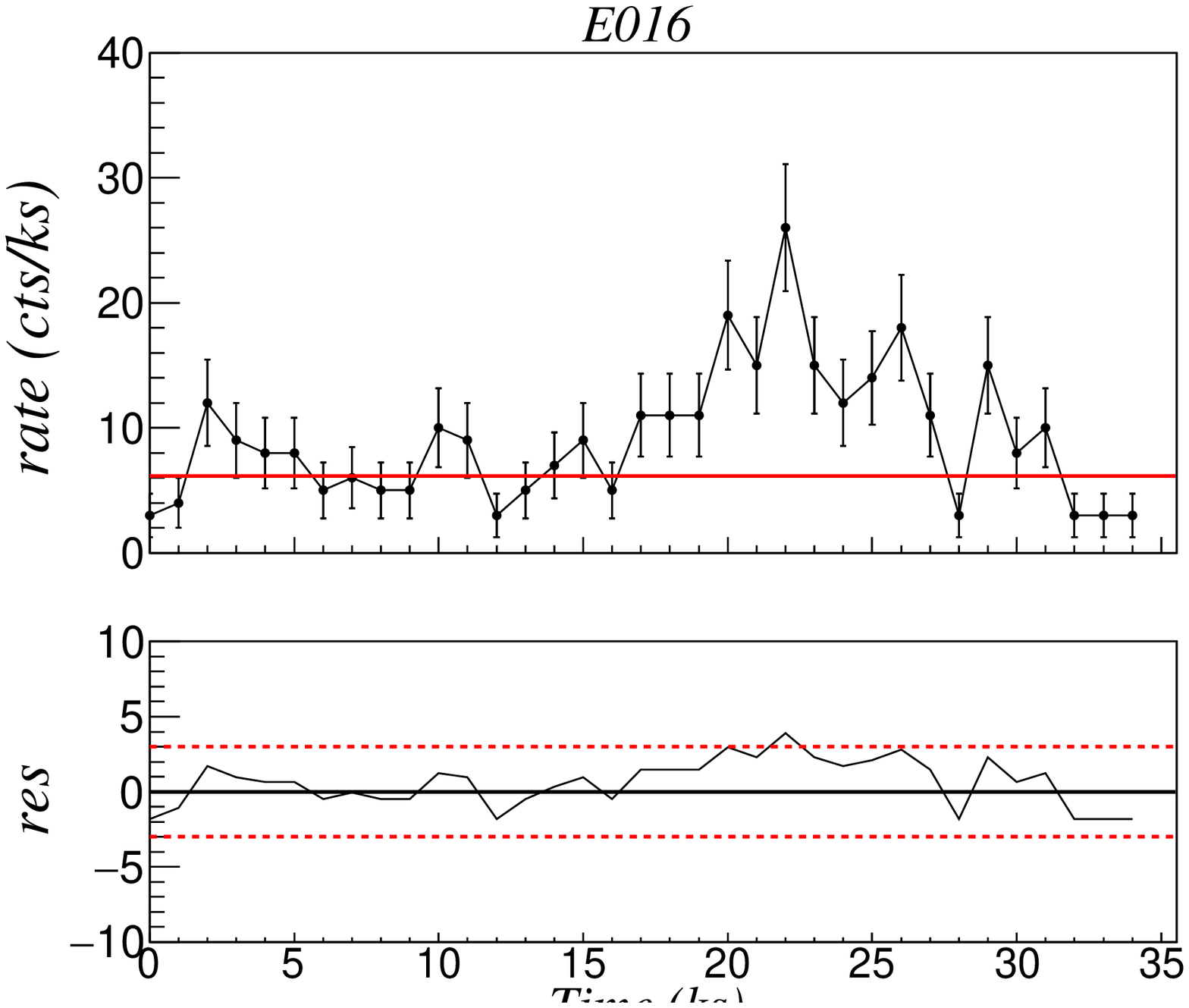}
\includegraphics[width=0.78\columnwidth, angle=0]{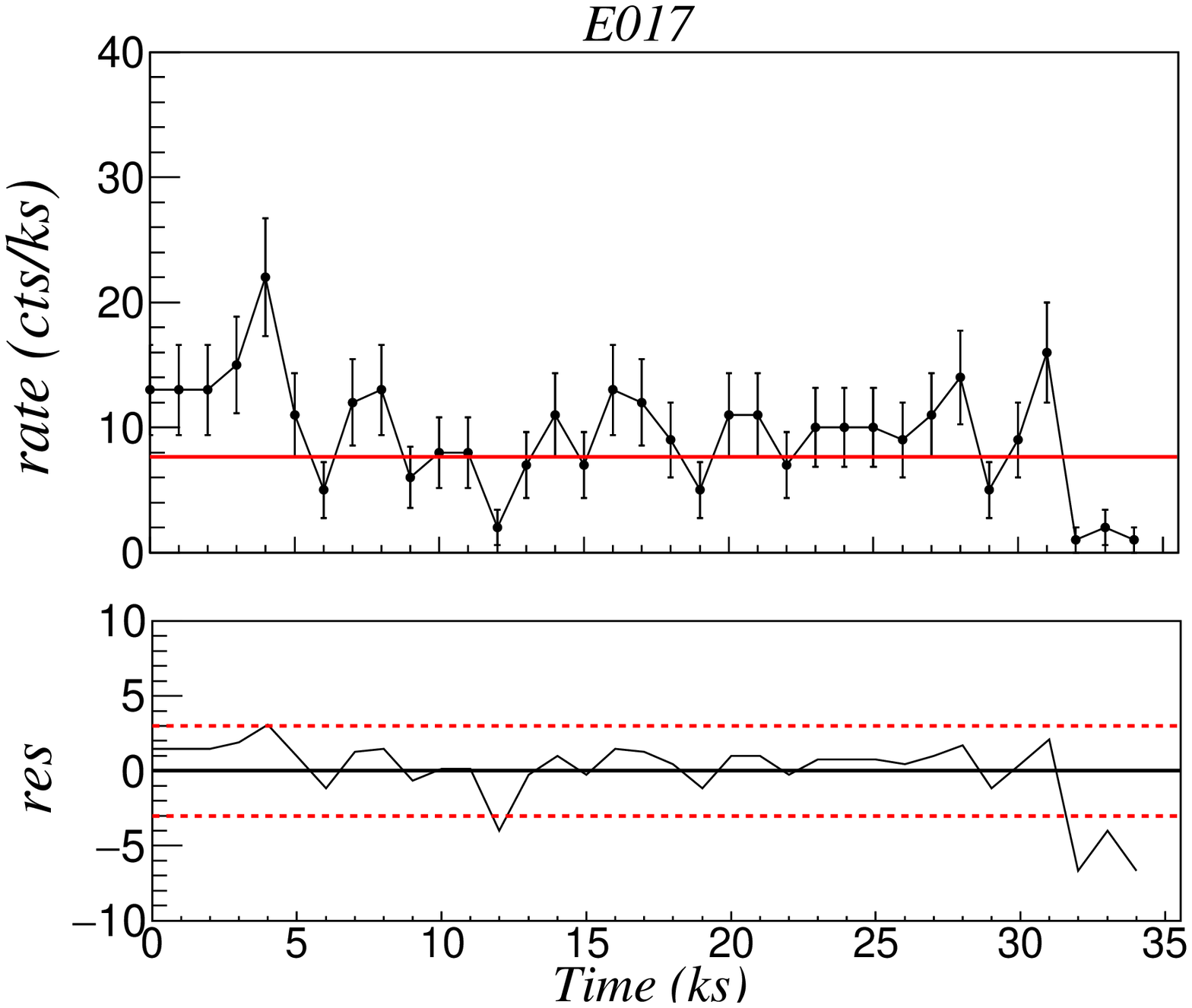}
\includegraphics[width=0.78\columnwidth, angle=0]{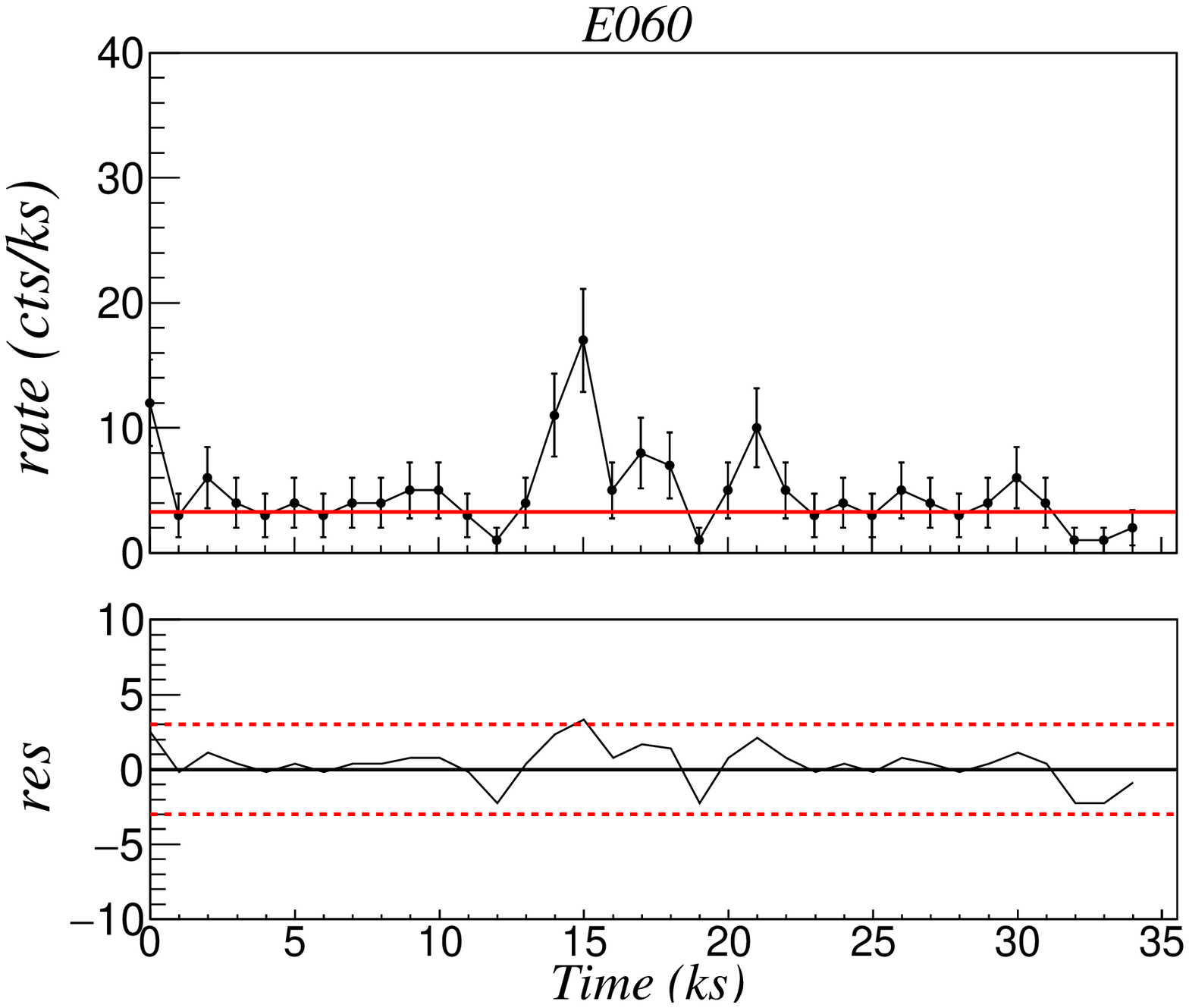}
\includegraphics[width=0.78\columnwidth, angle=0]{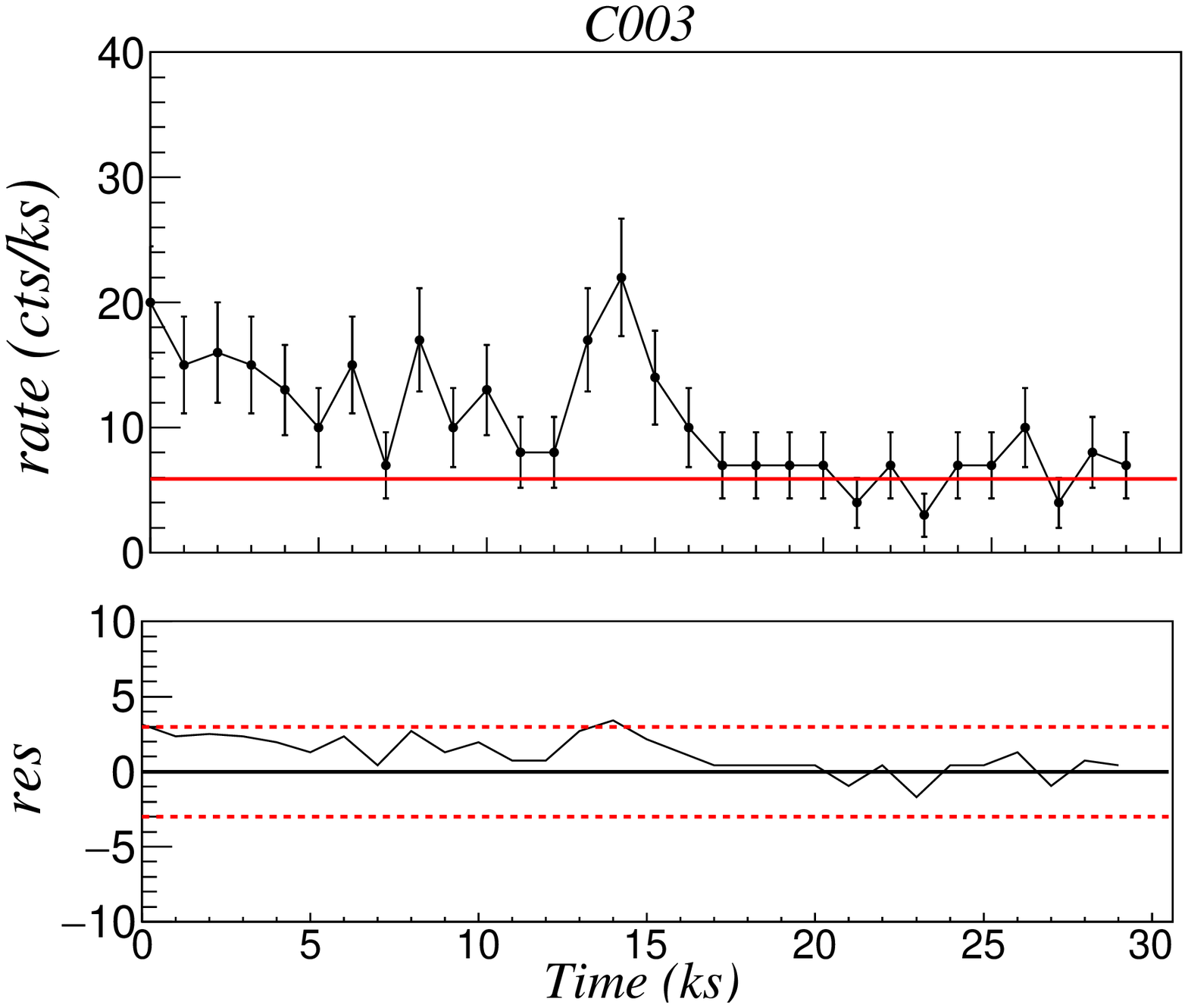}
\includegraphics[width=0.78\columnwidth, angle=0]{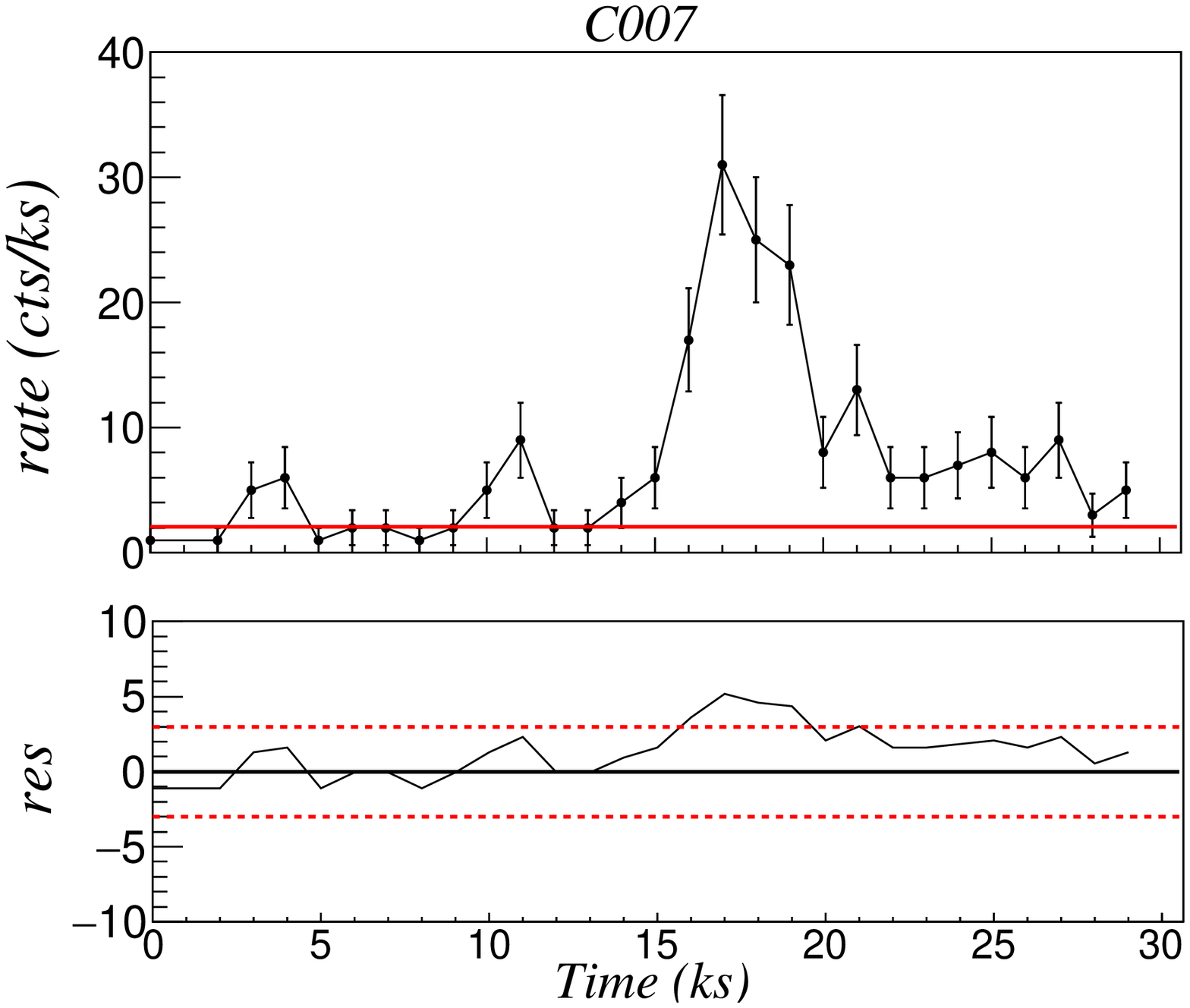}
\includegraphics[width=0.78\columnwidth, angle=0]{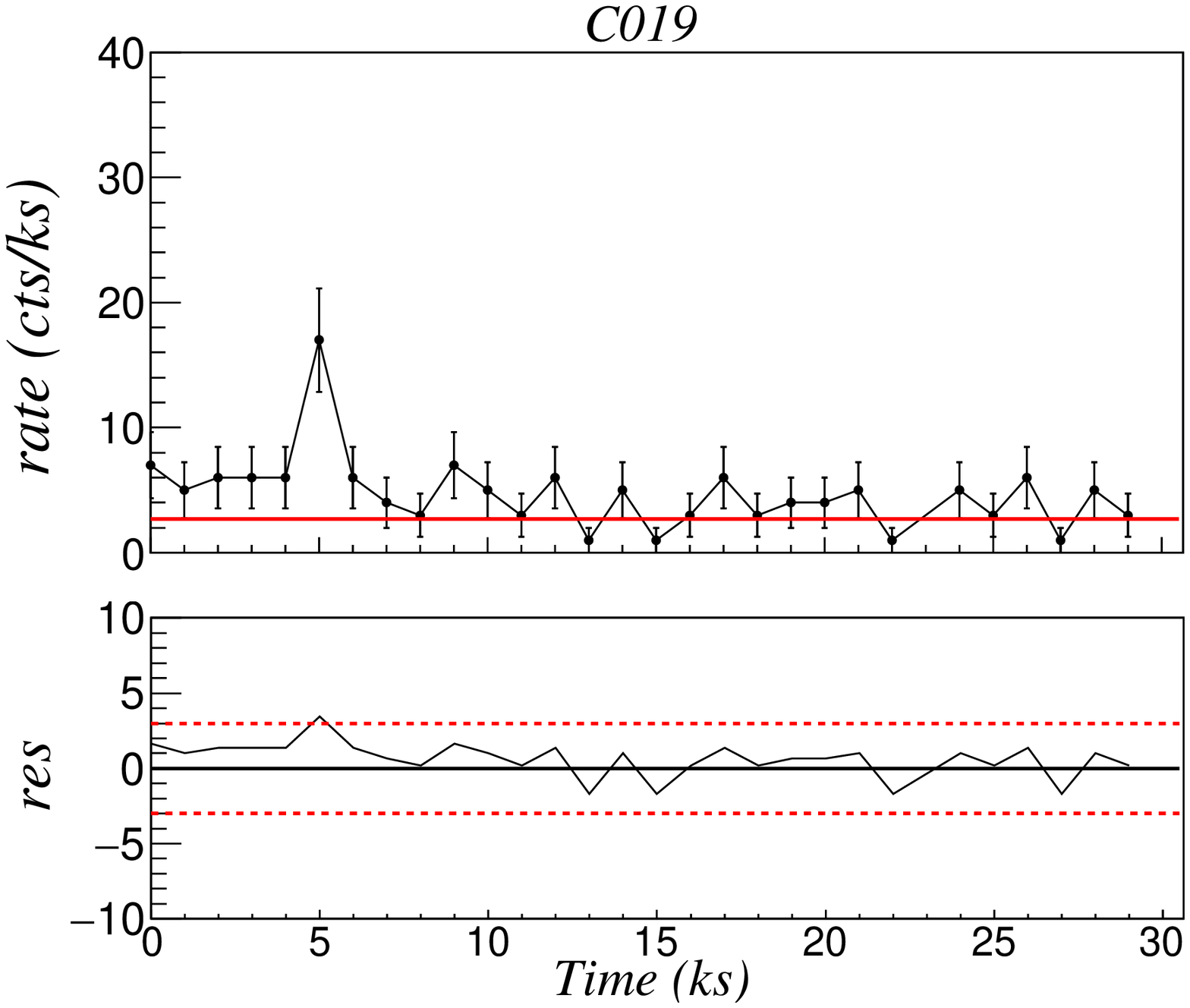}

\caption{{\it XMM-Newton} EPIC/pn light curves in the energy band 0.5 -7.3 keV of X-ray sources with flares. In bottom panels are presented the factor $res \ = \frac{C_{i}-C_{ch}}{\sigma_{i}}$. The time bins correspond to 1ks.}

\label{Lcurve0}
\end{center}
\end{figure*}

\begin{figure*}[ht]

\begin{center}
\includegraphics[width=0.78\columnwidth, angle=0]{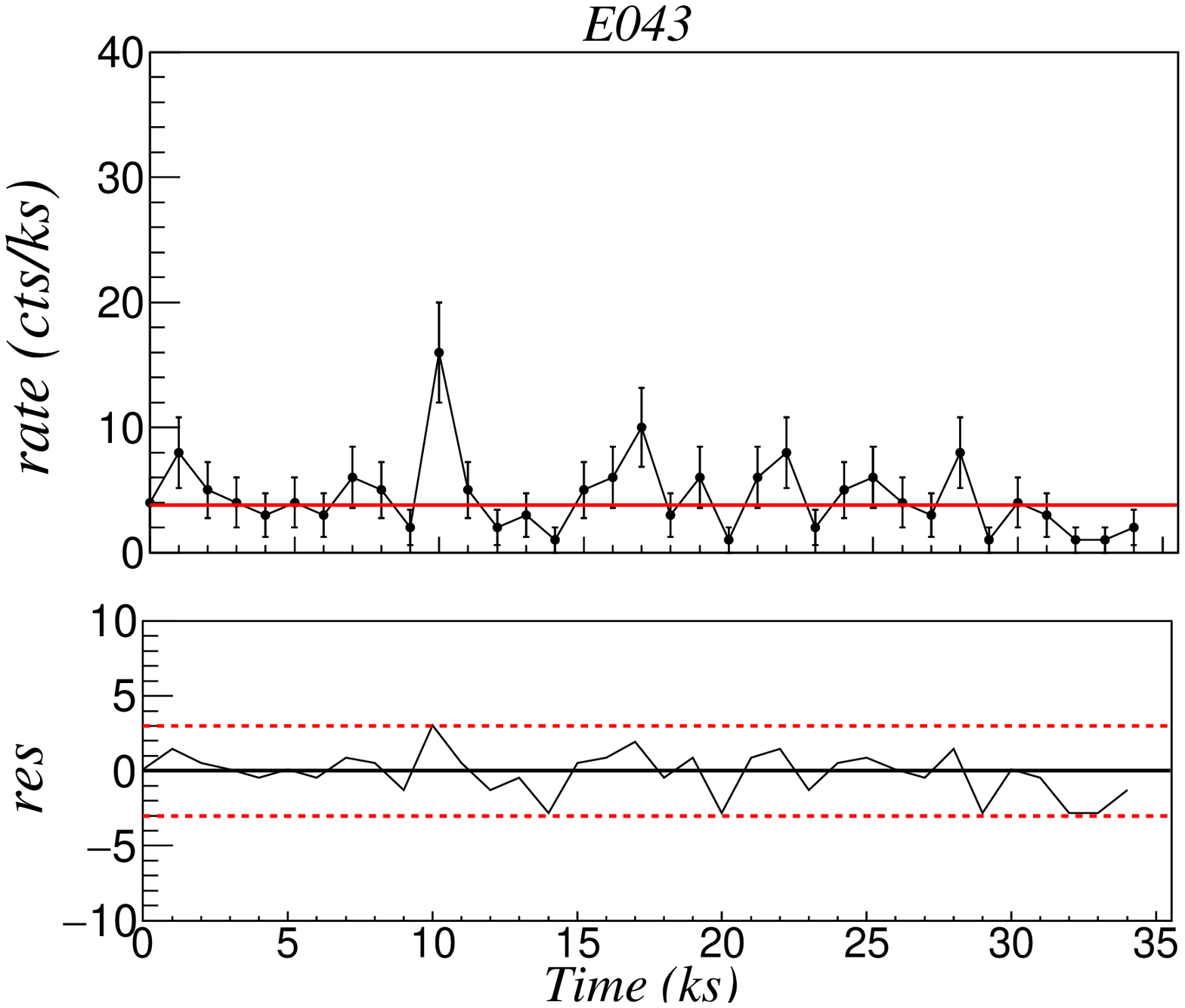}
\includegraphics[width=0.78\columnwidth, angle=0]{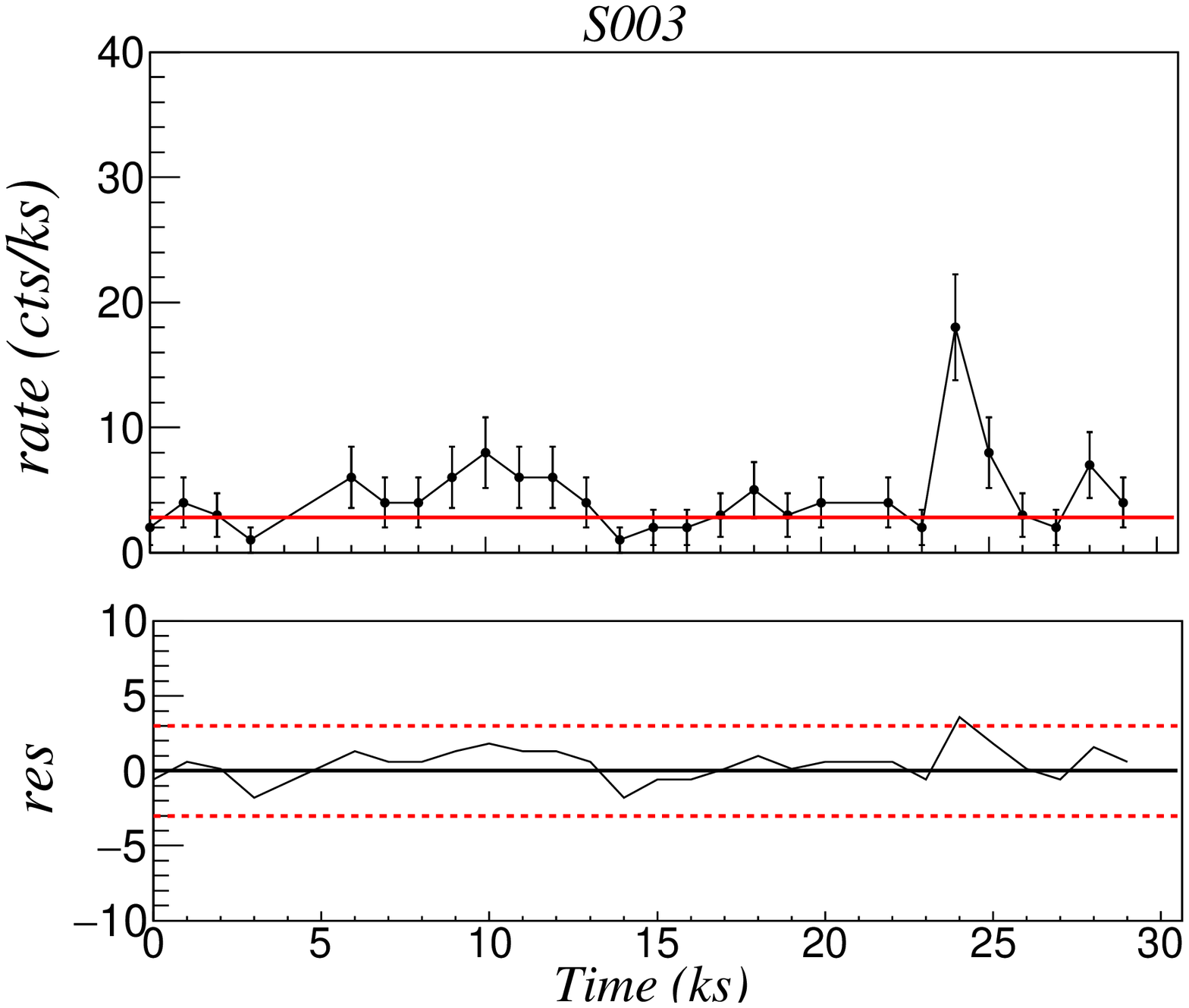}
\includegraphics[width=0.78\columnwidth, angle=0]{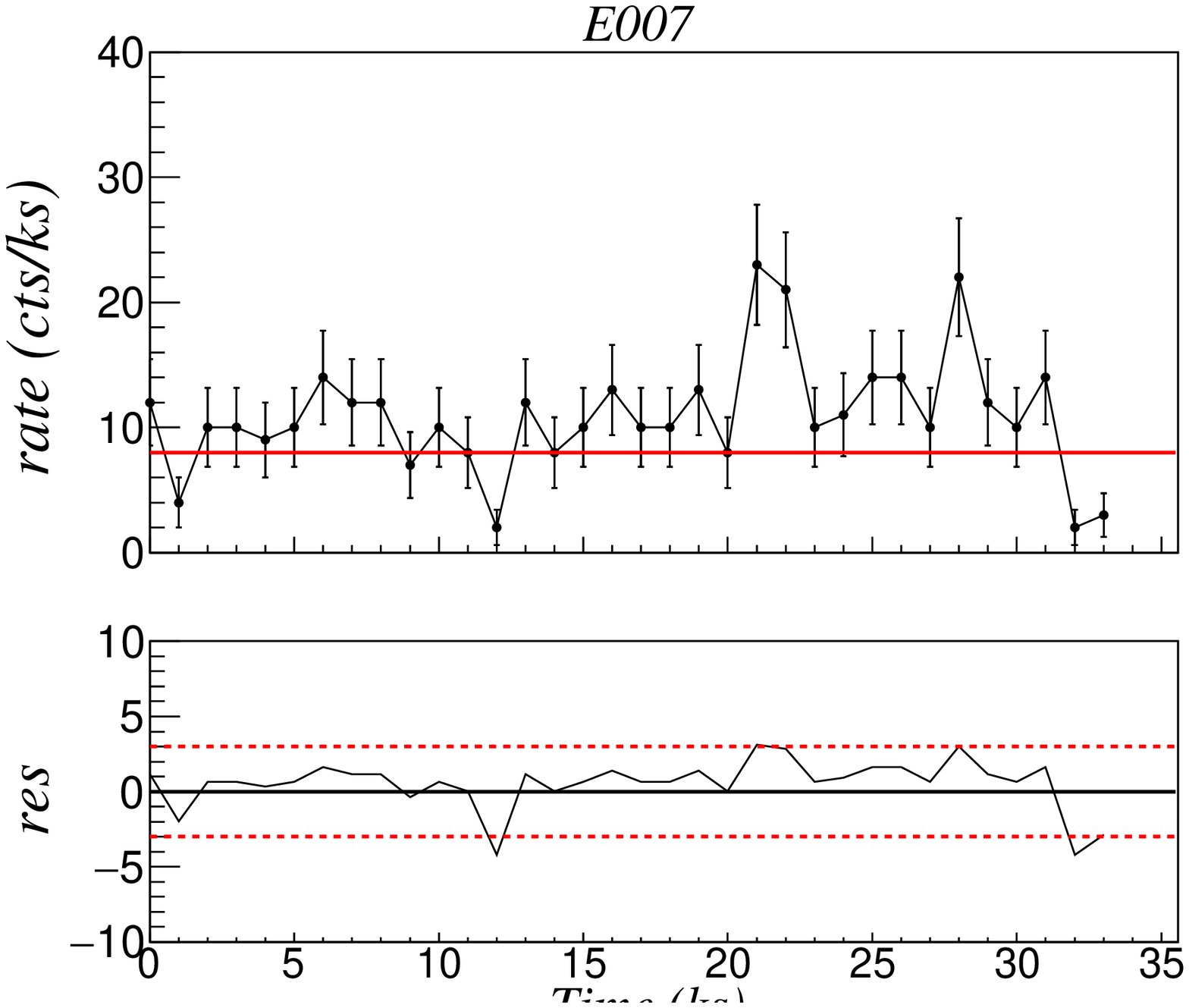}
\includegraphics[width=0.78\columnwidth, angle=0]{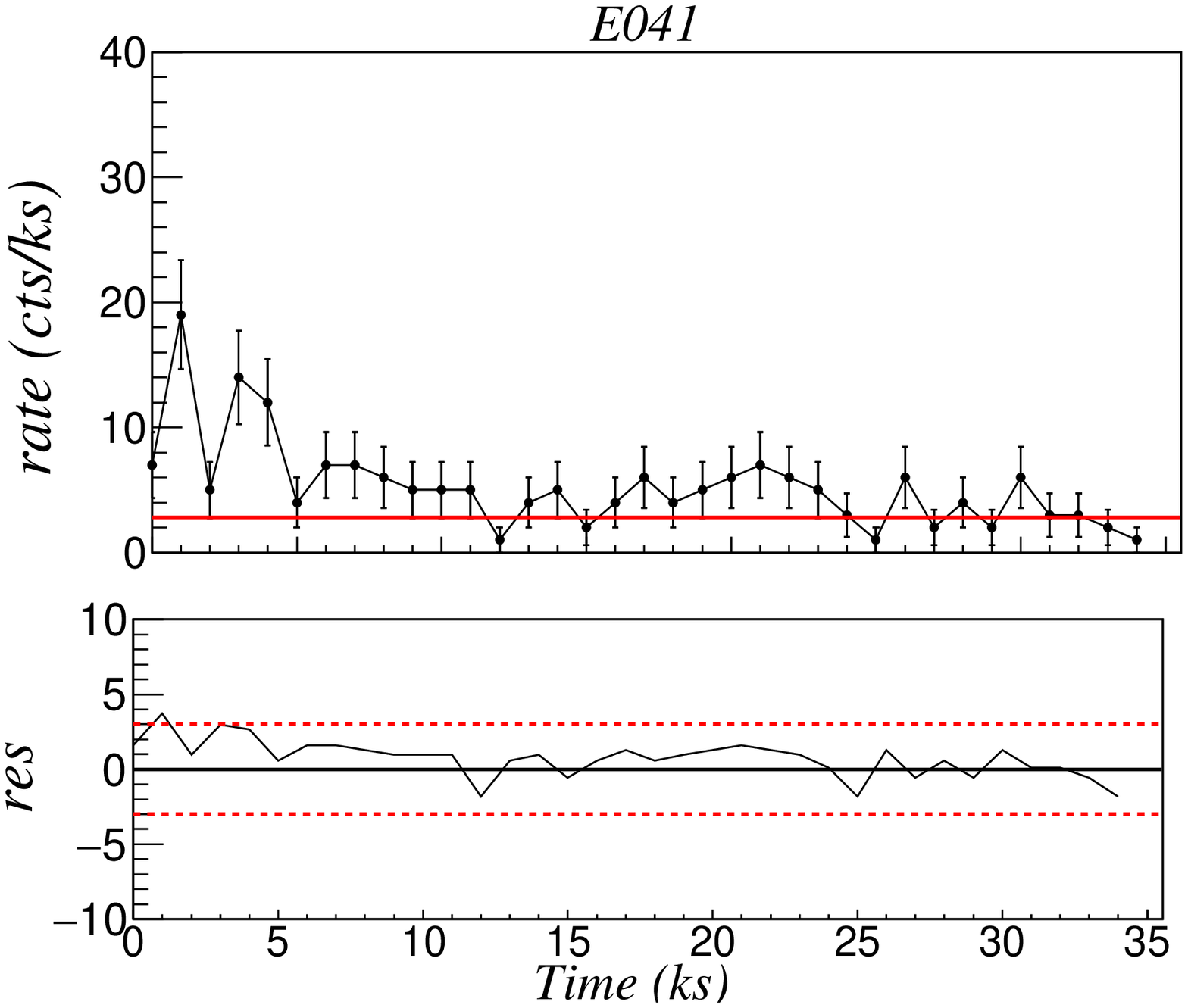}
\includegraphics[width=0.78\columnwidth, angle=0]{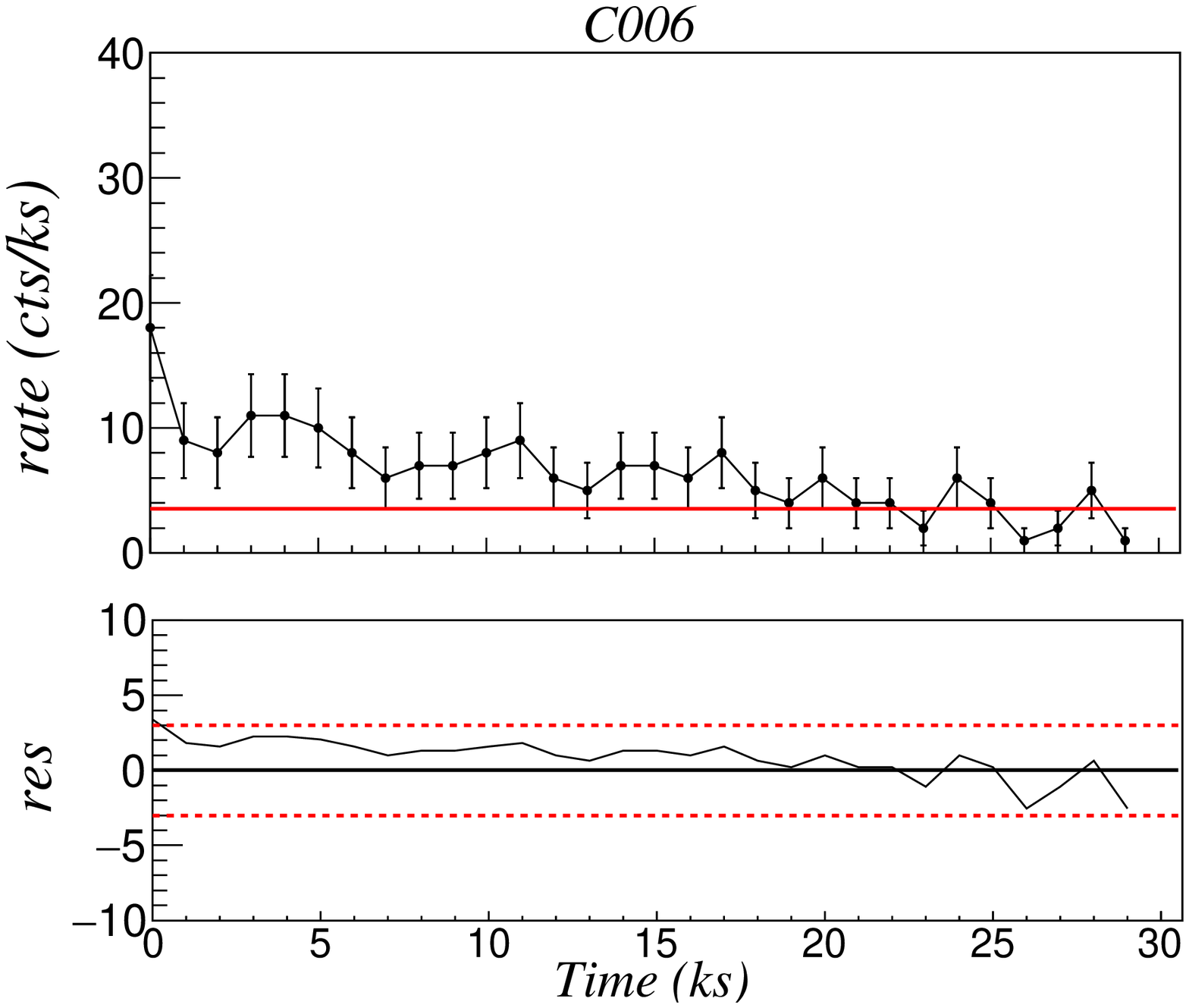}

\caption{{\it XMM-Newton} EPIC/pn light curves in the energy band 0.5 -7.3 keV of X-ray sources with flares and ``partial flares''. In bottom panels are presented the factor $res \ = \frac{C_{i}-C_{ch}}{\sigma_{i}}$. The time bins correspond to 1ks.}

\label{Lcurve2}
\end{center}
\end{figure*}


\subsection{Flares}


The extraction of light curves was made for 340 sources detected with the EPIC-PN camera (typical exposures $\sim$ 30 ks) in the 0.5 - 7.3 keV energy band by estimating the count rate in time bins of 1000s. For each source, the count rate was calculated over an area of the image, 
 the size of which is proportional to the intensity of the X-ray emission. 
The radius of the extraction region varies from r $=$ 10'', for faint sources 
(L$_{X}$ $\sim$ 3 x 10$^{29}$ erg/s) to r $=$ 40'' for the brightest source showing 
flare (L$_{X}$ $\sim$ 4 x 10$^{31}$ erg/s), but typical values were about 15''. 
Following standard routines from SAS\footnote[13] 
{http://xmm.esac.esa.int/sas/current/documentation/threads/timing.-shtml}, the 
background light curve, corresponding to  an area of the same size near the source 
(on the same detector), but free of other sources contamination, 
was subtracted from the  light curve measured at the source position. 
We searched for sources showing signs of variability by adopting the elevated levels $1.2\ \times  \ C_{ch} \ + \ 1.5 \sigma_{ch} \ < \  C_{i} \  < \  2.5 \ \times \ C_{ch} \ + \  1.5 \sigma_{ch}$  proposed by \citet{2005ApJS..160..423W}  to indicate peaks of X-ray emission ($C_{i}$, in a time bin (i)) detected above the characteristic count-rate ($C_{ch}\pm \sigma_{ch}$).  32 objects were found showing two or more peaks of intensity. In order to identify variations caused by flares, in a first step we estimated C$_{ch}$ by the linear fit of count-rates in the bins (1000s each) with lowest count-rates in a time interval.

We consider a {\it flare-like} candidate the event with an increase by factor $res \ = \frac{C_{i}-C_{ch}}{\sigma_{i}}$ higher than 3, where $C_{i}$ and $\sigma_{i}$ are the count-rate and its error in a time interval $i$, respectively. The {\it flare-like} events were confirmed using the definition of flare proposed by  \citet{2005ApJS..160..423W} and \citet{2007A&A...468..463S} that makes use of amplitude (A$_{i}$) and the derivative ($\Delta_{i}$) of a count-rate in a  time interval, both defined as:

\begin{equation}
A_{i} =\frac{C_{i} - 2\sigma_{i}}{C_{ch} + 2\sigma_{ch}}
\label{eq.3.2}
\end{equation}

\begin{equation}
\Delta_{i+1} =\frac{(C_{i+1} - C{i})/C{i}}{MIN[t_{i+1}, t_{i}]}[s^{-1}]
\label{eq.3.3}
\end{equation}

The confirmed flares have amplitude  $ A_{i}$ $>$ 1.5 in one or more consecutive bins and the maximum derivative in these intervals above the 
threshold is $\Delta_{i}$ $>$ 5 $\times$ 10$^{-5}$. 
We also derived the flare luminosity L$_{f}$ subtracting the characteristic luminosity 
L$_{ch}$ from the average value of source luminosity during the flare. The flare energies 
(E$_{f}$) were calculated by multiplying this L$_{f}$  by the time interval 
 during which the X-ray emission is larger than C$_{ch}$.

The light curves of the 13 sources with {\it flare-like} events are presented in Figs~\ref{Lcurve0} and \ref{Lcurve2}. The increasing factor is shown in the bottom panel of these figures. All parameters derived for these events are listed in Table \ref{tabA1}.  Sources E001, E003 and C007 have  shown a strong flare. Sources C006 and E041 shown a decay phase that we called  `partial' flare, while only the source E007 was not confirmed by \citet{2007A&A...468..463S} criteria.


\begin{table*}[ht]
\caption{Parameters of X-ray flares detected on CMa R1 sources}
\begin{center}
{
\begin{tabular}{|l|c|c|c|c|c|c|c|c|}

\hline 
Source 	&	T$^{a}$ 	&	log L$_{ch}$	$^{b}$&	res $^{c}$	&	log $L_{f}$ $^{d}$	&	log $E_{f}$ $^{e}$	&	A$_{f}$ $^{f}$	&	$\Delta_{f}$	 $^{g}$&	Flare 	\\
	&	(ks)	&	[erg/s]	&		&	[erg/s]	&	[erg]	&		&	$[10^{-5}cts/s^{2}]$ 	&		\\ \hline \hline

C003	&	5	&	30.1	&	3.4	&	30.0	&	33.7	&	1.7	&	2262.4	&	Confirmed	\\
C006	&	$>$7	&	29.8	&	3.4	&	$>$29.9	&	$>$33.7	&	-	&	-	&	Partial	\\
C007	&	7	&	29.6	&	4.6	&	30.2	&	34.1	&	7.1	&	5147.1	&	Confirmed	\\
C019	&	2	&	29.7	&	3.5	&	29.7	&	33.0	&	2.6	&	45833.3	&	Confirmed	\\
E001	&	26	&	30.7	&	11.5	&	31.1	&	35.5	&	5.2	&	4340.3	&	Confirmed	\\
E003	&	19	&	30.2	&	11.2	&	30.9	&	35.2	&	12.2	&	20122	&	Confirmed	\\
E007	&	3	&	30.2	&	3.1	&	29.5	&	33.0	&	-	&	-	&	Candidate	\\
E016	&	2	&	30.1	&	3.9	&	29.3	&	32.6	&	2.2	&	3492.1	&	Confirmed	\\
E017	&	4	&	30.0	&	3.6	&	30.3	&	33.9	&	2.2	&	15555.6	&	Confirmed	\\
E041	&	$>$15	&	29.7	&	3.7	&	$>$30.0	&	$>$34.2	&	2.9	&	-	&	Partial	\\
E043**	&	5	&	29.9	&	3.0	&	29.3	&	33.0	&	1.6	&	77777.8	&	Confirmed	\\
E060	&	7	&	29.8	&	3.3	&	29.7	&	33.5	&	2.1	&	3896.1	&	Confirmed	\\
S003**	&	4	&	29.7	&	3.6	&	29.2	&	32.8	&	2.7	&	34782.6	&	Confirmed	\\ \hline
\end{tabular}
}
\label{tabA1}
\end{center}
{\scriptsize
(a) Duration of flares (T);
(b) Characteristic luminosity (L$_{ch}$);
(c) Maximum factor of light curves;
(d) Flare luminosity (L$_f$);
(e) Flare Energy (E$_f$);
(f) Amplitude (A$_{f}$);
(g) Derivative$\Delta_{f}$;
** Sources with two NIR counterparts.

}
\end{table*}



\subsection{Spectra}


 In order to investigate the X-ray emitting plasma we analysed the low-resolution  EPIC 
 spectra integrated  over the whole exposure. They were grouped according to the 
 number of counts,  adopting a threshold of 10 counts per bin, after background 
 subtraction.  We tested fits for 1T and 2T temperature APEC models.  However, due to  intrinsically low luminosity, bad statistics or quality of spectra, only 1T model produced good fits with  Z $=$ 0.2 Z$_{\sun}$  (see Appendix A.1), and photoelectric absorption model (PHABS)  in the XSPEC environment.  The fits and the residual [$\sigma$(data $-$ model) $\times$ $\Delta$($\chi^{2}$)] of observed spectra are shown in Figs. A.7, A.8 and A.9. 
 Best-fits parameters were found by $\chi^{2}$ minimization. Only  for 21 sources 
 a good fit with 0.75 $<$ $\chi^{2}$ $<$ 1.3 was achieved. 
 The  fit parameters (N$_{H}$ and kT1), reduced $\chi^{2}$ and degrees of freedom (d.o.f) are presented in Table \ref{tab_expec}, along with the derived fluxes (and luminosities) in the 0.5 - 7.3 keV band.


\begin{table*} 

\caption{Spectral parameters of CMa R1 bright sources.} 

\begin{center}
\begin{tabular}{|l|c|c|c|c|c|c|}


\hline 

ID		 & 	 N$_{H}$			 & 	kT1			 &  	 $\chi^2$(d.o.f) & F$_{X}$ (10$^{-14}$)	& log (L$_{X}$)	\\
		 & 	(10$^{22}$cm$^{-2}$) & 	(keV)	 &  	& (erg/cm$^2$/s) 	&	 			\\ \hline\hline

C001	& 	0.36	$^{+	0.14	}_{	-0.08	}$ &  	1.25	$^{+	0.12	}_{	-0.26	}$ &  	1.20	(	39	) & 	4.7	$^{+	1.03	}_{	-0.78	}$ & 	30.8	\\
C005	& 	0.13	$^{+	0.05	}_{	-0.05	}$ &  	1.49	$^{+	0.36	}_{	-0.28	}$ &  	0.84	(	16	) & 	2.6	$^{+	0.56	}_{	-0.50	}$ & 	30.5	\\
E002	& 	0.14	$^{+	0.03	}_{	-0.03	}$ &  	2.04	$^{+	0.61	}_{	-0.27	}$ &  	0.97	(	63	) & 	11.9	$^{+	1.28	}_{	-1.21	}$ & 	31.2	\\
E004	& 	0.16	$^{+	0.05	}_{	-0.04	}$ &  	1.79	$^{+	0.33	}_{	-0.25	}$ &  	0.89	(	47	) & 	9.2	$^{+	1.35	}_{	-1.22	}$ & 	31.0	\\
E005	& 	0.07	$^{+	0.03	}_{	-0.03	}$ &  	1.55	$^{+	0.19	}_{	-0.28	}$ &  	1.25	(	47	) & 	5.8	$^{+	0.76	}_{	-0.76	}$ & 	30.8	\\
E015	& 	0.12	$^{+	0.07	}_{	-0.05	}$ &  	1.58	$^{+	0.49	}_{	-0.27	}$ &  	0.83	(	32	) & 	4.6	$^{+	0.83	}_{	-0.77	}$ & 	30.7	\\
E019	& 	0.34	$^{+	0.13	}_{	-0.10	}$ &  	2.10	$^{+	1.10	}_{	-0.49	}$ &  	1.09	(	25	) & 	3.7	$^{+	0.92	}_{	-0.86	}$ & 	30.7	\\
E020	& 	0.07	$^{+	0.05	}_{	-0.04	}$ &  	1.25	$^{+	0.29	}_{	-0.25	}$ &  	1.01	(	28	) & 	2.7	$^{+	0.62	}_{	-0.55	}$ & 	30.5	\\
E021	& 	0.30	$^{+	0.13	}_{	-0.15	}$ &  	0.70	$^{+	0.14	}_{	-0.10	}$ &  	0.91	(	15	) & 	2.7	$^{+	1.59	}_{	-1.16	}$ & 	30.5	\\
E022	& 	0.08	$^{+	0.06	}_{	-0.05	}$ &  	1.64	$^{+	1.86	}_{	-0.38	}$ &  	1.01	(	20	) & 	2.4	$^{+	0.56	}_{	-0.49	}$ & 	30.5	\\
E023	& 	0.08	$^{+	0.08	}_{	-0.05	}$ &  	1.01	$^{+	0.12	}_{	-0.09	}$ &  	1.02	(	24	) & 	2.7	$^{+	0.80	}_{	-0.60	}$ & 	30.5	\\
E026	& 	0.61	$^{+	0.17	}_{	-0.17	}$ &  	0.95	$^{+	0.22	}_{	-0.14	}$ &  	0.94	(	31	) & 	3.1	$^{+	1.37	}_{	-1.03	}$ & 	30.6	\\
E034	& 	0.08	$^{+	0.07	}_{	-0.05	}$ &  	1.91	$^{+	1.66	}_{	-0.72	}$ &  	1.31	(	15	) & 	2.0	$^{+	0.45	}_{	-0.45	}$ & 	30.4	\\
E036	& 	0.12	$^{+	0.20	}_{	-0.10	}$ &  	1.23	$^{+	0.28	}_{	-0.18	}$ &  	0.84	(	20	) & 	2.4	$^{+	1.21	}_{	-0.73	}$ & 	30.5	\\
E038	& 	0.13	$^{+	0.11	}_{	-0.09	}$ &  	1.59	$^{+	1.25	}_{	-0.33	}$ &  	1.10	(	14	) & 	4.5	$^{+	1.28	}_{	-1.09	}$ & 	30.7	\\
E044	& 	0.54	$^{+	0.24	}_{	-0.13	}$ &  	0.54	$^{+	0.09	}_{	-0.20	}$ &  	1.21	(	11	) & 	1.2	$^{+	3.10	}_{	-0.47	}$ & 	30.2	\\
E055	& 	0.11	$^{+	0.08	}_{	-0.06	}$ &  	1.25	$^{+	0.44	}_{	-0.51	}$ &  	1.08	(	18	) & 	1.9	$^{+	0.56	}_{	-0.49	}$ & 	30.4	\\
W001	& 	0.04	$^{+	0.01	}_{	-0.01	}$ &  	0.99	$^{+	0.02	}_{	-0.02	}$ &  	0.99	(	237	) & 	54.5	$^{+	2.36	}_{	-2.28	}$ & 	31.8	\\
W002	& 	0.04	$^{+	0.05	}_{	-0.04	}$ &  	1.18	$^{+	0.16	}_{	-0.17	}$ &  	1.14	(	17	) & 	2.4	$^{+	0.59	}_{	-0.51	}$ & 	30.4	\\
W003	& 	0.12	$^{+	0.06	}_{	-0.06	}$ &  	2.09	$^{+	6.09	}_{	-0.59	}$ &  	0.79	(	18	) & 	3.8	$^{+	0.75	}_{	-0.68	}$ & 	30.7	\\
W006	& 	0.12	$^{+	0.10	}_{	-0.07	}$ &  	1.27	$^{+	0.41	}_{	-0.23	}$ &  	1.12	(	13	) & 	1.5	$^{+	0.43	}_{	-0.37	}$ & 	30.2	\\ \hline
\end{tabular}

\label{tab_expec}
\end{center}
\scriptsize Parameters from APEC with PHABS model fit. 
\end{table*}


 The hydrogen  column density obtained from the spectral fits varies from 0.4 x 10$^{21}$cm$^{-2}$ to 5.4 x 10$^{21}$cm$^{-2}$, with mean value N$_{H}$ = 1.8 $\pm$ 1.5 x 10$^{21}$cm$^{-2}$ that corresponds to an extinction A$_{V}$~=~0.9~$\pm$~0.7~mag, adopting N$_{H}$ = 2.1 $\times$ 10$^{21}$ A$_{V}$ cm$^2$ \citep[e.g.,][]{2003A&A...408..581V}.  
The coronal temperatures in Table \ref{tab_expec} vary from 0.7 to 2.1 keV, with average value of 1.4 $\pm$ 0.4 keV.


\begin{figure*}[ht]
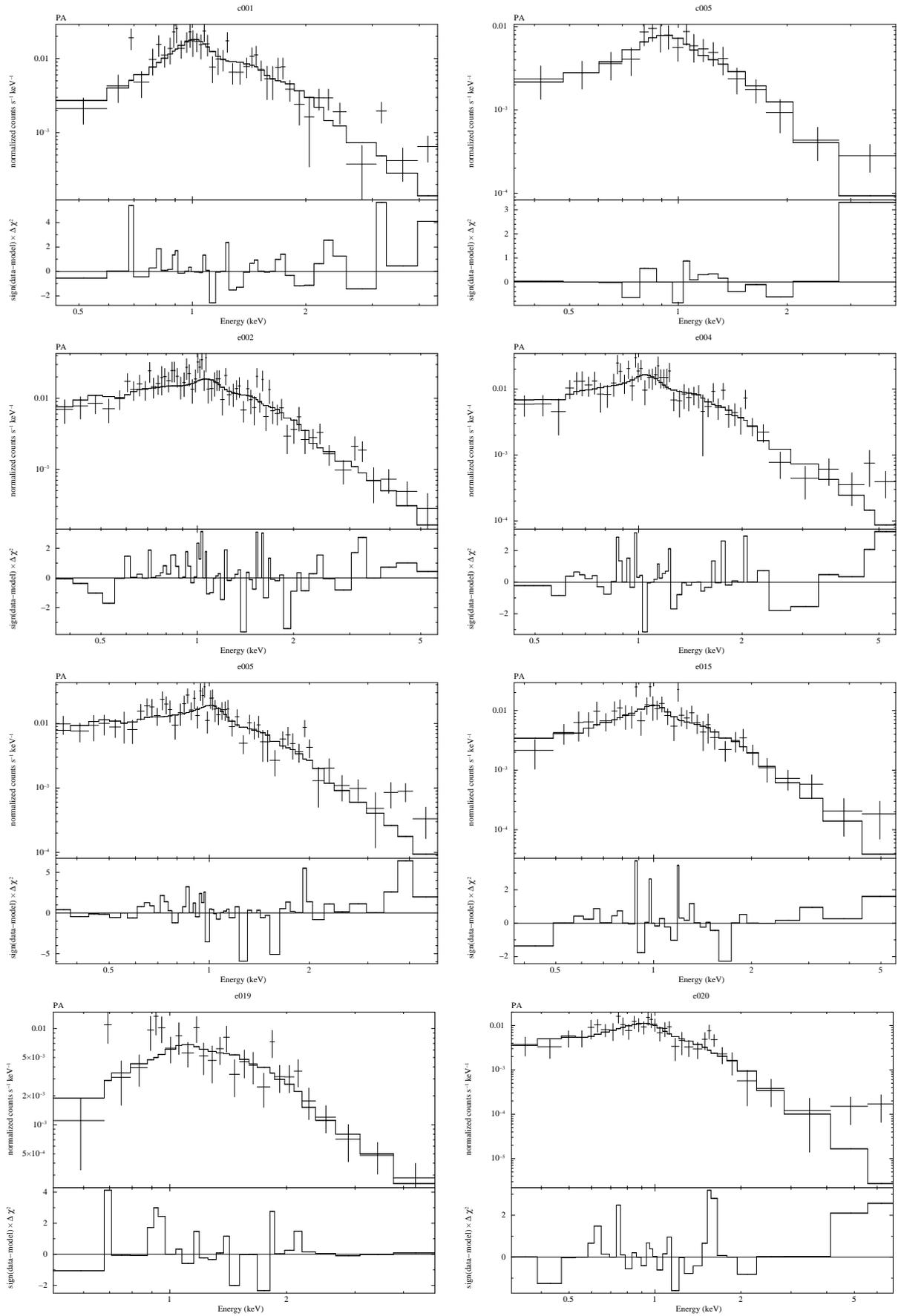

\begin{center}
\includegraphics[width=0.65\columnwidth, angle=270]{c001_PN_pa.eps}
\includegraphics[width=0.65\columnwidth, angle=270]{c005_PN_pa.eps}
\includegraphics[width=0.65\columnwidth, angle=270]{e002_PN_pa.eps}
\includegraphics[width=0.65\columnwidth, angle=270]{e004_PN_pa.eps}
\includegraphics[width=0.65\columnwidth, angle=270]{e005_PN_pa.eps}
\includegraphics[width=0.65\columnwidth, angle=270]{e015_PN_pa.eps}
\includegraphics[width=0.65\columnwidth, angle=270]{e019_PN_pa.eps}
\includegraphics[width=0.65\columnwidth, angle=270]{e020_PN_pa.eps}
{ \scriptsize
\caption{Best-fits and residual (see Sect. A.3) obtained for spectra of bright X-ray sources.
}
}\label{figAespec1}
\end{center}
\end{figure*}     


\begin{figure*}[ht]
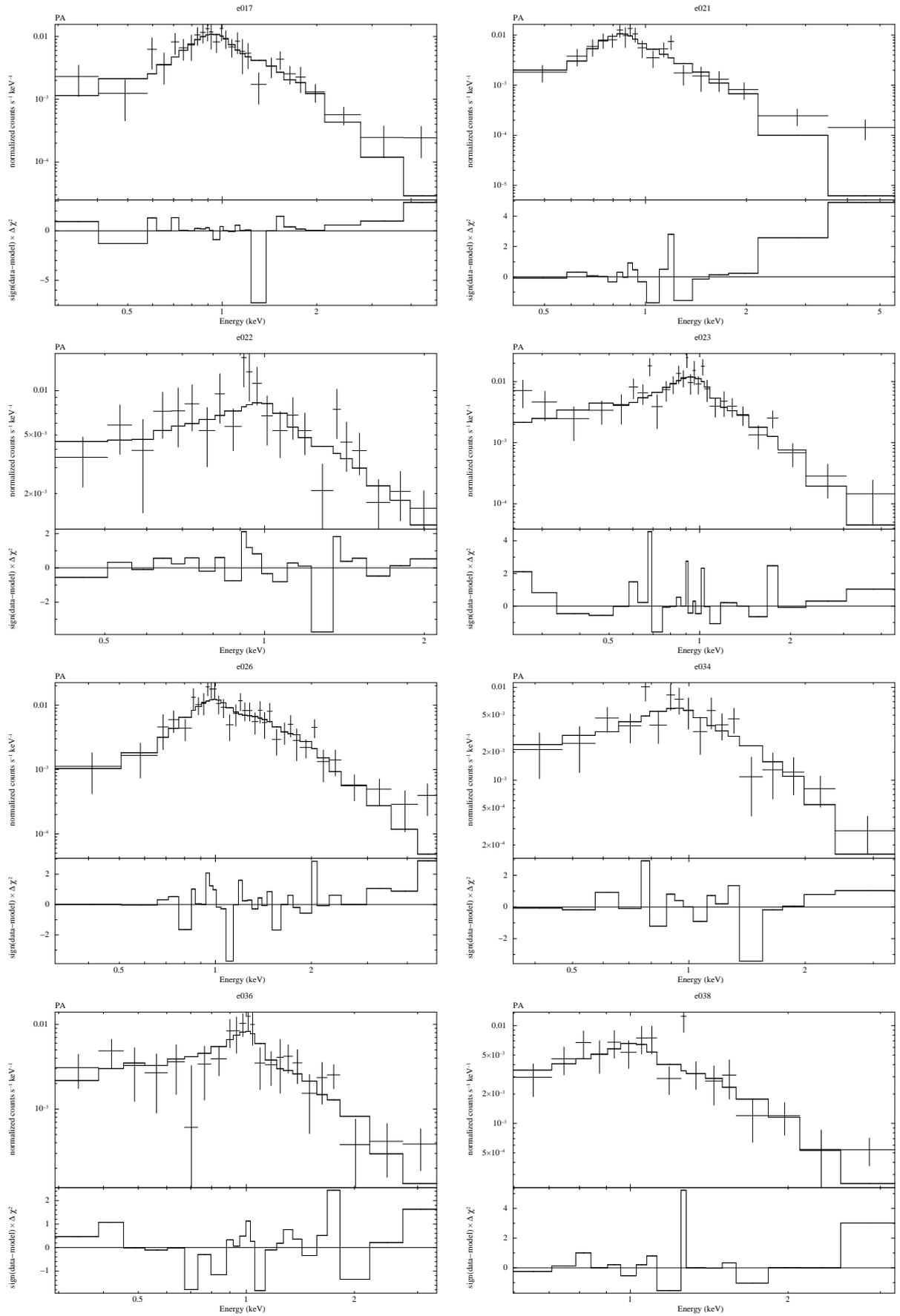

\begin{center}

\includegraphics[width=0.65\columnwidth, angle=270]{e017_PN_pa.eps}
\includegraphics[width=0.65\columnwidth, angle=270]{e021_PN_pa.eps}
\includegraphics[width=0.65\columnwidth, angle=270]{e022_PN_pa.eps}
\includegraphics[width=0.65\columnwidth, angle=270]{e023_PN_pa.eps}
\includegraphics[width=0.65\columnwidth, angle=270]{e026_PN_pa.eps}
\includegraphics[width=0.65\columnwidth, angle=270]{e034_PN_pa.eps}
\includegraphics[width=0.65\columnwidth, angle=270]{e036_PN_pa.eps}
\includegraphics[width=0.65\columnwidth, angle=270]{e038_PN_pa.eps}

{\scriptsize
\caption{The same as Fig. A.7.
}
}\label{figAespec2}
\end{center}
\end{figure*}     


\begin{figure*}[ht]
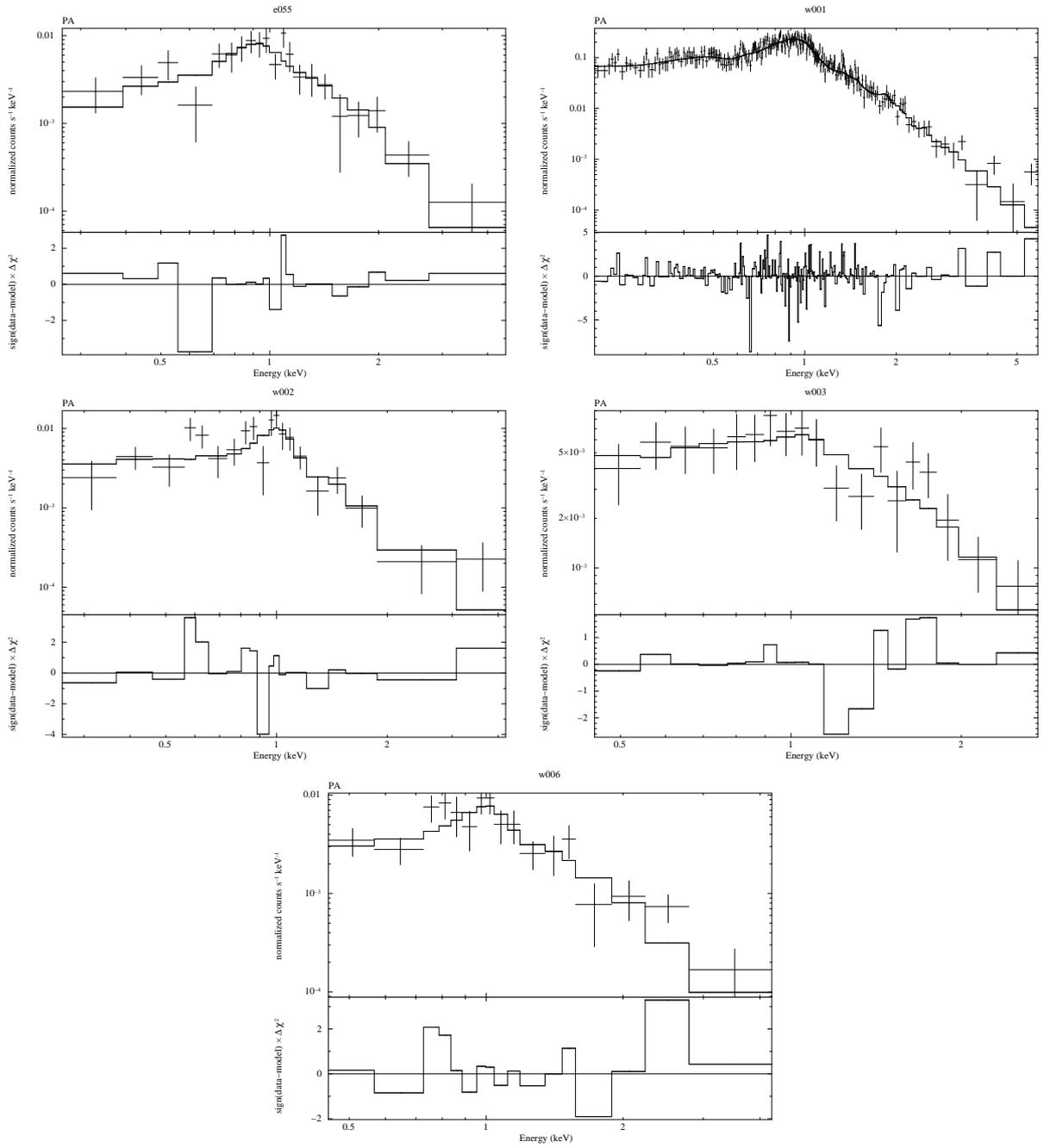

\begin{center}

\includegraphics[width=0.65\columnwidth, angle=270]{e055_PN_pa.eps}
\includegraphics[width=0.65\columnwidth, angle=270]{w001_PN_pa.eps}
\includegraphics[width=0.65\columnwidth, angle=270]{w002_PN_pa.eps}
\includegraphics[width=0.65\columnwidth, angle=270]{w003_PN_pa.eps}
\includegraphics[width=0.65\columnwidth, angle=270]{w006_PN_pa.eps}

{ \scriptsize
\caption{The same as Fig. A.7.
}
}\label{figAespec3}
\end{center}
\end{figure*}     

\onecolumn


\section{Catalogue of X-ray sources}


In this section we present all the 387 X-ray sources detected with EPIC cameras PN, MOS1 and MOS2  in tabular form (Table B.1). The infrared counterparts candidates of these sources (2MASS and {\it WISE}) are presented in Table B.2.



\begin{center}
\scriptsize
\begin{longtable}{|l|c|c|c|ccc|c|c|c|c|}

\caption{X-ray sources detected in the combined images of the EPIC cameras PN, MOS1 and MOS2.
{\scriptsize Column description: (1) X-ray source identification; (2, 3) J$_{2000}$ coordinates (4) Maximum likelihood; (5,6 and 7) Count Rates detected in each EPIC camera PN, MOS1 and MOS2 individually ; (8, 9, 10 and 11) PN hardness ratios.}}\\

\hline

\hline
ID & RA (J2000) & DEC (J2000) & ML &  & CR (cts/ks) & & HR1$_{1}^{(8)}$ & HR2$_{1}^{(9)}$ & HR1$_{2}^{(10)}$ & HR2$_{2}^{(11)}$\\
CMaX$^{(1)}$ & (2) & (3) & (4) & PN$^{(5)}$ & MOS1$^{(6)}$ & MOS2$^{(7)}$ & 0.5 -1.0-2.0   keV & 1.0-2.0-7.3   keV & 0.2 -0.5-2.0   keV & 0.5-2.0-4.5   keV\\ \hline\hline
\endfirsthead

\caption{Continued.
{\scriptsize Column description: (1) X-ray source identification; (2, 3) J$_{2000}$ coordinates (4) Maximum likelihood; (5,6 and 7) Count Rates detected in each EPIC camera PN, MOS1 and MOS2 individually ; (8, 9, 10 and 11) PN hardness ratios.}} \\
\hline
ID & RA (J2000) & DEC (J2000) & ML &  & CR (cts/ks) & & HR1$_{1}^{(8)}$ & HR2$_{1}^{(9)}$ & HR1$_{2}^{(10)}$ & HR2$_{2}^{(11)}$\\
CMaX$^{(1)}$ & (2) & (3) & (4) & PN$^{(5)}$ & MOS1$^{(6)}$ & MOS2$^{(7)}$ & 0.5 -1.0-2.0   keV & 1.0-2.0-7.3   keV & 0.2 -0.5-2.0   keV & 0.5-2.0-4.5   keV\\ \hline\hline
\hline
\endhead
\hline
\endfoot
\hline
\endfoot
W066 & 07 00 28.5 & -11 21 36.40 & 26 & - & 2.2 $\pm$ 0.6 & 2.7 $\pm$ 0.6 & - & - & - & -\\
W017 & 07 00 37.0 & -11 23 46.28 & 111 & 7.2 $\pm$ 1.0 & 2.8 $\pm$ 0.5 & 3.2 $\pm$ 0.6 & 0.65 $\pm$ 0.19 & 0.23 $\pm$ 0.14 & 1.00 $\pm$ 0.11 & 0.01 $\pm$ 0.15\\
W001 & 07 00 37.4 & -11 14 45.07 & 18.701 & 412.1 $\pm$ 7.1 & 117.6 $\pm$ 3.0 & 117.7 $\pm$ 3.1 & -0.28 $\pm$ 0.02 & -0.75 $\pm$ 0.02 & 0.63 $\pm$ 0.01 & -0.91 $\pm$ 0.01\\
W040 & 07 00 40.2 & -11 13 48.51 & 17 & 4.6 $\pm$ 1.0 & - & - & -0.13 $\pm$ 0.22 & -0.46 $\pm$ 0.33 & - & -\\
W002 & 07 00 43.3 & -11 17 15.10 & 443 & 16.8 $\pm$ 1.3 & 4.3 $\pm$ 0.6 & 4.4 $\pm$ 0.6 & -0.25 $\pm$ 0.07 & -0.76 $\pm$ 0.12 & 0.82 $\pm$ 0.05 & -0.85 $\pm$ 0.05\\
W048 & 07 00 47.1 & -11 22 02.30 & 35 & 2.4 $\pm$ 0.5 & - & - & -0.52 $\pm$ 0.21 & -1.00 $\pm$ 0.50 & 0.46 $\pm$ 0.17 & -0.95 $\pm$ 0.11\\
W042 & 07 00 48.3 & -11 12 37.90 & 35 & 3.9 $\pm$ 0.8 & - & - & 0.03 $\pm$ 0.21 & -0.56 $\pm$ 0.30 & 0.48 $\pm$ 0.17 & -0.85 $\pm$ 0.15\\
W078 & 07 00 55.7 & -11 26 24.39 & 17 & - & - & 1.2 $\pm$ 0.3 & - & - & - & -\\
\hline \hline

\end{longtable}
\end{center}
 \label{tab_sources}



\newpage

\begin{center}
\scriptsize

\begin{longtable}{|l|c|c|c|c|c|c|c|c|c|c|c|c|c|}

\caption{  Infrared {\it 2MASS} photometry, mass, age and infrared classification of members of CMa R1 region and counterparts to X-ray sources. A distance of 1 kpc is assumed for them. 
  {\scriptsize Column description: (1) X-ray source identification (Table B.1); (2) Distance from X-ray source center; (3) {\it 2MASS} identification (4, 5 and 6) J, H and K magnitudes; (7) {\it 2MASS} Quality flag; (8) Mass; (9) Age; (10) {\it WISE} classification. Note:(*) Counterpart candidate not considered, due to distance incompatible with the cloud (see Sect. 4.1).}}\\
\hline

ID$^{(1)}$ & D & ID$^{(3)}$ & J$^{(4)}$ & $\sigma_{J}$ & H$^{(5)}$ & $\sigma_{H}$ & K$^{(6)}$ & $\sigma_{K}$ & Qflg$^{(7)}$ & Mass$^{(8)}$ & Age$^{(9)}$ & {\it WISE}$^{(10)}$ \\                                 
CMaX & (") & {\it 2MASS} & mag & mag & mag & mag & mag & mag  &  & M$_{\odot}$ &  Myr &  \\ \hline \hline

\hline\hline
\endfirsthead

\caption{Continued. {\scriptsize Column description: (1) X-ray source identification (Table B.1); (2) Distance from X-ray source center; (3) {\it 2MASS} identification (4, 5 and 6) J, H and K magnitudes; (7) {\it 2MASS} Quality flag; (8) Mass; (9) Age; (10) {\it WISE} classification.Note:(*) Counterpart candidate not considered, due to distance incompatible with the cloud (see Sect. 4.1).}} \\
\hline
ID$^{(1)}$ & D & ID$^{(3)}$ & J$^{(4)}$ & $\sigma_{J}$ & H$^{(5)}$ & $\sigma_{H}$ & K$^{(6)}$ & $\sigma_{K}$ & Qflg$^{(7)}$ & Mass$^{(8)}$ & Age$^{(9)}$ & {\it WISE}$^{(10)}$ \\                                 
CMaX & (") & {\it 2MASS} & mag & mag & mag & mag & mag & mag  &  & M$_{\odot}$ &  Myr &  \\ \hline \hline    

\hline
\endhead
\hline
\endfoot
\hline
\endlastfoot

W066	&	--	&	--	&	--	&	--	&	--	&	--	&	--	&	--	&	--	&	--	&	--	&	--	\\	
W017*	&	9	&	07003741-1123529		& 13.554 &	0.024 &	13.231 & 	0.026 &	13.211 & 	0.035 &	AAA 	&	--	&	--	&	--	\\	
W001	&	2	&	07003760-1114451	&	6.575	&	0.023	&	6.032	&	0.049	&	5.905	&	0.023	&	AAA	&	??	&	??	&	Class III	\\	
W040	&	--	&	--	&	--	&	--	&	--	&	--	&	--	&	--	&	--	&	--	&	--	&	--	\\	
W002	&	3	&	07004336-1117153	&	12.883	&	0.026	&	12.312	&	0.024	&	12.24	&	0.024	&	AAA	&	1.45	&	12	&	??	\\	
W048	&	2	&	07004705-1122044	&	15.438	&	0.048	&	14.799	&	0.083	&	14.385	&	0.079	&	AAA	&	0.28	&	4.5	&	Class III	\\	
W042	&	4	&	07004851-1112380	&	13.901	&	0.03	&	13.225	&	0.037	&	13.114	&	0.041	&	AAA	&	1.18	&	14	&	Class III	\\	
W078	&	3	&	07005585-1126246	&	13.542	&	0.024	&	12.826	&	0.034	&	12.58	&	0.026	&	AAA	&	0.7	&	1.8	&	Class III	\\
\hline

\end{longtable}
\end{center}
\label{2masources}


\end{appendix}

\end{document}